\documentclass[a4paper,11pt]{article}

\pdfoutput=1
\usepackage{comment}
\usepackage{fixltx2e}
\usepackage[table,xcdraw]{xcolor}
\usepackage{jcappub} 
\usepackage{float} 
\usepackage{graphicx} 
\usepackage[scr=boondox]{mathalfa}
\usepackage{bm}
\usepackage{stmaryrd}
\usepackage{bbold}
\usepackage[normalem]{ulem}
\usepackage{amsmath}
\usepackage{bbm}
\usepackage{changepage}
\usepackage{geometry}
\usepackage{cancel}
\usepackage{xcolor}

\def\be{\begin{equation}}
\def\ee{\end{equation}}
\def\bea{\begin{eqnarray}}
\def\eea{\end{eqnarray}}
\newcommand{\vs}{\nonumber\\}
\def\ba#1\ea{\begin{align}#1\end{align}}
\def\bg#1\eg{\begin{gather}#1\end{gather}}

\newcommand{\reffig}[1]{Fig.~\ref{fig:#1}}

\newcommand{\reftab}[1]{Tab.~\ref{tab:#1}}          
\newcommand{\refsec}[1]{Sec.~\ref{sec:#1}}          
\newcommand{\refapp}[1]{App.~\ref{app:#1}}

\newcommand{\AB}{LG-scheme}

\newcommand*{\non}  {\nonumber}

\def\O{\mathcal{O}}

\newcommand{\ea}{\end{aligned}\]}

\newcommand{\nn}{\nonumber}

\renewcommand{\emph}[1]{\textit{#1}}

\def\O{\mathcal{O}}

\newcommand{\perm}[1]{ \expandafter\ifstrempty\expandafter{#1} {\mbox{perm.}} {\mbox{$#1$ perm.}} }

\makeatletter
\newlength{\apb@width}
\newcommand{\autoparbox}[2][c]{\settowidth{\apb@width}{#2}\parbox[#1]{\apb@width}{#2}}

\makeatother

\usepackage[T1]{fontenc}
\usepackage{graphbox}
\usepackage{subfig}
\usepackage{booktabs}
\subheader{
TUM-HEP-1608/26\\
RBI-ThPhys-2026-23}
\title{
Galaxy bias renormalization: Two-loop Power Spectrum, One-loop Trispectrum and Bispectrum
}

\usepackage{amsmath}
\author[a]{Thomas Bakx,}
\author[b]{Mathias Garny,}
\author[c,d,e]{Henrique Rubira,}
\author[f]{Zvonimir Vlah}
\affiliation[a]{Institute for Theoretical Physics, Utrecht University, Princetonplein 5, 3584 CC, Utrecht, The Netherlands}
\affiliation[b]{Physik Department T31, School of Natural Sciences, Technische Universit\"at M\"unchen,\\
James-Franck-Stra{\ss}e 1, D-85748 Garching, Germany}

\affiliation[c]{University Observatory, Faculty of Physics, Ludwig-Maximilians-Universit\"at, Scheinerstr. 1, D-81679 München, Germany}
\affiliation[d]{Kavli Institute for Cosmology Cambridge, Madingley Road, Cambridge CB3 0HA, UK}
\affiliation[e]{Centre for Theoretical Cosmology, Department of Applied Mathematics and Theoretical Physics
University of Cambridge, Wilberforce Road, Cambridge, CB3 0WA, UK}
\affiliation[f]{Division of Theoretical Physics, Ru\dj er Bo\u skovi\' c Institute, 10000 Zagreb, Croatia,}%

\abstract{We present a complete treatment of fifth-order renormalized galaxy bias at the one- and two-loop level in real space, including gradient corrections to deterministic bias operators at next-to-leading order. We then provide a complete computation of the two-loop power spectrum as well as the one-loop bispectrum and trispectrum of biased tracers, and demonstrate how to jointly model these statistics in a fully renormalized framework. These statistics also require stochastic renormalization of products of two, three or four operators at coincidence, which we include at leading order in gradients by means of an operator product expansion. We verify that all UV limits of loop integrals are absorbed by the counterterms we consider. Upon solving the resulting renormalization group equations, we find a pronounced scale-dependence of higher-gradient bias coefficients.  Since our renormalization prescription is performed manifestly at the operator level, our results can also easily be extended to higher $N$-point functions, higher loop orders and field-level analyses. 
} 


\begin{document}

\maketitle
\flushbottom

\section{Introduction}\label{sec:intro}

The next generation of galaxy surveys such as DESI, Euclid, the Vera C. Rubin Observatory (LSST), Roman and SPHEREx map the large-scale structure of the Universe with unprecedented statistical precision \cite{DESI:2024jxi,Euclid:2023tog,SPHEREx:2014bgr,akeson2019widefieldinfraredsurvey,LSST:2008ijt}. Exploiting these datasets to constrain cosmological parameters at the percent level requires equally precise theoretical models of galaxy clustering. Perturbation theory (PT) within the effective field theory (EFT) approach to large-scale structure (LSS) formation provides a systematic framework for computing the statistics of biased tracers, such as galaxies and halos, in terms of an expansion in the linear density field and its spatial derivatives, while encoding the effects of small-scale, nonlinear physics through a finite set of bias and counterterm coefficients \cite{Kaiser:1984sw, Fry:1992vr, McDonald:2009dh, Matsubara:2011ck, Chan:2012jj, Mirbabayi:2014zca, Senatore:2014eva, Eggemeier:2018qae}, see~\cite{Desjacques:2016bnm} for a review.

In the EFT approach, the overdensity of biased tracers is written as a sum over scalar operators built from the tidal tensor and its convective time derivatives, organized in a double expansion: perturbative order in the linear density contrast and number of spatial gradients. The coefficients of these operators are not predicted from first principles but must either be measured from simulations or marginalized over when comparing to data.  A central challenge is that naive perturbative loop integrals are sensitive to UV modes beyond the validity of the perturbative expansion, and a systematic renormalization procedure is required to absorb these UV divergences into redefinitions of the bias coefficients and to render predictions for observable correlators finite and well-defined.

A systematic renormalization scheme for biased tracers was introduced in~\cite{McDonald:2006mx,Assassi:2014fva} (see also~\cite{Eggemeier:2018qae} for a complementary approach), determined by  renormalization conditions requiring that renormalized bias parameters match the tree-level  prediction in the large-scale limit of $N$-point functions. This condition ensures consistency  with the separate-universe prediction~\cite{Lazeyras:2015lgp,Baldauf:2015vio,Li:2015jsz} and provides a direct physical interpretation for the renormalized bias coefficients. It furthermore ensures a consistent formulation applicable to analyses combining various summary statistics such as the power- and bispectrum, or even at the field level. This framework was systematically extended in \cite{Bakx:2025cvu} (hereafter \hyperlink{cite.Bakx:2025cvu}{Paper~I}) to include the complete non-redundant basis of leading-gradient (LG) scalar bias operators up to fifth perturbative order, comprising 29 operators. Paper~I derived the full one-loop and two-loop   renormalization matrices for LG operators, established commutation relations between the double- and single-hard limits of bias kernels, and derived the coupled renormalization group equations (RGEs) governing the dependence on the UV-cutoff, as introduced at one-loop level in \cite{Rubira:2023vzw}. Paper~I furthermore covers stochastic noise contributions in terms of contact terms arising from renormalization of products of  bias operators (see also \cite{Assassi:2014fva,Rubira:2024tea}). 

However, two-loop predictions for the galaxy power spectrum require not only the renormalization of LG operators, but also the inclusion of \emph{next-to-leading-gradient} (NLG) operators—those carrying two additional spatial derivatives, which are suppressed by $(kR_L)^2$ relative to the leading-gradient contributions, where $R_L$ is expected to be of the order of the Lagrangian radius of the tracer.  At two-loop order, UV-sensitive contributions from LG operators mix with NLG operators, and a complete renormalization requires the full set of second-gradient counterterms. Similarly, the computation of the renormalized one-loop bispectrum and trispectrum calls for a systematic treatment of second-gradient contributions. This extension is the main subject of the present paper.

In this work, we generalize the renormalization scheme from \cite{McDonald:2006mx, Assassi:2014fva} to incorporate the full basis of NLG bias operators at second gradient order. We derive the complete block structure of the one and two-loop renormalization matrix and establish new commutation relations connecting the double-hard and single-hard limits of the full fifth-order bias basis. We then construct the renormalized two-loop power spectrum, one-loop bispectrum, and one-loop trispectrum, identifying which NLG operators yield distinct, non-degenerate contributions at each loop order. The computation of the renormalized bispectrum and trispectrum also requires us to carry out a generalization of the renormalization prescription for contact operators introduced in \hyperlink{cite.Bakx:2025cvu}{Paper~I} to products of three or four operators at coincidence. We renormalize these products only at leading order in gradients, leaving the extension to second-gradient stochastic contributions to future work. Finally, we derive the full RGEs governing the coupled running of LG and NLG bare bias coefficients in the deterministic sector, and solve them at one- and two-loop order.

This paper is organized as follows. In \refsec{basis} we construct the non-redundant basis of NLG scalar bias operators at second gradient order. \refsec{renorm} develops the generalized renormalization scheme and derives the one- and two-loop renormalization matrices, commutation-of-limit relations (generalizing \hyperlink{cite.Bakx:2025cvu}{Paper~I}), and renormalized operator expressions. \refsec{stochastic} focuses on the renormalization of stochastic bias terms. In \refsec{correlators} we translate the renormalized operator expressions into renormalized two-loop power spectra and one-loop bispectra and trispectra. \refsec{RGE} derives the renormalization group equations for the full LG+NLG operator basis and discusses their solutions. We conclude in \refsec{conclusion}. Several appendices contain supplementary material: 
a derivation of the NLG bias basis (\refapp{nlg_basis}), 
commutation of sequential loop limits (\refapp{consistency}), and extended tables of renormalization coefficients (\refapp{tables}).

\newpage
\section{Basis of operators at second gradient order} \label{sec:basis}

Here we list the complete non-redundant basis of bias operators at second gradient order and third perturbative order and address the relation to previous literature. Bias operators can be classified in terms of a double expansion (e.g.~\cite{Sheth:2012fc,Desjacques:2016bnm}):
\begin{itemize}
\item Perturbative expansion in $\delta_L$: Minimal order $n$ in the expansion in the initial linear density contrast $\delta_L(\bm  k)$ at which they are non-zero.
\item Gradient or power expansion: Number of (extra) spatial gradients, leading to a parametric suppression in powers of $kR_L$. 
\end{itemize}
Throughout this work, we adopt the familiar power counting rule where one power of $kR_L$ counts as one power of $\delta_L$, so that we require second-gradient operators at most at third order in perturbations. We refer to operators with no derivatives as {\it leading order in gradients} (LG), and those with two extra gradients as {\it next-to-leading in gradients} (NLG). It is convenient to consider separately the leading and subleading gradient operators, denoted by  ${\cal O}_{a_m}^{[n]}$ at $m$-th gradient order. The superscript denotes the lowest order $n$ in \textit{the perturbative expansion in $\delta_L$} where a given operator contributes. When omitting the superscript, the equations apply for any perturbative order. For brevity, we also often omit the subscript ${\cal O}_{a_0}\equiv {\cal O}_a$
for operators at leading gradient order. The following table thus summarizes our notation:
\be
\begin{array}{l|ccccc}
 & \delta_L^{n\geq 1} &\delta_L^{n\geq 2} &\delta_L^{n\geq 3} &\delta_L^{n\geq 4} &\delta_L^{n\geq 5} \\
\text{LG} & {\cal O}_a^{[1]} & {\cal O}_a^{[2]} & {\cal O}_a^{[3]} & {\cal O}_a^{[4]} & {\cal O}_a^{[5]} \\
\text{NLG} & {\cal O}_{a_2}^{[1]} & {\cal O}_{a_2}^{[2]} & {\cal O}_{a_2}^{[3]} &  
\end{array}
\ee
When referring to the `order in perturbation theory' we mean the order in $\delta_L$. The $\nabla^4\delta_L$ ({\it next-to-next-to-leading in gradients}, NNLG) operator has to be included when considering the two-loop power spectra. The complete non-redundant set of LG scalar bias operators comprises $1+2+4+8+14=29$ operators up to order $5$. We use the basis from~\cite{Mirbabayi:2014zca,Desjacques:2016bnm,Schmidt:2020tao,Bakx:2025cvu} (see also~\cite{Sheth:2012fc,Eggemeier:2018qae,Donath:2023sav}, and \cite{Fujita:2020xtd,DAmico:2021rdb,Ansari:2025nsf} for a bootstrap determination of the bias basis), that is constructed from the building blocks $\Pi^{[n]}_{ij}$ defined recursively via~\cite{Mirbabayi:2014zca} 
\be\label{eq:Pindef}
  \Pi^{[n]}_{ij} \equiv \frac{1}{(n-1)!}(D_\eta-(n-1))\Pi^{[n-1]}_{ij},
  \qquad \Pi^{[1]}_{ij} \equiv \frac{\nabla_i\nabla_j}{\nabla^2}\delta\,,
\ee
in terms of convective time derivatives $D_\eta=\partial_\eta-\left(\frac{\nabla_i\theta}{\nabla^2}\right)\nabla_i$ acting on the tidal tensor $\Pi^{[1]}_{ij}$. Here $\eta\equiv \ln(D(z))$, $D(z)$ is the linear growth factor, $\delta$ the non-linear density field, and $\theta$ the non-linear velocity divergence, normalized such that $\theta_L=\delta_L$ in linear approximation.
Throughout this work, as in \hyperlink{cite.Bakx:2025cvu}{Paper~I}, we assume an Einstein - de Sitter (EdS) time dependence, where an $n$-th order contribution in perturbation theory has time dependence $D^n$ where $D$ is the linear growing mode. In this case, the basis from \hyperlink{cite.Bakx:2025cvu}{Paper~I} is complete at leading order in gradients. If one relaxes this assumption, more terms would be needed at fourth and fifth order, see App.~B of \cite{Desjacques:2016bnm}. Using the perturbative expansion of $\delta$ and $\theta$ in terms of the usual EdS kernels $F_n({\bm q}_1,\dots,{\bm q}_n)$ and $G_n({\bm q}_1,\dots,{\bm q}_n)$~\cite{Bernardeau:2001qr} any operator ${\cal O}_A^{[n]}$ can be expanded perturbatively in Fourier space in terms of its kernels $K_A^{(m)}({\bm q}_1,\dots,{\bm q}_m)$ for $m\geq n$ as
\be
  {\cal O}_A^{[n]}({\bm k}) = \sum_{m=n}^\infty\int_{q_1\cdots q_m}(2\pi)^3\delta_D({\bm k}-\sum_{i=1}^m{\bm q}_i)\, D^m K_A^{(m)}({\bm q}_1,\dots,{\bm q}_m)\delta_L({\bm q}_1)\cdots\delta_L({\bm q}_m)\,,
\ee
where we use a notation for which a capital index $A$ encompasses both LG ($a$) and NLG ($a_2$) bias operators, and $\int_{q_1\cdots q_m}\equiv \int d^3q_1/(2\pi)^3 \cdots d^3q_m/(2\pi)^3$.

Consider bias renormalization at higher order in gradient expansion. There are no scalar operators with a single gradient due to rotational invariance.
Thus the first subleading order of the derivative expansion for the galaxy density bias expansion starts at operators suppressed by {\it two} gradients. 
We schematically write the deterministic bias expansion up to second gradient order as 
\be
  \delta_g = \sum b_{a_0}{\cal O}_{a_0} + 
  \sum b_{a_2}{\cal O}_{a_2}\,,
\ee
where the subscript on the index refers to the leading ($a_0$) and next-to-leading ($a_2$) order in gradient expansion.
The second-gradient contributions $b_{a_2}$ have the dual purpose of canceling the UV-dependent parts of loop integrals via counterterms as well as encoding physical non-locality effects of order $(kR_L)^2$.

Naively, we expect a generic structure of higher-derivative operators ${\cal O}_{a_2}$ of the form~\cite{Desjacques:2016bnm}
\bea\label{eq:NLGgeneric}
&&  (\nabla_k\Pi^{[n_1]}_{i_1j_1})\times (\nabla_l\Pi^{[n_2]}_{i_2j_2})\times \Pi^{[n_3]}_{i_3j_3}\times \Pi^{[n_4]}_{i_4j_4}\cdots\nn\,,\\
&&  (\nabla_k\nabla_l\Pi^{[n_1]}_{i_1j_1})\times \Pi^{[n_2]}_{i_2j_2}\times \Pi^{[n_3]}_{i_3j_3}\times \Pi^{[n_4]}_{i_4j_4}\cdots\,,
\eea
with indices contracted in all possible ways. Here gradients act only on the expression inside the round brackets.
As we will see below, some additional types of operators need to be considered eventually. Let us start by classifying all operators of the
form above. We first consider a subset of second gradient operators for which both gradients act on the entire operator, of the form
\be\label{eq:NLGoverall}
  \nabla_k\nabla_l\left[\Pi^{[n_1]}_{i_1j_1}\times \Pi^{[n_2]}_{i_2j_2}\times \Pi^{[n_3]}_{i_3j_3}\times \Pi^{[n_4]}_{i_4j_4}\cdots\right] \qquad \text{(overall-gradient operators)}
\ee
Here the two gradients act on the entire expression inside the square brackets. Due to rotational symmetry, the square bracket must correspond to a tensor-bias operator ${\cal O}_{kl}$ at leading gradient order. However, we find that not all linearly independent tensor operators ${\cal O}_{kl}$ (at a given order in perturbation theory) are needed to obtain a complete linearly independent basis of
overall-gradient operators $\nabla_k\nabla_l{\cal O}_{kl}$. This can for example be easily seen for the first order, where two independent tensor operators $\mbox{TF}(\Pi^{[1]}_{kl})$ and $\delta_{kl}^K\mbox{tr}(\Pi^{[1]})$ exist, but $\nabla_k\nabla_l\mbox{TF}(\Pi^{[1]}_{kl})=\frac23\nabla_k\nabla_l\delta_{kl}^K\mbox{tr}(\Pi^{[1]})$. Here TF denotes the trace-free part and $\delta^K_{ij}$ the Kronecker symbol. Taking such relations into account (which, unlike in the example above, do not necessarily hold in general  at all orders in perturbation theory - we comment more on this below), we find the
following non-redundant basis of $1 + 3 + 8 = 12$ overall-gradient operators up to third order (see also \cite{Bertolini:2015fya,Bertolini:2016bmt}) in perturbations:
\bea\label{eq:NLGoverallgradients}
\text{1st} && \nabla^2\mbox{tr}(\Pi^{[1]})\nn\\
\text{2nd} && \nabla^2\mbox{tr}[(\Pi^{[1]})^2],\quad \nabla^2[\mbox{tr}(\Pi^{[1]})]^2,\quad \nabla_k\nabla_l[\Pi^{[1]}_{kl}\mbox{tr}(\Pi^{[1]})]\nn\\
\text{3rd} && \nabla^2[\mbox{tr}(\Pi^{[1]})]^3,\quad \nabla^2[\mbox{tr}[(\Pi^{[1]})^2]\mbox{tr}(\Pi^{[1]})],\quad  \nabla^2\mbox{tr}[(\Pi^{[1]})^3], \nn\\
           && \nabla_k\nabla_l[\Pi^{[1]}_{kl}\mbox{tr}[(\Pi^{[1]})^2]],\quad
              \nabla_k\nabla_l[\Pi^{[1]}_{kl}[\mbox{tr}(\Pi^{[1]})]^2],\quad
              \nabla_k\nabla_l[(\Pi^{[1]}\Pi^{[1]})_{kl}\mbox{tr}(\Pi^{[1]})],\nn\\
           && \nabla^2\mbox{tr}[\Pi^{[1]}\Pi^{[2]}],\quad
              \nabla_k\nabla_l[(\Pi^{[1]}\Pi^{[2]})_{kl}] \,.
\eea
The first one corresponds to the well-known higher-derivative bias, or `$c_s^2$ counterterm' in the context of EFT corrections to the density field.
The second-order operators correspond to the set of three additional bispectrum counterterms for the matter density bispectrum within the EFT~\cite{Baldauf:2014qfa}, in addition to the
second-order contribution obtained from evaluating the first line at second order in perturbations (see e.g. Eq.~17 in~\cite{Baldauf:2021zlt} and Eq.~56 in~\cite{Garny:2022kbk}).\footnote{The choice of operators from Eq.~17 in~\cite{Baldauf:2021zlt} is equivalent, which can be seen using $\nabla_k[\Pi^{[1]}_{kl}\nabla_l\mbox{tr}(\Pi^{[1]})] = \nabla_k\nabla_l[\Pi^{[1]}_{kl}\mbox{tr}(\Pi^{[1]})]-\frac12\nabla^2[\mbox{tr}(\Pi^{[1]})]^2$. This follows using $\nabla_l\Pi^{[1]}_{kl}=\nabla_k\mbox{tr}(\Pi^{[1]})$ for $\Pi^{[1]}$.}

Next, there could be operators for which one of the two gradients is an overall derivative acting on the entire operator, and the other one acts only on a subset of its building blocks. We find that those are all redundant up to third order.

Finally, there are second-gradient operators without any total, overall derivative, starting at second order in perturbations. These operators are only required for biased tracers. They are not needed for renormalizing the density field, due to momentum conservation.  We find the following non-redundant set of $1 + 4 = 5$ operators up to third order:
\bea\label{eq:NLGnonoverallgradients}
\text{2nd} && \mbox{tr}(\Pi^{[1]})\nabla^2\mbox{tr}(\Pi^{[1]})\nn\\
\text{3rd} && [\mbox{tr}(\Pi^{[1]})]^2\nabla^2\mbox{tr}(\Pi^{[1]}),\quad \mbox{tr}(\Pi^{[1]})\mbox{tr}[\Pi^{[1]}\nabla^2\Pi^{[1]}],\quad [\nabla_k\Pi^{[1]}_{kl}]\times \mbox{tr}[\Pi^{[1]}\nabla_l\Pi^{[1]}],\nn\\
           && \mbox{tr}[\Pi^{[2]}\nabla^2\Pi^{[1]}]  \,.
\eea
This type of operators starts at second order, and the single additional one at second order is consistent with the number of higher-derivative bias terms for the galaxy bispectrum from
Table I in~\cite{Eggemeier:2018qae} (specifically, their operator $[\nabla_k\mbox{tr}(\Pi^{[1]})]\times[\nabla_k\mbox{tr}(\Pi^{[1]})]=-\mbox{tr}(\Pi^{[1]})\nabla^2\mbox{tr}(\Pi^{[1]}) + \nabla^2[\mbox{tr}(\Pi^{[1]})]^2/2$ is equivalent to the basis chosen here). The second-order basis from~\cite{Eggemeier:2018qae} has also been used in~\cite{Philcox:2022frc} and \cite{DAmico:2022ukl}. 

Up to third order, we checked that the basis given above exhausts all possible operators of the form of Eq.~\eqref{eq:NLGgeneric}. While there are many more possibilities to write down some second-gradient operators at third order, they are either found to be expressible in terms of the ones given above, or redundant when evaluating them at third order in perturbation theory. We checked the latter statement by computing the third-order kernels and evaluating them with three generic wavenumber arguments ${\bf k}_1,{\bf k}_2,{\bf k}_3$. We note that there exist additional operators that start at second or third order, and that are redundant at up to third order, but could potentially be non-redundant at fourth and higher order. We checked that 
$\nabla_k\nabla_l\Pi^{[n]}_{kl}$ is redundant in general, but also find more relations showing that e.g. $\nabla_k\nabla_l[\Pi^{[2]}_{kl}\mbox{tr}(\Pi^{[1]})]$ is redundant at third order in perturbations. For some explicit relations of this form we refer to App. \ref{app:nlg_basis}. 

For renormalizing the two-loop power spectrum, only a subset of the above $12 + 5 = 17$ operators is required, since we only require the (angle-averaged) second-gradient operators in the configuration $\bm k, \bm p, -\bm p$. We defer a discussion of these redundancies to \refsec{correlators}. 

Nevertheless, the basis of these $17$ operators, which comes from the `naive' ansatz in Eq. \eqref{eq:NLGgeneric} is \textit{not sufficient} to renormalize the second-gradient parts of the UV limits of the
fifth-order kernels $K_{a_0}({\bf k}_1,{\bf k}_2,{\bf k}_3,{\bf p},-{\bf p})$ of the 29 bias operators ${\cal O}_{a_0}$,
i.e.~the $1/p^2$ terms in the limit of those kernels for large $p=|{\bf p}|$ (averaged over the direction of ${\bf p}$) cannot in general be written as a linear combination of the third-order kernels $K_{a_2}({\bf k}_1,{\bf k}_2,{\bf k}_3)$ for the set of operators ${\cal O}_{a_2}$ given above. We will introduce this renormalization procedure in detail in \refsec{renorm}. We find that this is true in particular for ${\cal O}_{a_0}=\delta$ and the operators $\mbox{tr}(\Pi^{[1]}\Pi^{[n]})$ for $n=1,2,3,4$, while for all other 24 bias operators ${\cal O}_{a_0}$ the ${\cal O}_{a_2}$ basis given above is sufficient to renormalize their $1/p^2$ part.

For the density field, the `missing' terms can be traced back to contributions that are generated from `back-reaction' of the counterterm-correction to the density field on the tidal tensor, and the corresponding induced density-correction at higher order~\cite{Anastasiou:2025jsy} (see App.~\ref{app:nlg_basis} for details). 

For a general biased tracer, the EFT operator basis should follow from allowing for the most general terms admitted by the assumed symmetries. While providing such a general analysis goes beyond the scope of the present work, we expect that NLG operators can, like those at LG order, be assembled from a given set of `building blocks', possibly with an additional class of such building blocks being complemented to the `standard' building blocks $\Pi^{[n]}_{ij}$ when going from LG to NLG order in the gradient expansion. This is similar to what occurs in other known EFTs. For example, within Soft-Collinear Effective Theory, the analog role of NLG bias operators is so-called power-suppressed operators, for which a classification purely within the EFT is possible~\cite{Beneke:2017ztn}. In this case, additional building blocks  occur for `power suppressed' EFT operators, being related to non-locality along the light-cone. 

Leaving an analog investigation for the EFT description of biased tracers for future work (possibly with some ramifications related to non-locality in time), we adopt in the following a working hypothesis about which additional building blocks are needed at NLG order. We then show that adopting this hypothesis for the moment yields a basis of NLG EFT operators that is sufficient to consistently renormalize the deterministic sector of the one-loop bi- and trispectrum for general shapes of wavenumber configurations as well as the two-loop power spectrum of a general biased tracer.
Specifically, in the construction of general NLG operators we allow for additional building blocks  of the form\footnote{We checked that allowing the more general set of building blocks $\frac{\nabla_k\nabla_l}{\nabla^2} {\cal O}^{[n]}_{kl}$ with any LG tensor bias operators ${\cal O}^{[n]}_{kl}$ does not lead to any differences in the resulting NLG operator basis at up to third order in perturbation theory.}
\be\label{eq:extrabuildingblock}
  \frac{\nabla_k\nabla_l}{\nabla^2} \Pi^{[n]}_{kl} \,.
\ee
These may occur inside NLG operators, i.e.~featuring two additional gradients as compared to LG order, according to our hypothesis. Thus, they are {\it not} required to construct the complete set of LG EFT operators, for which the `standard' building blocks $\Pi^{[n]}_{ij}$ suffice.

In the following, we show that this extra building block is all that is needed to construct a closed basis of NLG deterministic bias operators applicable to general biased tracers sufficient for describing the combined two-loop power spectrum, one-loop bispectrum, and one-loop trispectrum within the EFT framework.
We stress again that this hypothesis (or a suitable generalization applicable at any order in perturbation theory) should conceptually follow from within the EFT itself, deferring further investigation of this point to future work.

Let us derive which novel operators this hypothesis yields, in addition to those featuring the `standard structure', i.e.~in addition
to the set of operators given in Eq.~\eqref{eq:NLGoverallgradients} and Eq.~\eqref{eq:NLGnonoverallgradients}.

For $n=1$ we have 
\be\label{eq:extraPi1redundant}
  \frac{\nabla_k\nabla_l}{\nabla^2} \Pi^{[1]}_{kl} =\text{tr}(\Pi^{[1]})\,,
\ee
which is already included in the operator construction. This has two implications: first, even when allowing building blocks of the form in Eq.~\eqref{eq:extrabuildingblock}, no extra linear-order counterterms arise. Second, we do not need to consider the building block $\frac{\nabla_k\nabla_l}{\nabla^2} \Pi^{[1]}_{kl}$ at all when constructing NLG operators beyond linear order.

For $n=2$, the only additional operator starting at second order, with two gradients, is $\nabla^2\frac{\nabla_k\nabla_l}{\nabla^2} \Pi^{[2]}_{kl} =\nabla_k\nabla_l\Pi^{[2]}_{kl}$ which is redundant with the set of operators in Eq. \eqref{eq:NLGoverall}.
Thus, the first new terms appear at 3rd order in perturbations.

Following the hypothesis from above, there may be two types of contributions: {\it (i)} operators involving products of $\frac{\nabla_k\nabla_l}{\nabla^2} \Pi^{[n]}_{kl}$ and $\Pi^{[m]}_{ij}$ building blocks, and {\it (ii)} operators arising from acting with  convective time derivatives $D_\eta$ on $\frac{\nabla_k\nabla_l}{\nabla^2} \Pi^{[n]}_{kl}$. The latter can be viewed as the analog to the construction of the $\Pi^{[n]}_{kl}$ from $\Pi^{[1]}_{kl}$. Notably, $D_\eta$ cannot be interchanged with $\frac{\nabla_k\nabla_l}{\nabla^2}$, even not up to terms that yield contributions included in the NLG operator basis. This is different from the case when interchanging $D_\eta$ with usual gradients $\nabla_{i_1}\nabla_{i_2}\cdots$. Therefore, operators of the form {\it (ii)} need to be included explicitly, see below for more details.

At third order in perturbations, the only contributions of the type {\it (i)} are those involving a product of a single $\frac{\nabla_k\nabla_l}{\nabla^2} \Pi^{[2]}_{kl}$ with one $\Pi^{[1]}_{ij}$ operator. This is because all other possible combinations of type {\it (i)} would be degenerate with the `standard' NLG operators.\footnote{This is because the product of $\frac{\nabla_k\nabla_l}{\nabla^2} \Pi^{[1]}_{kl}$ with e.g. $\Pi^{[2]}_{ij}$ would yield redundant operators due to Eq.~\eqref{eq:extraPi1redundant}, and the only NLG operator involving $\frac{\nabla_k\nabla_l}{\nabla^2} \Pi^{[3]}_{kl}$ would be $\nabla^2\frac{\nabla_k\nabla_l}{\nabla^2} \Pi^{[3]}_{kl}=\nabla_k\nabla_l\Pi^{[3]}_{kl}$ which is  redundant with Eq.~\eqref{eq:NLGoverallgradients}.} Taking all possible index contractions as well as the various possibilities to add the two extra gradients into account, we find two `non-standard' second-gradient operators of the type {\it (i)} at 3rd order in perturbations which are non-redundant with the `standard' operators from Eq.~\eqref{eq:NLGoverallgradients} and Eq.~\eqref{eq:NLGnonoverallgradients}
($B$ for backreaction):\footnote{We checked that similar operators but without {\it overall} gradients are redundant at 3rd order,
specifically e.g. $(\nabla^2\text{tr}\Pi^{[1]})\frac{\nabla_k\nabla_l}{\nabla^2} \Pi^{[2]}_{kl} $ and $\Pi^{[1]}_{ij}\nabla_i\nabla_j\frac{\nabla_k\nabla_l}{\nabla^2} \Pi^{[2]}_{kl} $.}
\be\label{eq:PiB12}
  \Pi_{B1}\equiv\nabla^2\left[\text{tr}(\Pi^{[1]})\frac{\nabla_k\nabla_l}{\nabla^2} \Pi^{[2]}_{kl} \right],\qquad
  \Pi_{B2}\equiv\nabla_i\nabla_j\left[\Pi^{[1]}_{ij}\frac{\nabla_k\nabla_l}{\nabla^2} \Pi^{[2]}_{kl} \right]\,.
\ee
These need to be complemented with the `non-standard' operators of type {\it (ii)}, i.e.~of the schematic form $(D_\eta)^m\frac{\nabla_k\nabla_l}{\nabla^2} \Pi^{[n]}_{kl}$. In analogy to the construction of the $\Pi^{[n]}_{ij}$ operators, it is sufficient to consider $m+n\leq 3$ at 3rd order in perturbations, which we are interested in. Given that we can take $n\geq 2$ due to Eq.~\eqref{eq:extraPi1redundant}, it is thus sufficient to consider a single convective derivative, i.e.~operators with $m=1$, within category ${\it (ii)}$. Focusing therefore on $m=1$, we define the class of operators
\be\label{eq:hatpin}
  \hat\Pi^{[n]} \equiv \frac{1}{(n-1)!} \nabla^2\left[(D_\eta-(n-1))  \frac{\nabla_k\nabla_l}{\nabla^2} \Pi^{[n-1]}_{kl} \right] - \nabla_k\nabla_l\Pi^{[n]}_{kl} \,,
\ee
for $n\geq 2$.
The difference $D_\eta-(n-1)$ ensures that $\hat\Pi^{[n]}$ starts at order $n$ in perturbation theory. Furthermore, we find it convenient to
include the subtraction term $\nabla_k\nabla_l\Pi^{[n]}_{kl}$, since this ensures that $\hat\Pi^{[n]}$ has no displacement poles.
This can be seen since a displacement pole term would require that both of the two overall derivatives act on $\frac{\nabla_k\nabla_l}{\nabla^2} \Pi^{[n-1]}_{kl}$.\footnote{
Note that if a $\nabla$ acts on the velocity field contained in $D_\eta$, it does not have a displacement pole, but a structure degenerate with the usual $\Pi$ terms.
Thus, displacement poles occur only for those contributions for which the gradient operators do not act on $D_\eta$.
}
However, inserting the definition Eq.~\eqref{eq:Pindef}
in the last term, one sees that the displacement poles cancel among the first and second terms. 

These operators start by definition at $n\geq 2$. However, for $n=2$, using Eq.~\eqref{eq:extraPi1redundant} and Eq.~\eqref{eq:Pindef} yields
\be
  \hat\Pi^{[2]} = \nabla^2\text{tr}(\Pi^{[2]}) - \nabla_k\nabla_l\Pi^{[2]}_{kl} \,,
\ee
which shows that $\hat\Pi^{[2]}$ is redundant with the `standard' set in Eq.~\eqref{eq:NLGoverall} of NLG operators. We thus consider the additional operator $\hat\Pi^{[3]}$ at $n=3$. 

In summary, we find that {\it (i)} $\hat\Pi^{[3]}$ and the operators $\Pi_{B1}$ and $\Pi_{B2}$ from Eq.\,\eqref{eq:PiB12} are non-redundant for generic wavenumbers $\bm k_1, \bm k_2, \bm k_3$; {\it (ii)} in combination with the operators from Eq.~\eqref{eq:NLGoverallgradients} and Eq.~\eqref{eq:NLGnonoverallgradients}, they are  sufficient to renormalize the matter kernel $F_5(\bm k_1, \bm k_2, \bm k_3, \bm p, -\bm p)$ at subleading order in gradients.
The same holds for the renormalization of the second-gradient-contributions to all fifth-order kernels $K^{(5)}_a(\bm k_1, \bm k_2, \bm k_3, \bm p, -\bm p)$ for all 29 fifth order bias operators. This brings us to a total of $17 + 3 = 20$ second-gradient operators at up to third order in perturbations, of which $15$ start at third order in perturbations. 

In anticipation of the explicit calculation of UV limits in \refsec{renorm}, we do however find that while $\Pi_{B1}$ and $\Pi_{B2}$ are in general non-redundant, only the specific combination 
\begin{equation}
    \Pi_{B1}-\Pi_{B2} \,,
\end{equation}
is needed to renormalize $K^{(5)}_a(\bm k_1, \bm k_2, \bm k_3, \bm p, -\bm p)$ for any $a$. However, for generality, we keep both when discussing the general single-hard renormalization. We will see that this set of operators is a superset of the operators discussed in the context of the two-loop matter power spectrum from \cite{Anastasiou:2025jsy}. We discuss this in more detail in \refsec{correlators}. We also note that allowing for the building block from Eq.~\eqref{eq:extrabuildingblock} yields an extra NLG operator $(\hat z\cdot\nabla)^2\frac{\nabla_a\nabla_b}{\nabla^2}\Pi^{[2]}_{ab}$ in redshift space already at second order in perturbation theory. This is in line with the findings in~\cite{DAmico:2022ukl}.\footnote{Specifically, the operator from Eq.~(1.5) in~\cite{DAmico:2022ukl} is equivalent to $(\hat z\cdot\nabla)^2\frac{\nabla_a\nabla_b}{\nabla^2}\Pi^{[2]}_{ab}$ up to `standard' operators, as can be seen using e.g. Eq.~(C.27) from~\cite{Desjacques:2016bnm}.}

In anticipation of the discussion of the matter renormalization, we find it convenient to split the higher-gradient operators into those with two overall gradients and those without, that is 
\be
  \sum b_{a_2}{\cal O}_{a_2} = \sum b_{a_2}^{\nabla^2}{\cal O}^{\nabla^2}_{a_2} + \sum b_{a_2}^{\text{non-}\nabla^2}{\cal O}^{\text{non-}\nabla^2}_{a_2}\,,
\ee
where 
\bea\label{eq:ctbar}
{\cal O}_{a_2}^{\text{non-}\nabla^2} &= \{\mbox{tr}(\Pi^{[1]})\nabla^2\mbox{tr}(\Pi^{[1]}),\quad [\mbox{tr}(\Pi^{[1]})]^2\nabla^2\mbox{tr}(\Pi^{[1]}),\quad \mbox{tr}(\Pi^{[1]})\mbox{tr}[\Pi^{[1]}\nabla^2\Pi^{[1]}],\nn\\
& [\nabla_k\Pi^{[1]}_{kl}]\times \mbox{tr}[\Pi^{[1]}\nabla_l\Pi^{[1]}],
           \mbox{tr}[\Pi^{[2]}\nabla^2\Pi^{[1]}] \}\,,
\eea
are the five operators from Eq. \eqref{eq:NLGnonoverallgradients} which do not occur as counterterms in the Euler equation and 
\ba 
{\cal O}_{a_2}^{\nabla^2} = {\cal O}_{a_2} \setminus {\cal O}_{a_2}^{\text{non-}\nabla^2}.
\ea

\newpage
\section{Deterministic renormalization with higher-derivative operators} \label{sec:renorm}

We consider a generalization of the leading-in-gradients renormalization scheme (LG-scheme) introduced in~\cite{McDonald:2006mx,Assassi:2014fva} (and further extended in \hyperlink{cite.Bakx:2025cvu}{Paper~I}) that also includes second-gradient operators.

\subsection{Renormalization conditions}

The biased tracer density contrast can be equivalently expanded in terms of bare or renormalized operators, with corresponding bare or renormalized bias coefficients,  
\be
  \delta_g = b_A{\cal O}_A = b_{[A]}[{\cal O}_{A}]\,.
\ee
For any operator ${\cal O}_A$ (once more, including both leading and subleading order operators in gradient expansion) we define a renormalized operator  $[{\cal O}_A]$ by
\be\label{eq:Oren}
  [{\cal O}_A] = {\cal O}_A + Z_{AB}{\cal O}_B \,,
\ee
with some (scheme- and regulator-dependent) renormalization constants $Z_{AB}$ and the summation over $B$ implicit, including both leading and subleading gradient operators. Note that operators at different derivative orders have different dimensions, leading also to dimensionful entries for $Z_{AB}$. For the operators we consider in the renormalization, the indices $A,B$ run over \textit{both} leading gradient operators (indices $a,b$) \textit{and} second-order-in-gradient operators (indices $a_2,b_2$).
The renormalization matrix can be split into the four blocks
\be
  Z_{AB} = \left(\begin{array}{cc} Z_{ab} & Z_{ab_2} \\ Z_{a_2b} & Z_{a_2b_2}\end{array}\right)\,.
\ee
The $Z$ matrix is fixed by a choice of renormalization condition. To define the choice adopted in this work, we consider the correlators 
\be\label{eq:fAdef}
  f_A({\bm k}_1,\dots,{\bm k}_n) \equiv \langle [{\cal O}_A] \delta_L({\bm k}_1)\cdots\delta_L({\bm k}_n)\rangle'\,, 
\ee
where the prime indicates that the overall-momentum Dirac-delta and factors of $P^\text{lin}(k_i)$ have been dropped.
We consider a rescaling of all wavenumber arguments by a common factor, ${\bm k}_i\mapsto \alpha{\bm k}_i$. Then the \AB{} renormalization condition outlined in \cite{Assassi:2014fva} reads 
\be\label{eq:rencond}
  f_A(\alpha{\bm k}_1,\dots,\alpha{\bm k}_n)\Big|_{\alpha\to 0} = f^\text{tree}_A(\alpha{\bm k}_1,\dots,\alpha{\bm k}_n)\Big|_{\alpha\to 0} \,.\quad(\text{\AB})
\ee
This condition ensures that the leading-in-derivative renormalized bias values match the values obtained by measurements using the tree-level low-$k_i$ regime of $n$-point functions and therefore the separate Universe prediction \cite{Lazeyras:2015lgp, Baldauf:2015vio, Li:2015jsz}. Here, we generalize this condition, by requiring in addition
\be\label{eq:rencond2}
  \boxed{\left( \frac{d^2}{d\alpha^2}\right)^mf_A(\alpha{\bm k}_1,\dots,\alpha{\bm k}_n)\Big|_{\alpha\to 0} = \left( \frac{d^2}{d\alpha^2}\right)^mf^\text{tree}_A(\alpha{\bm k}_1,\dots,\alpha{\bm k}_n)\Big|_{\alpha\to 0} \,,}
\ee
for different integer values $m$. In this work, we consider only the case of second-order in gradient operators, such that we consider simultaneously $m = 0$ and $m = 1$.
This second condition requires that, in a Taylor expansion around ${\bm k}_i=0$, also the loop contributions that scale quadratically in the external wavenumbers are renormalized, and required to match the
tree-level contribution. This requires including operators at second order in gradient expansion in the sum over ${\cal O}_B$ from above. 
Note that any odd number of derivatives with respect to the scaling parameter $\alpha$ would not yield a new condition for the considered scalar bias operators, since, due to isotropy, the Taylor expansion in $\alpha$ has only even powers. One could in principle consider a NNLG scheme by including the $\nabla^4\delta_L$ and $b_{a_4}$ operator and also consider the $m=2$ case. We include this operator when renormalizing the two-loop power spectra, but not in the renormalization scheme. 

We can further expand the $Z_{AB}$ in loops $Z_{AB}=Z_{AB}^\text{1L}+\dots$ with the $\ell$-loop contribution being determined by evaluating the correlation functions involved in the renormalization condition
at $\ell$-loop order \cite{Bakx:2025cvu},
\bea\label{eq:fA}
  f^\text{tree}_A({\bm k}_1,\dots,{\bm k}_n) &=& n! K_A^{(n)}({\bm k}_1,\dots,{\bm k}_n) \,,\nn\\
  f^\text{1L}_A({\bm k}_1,\dots,{\bm k}_n) &=& \frac{(n+2)!}{2}\int_q P^\text{lin}(q)K_A^{(n+2)}({\bm k}_1,\dots,{\bm k}_n,{\bm q},-{\bm q}) +   Z_{AB}^\text{1L}f^\text{tree}_B({\bm k}_1,\dots,{\bm k}_n) \,,\nn\\
  f^\text{2L}_A({\bm k}_1,\dots,{\bm k}_n) &=& \frac{(n+4)!}{2\cdot 2^2}\int_{pq} P^\text{lin}(p)P^\text{lin}(q)K_A^{(n+4)}({\bm k}_1,\dots,{\bm k}_n,{\bm p},-{\bm p},{\bm q},-{\bm q}) \nn\\
  && {} +  Z_{AB}^\text{1L}f^\text{1L}_B({\bm k}_1,\dots,{\bm k}_n)  +  (Z_{AB}^\text{2L}-Z_{AC}^\text{1L}Z_{CB}^\text{1L})f^\text{tree}_B({\bm k}_1,\dots,{\bm k}_n) \,,
\eea
where factors $P^\text{lin}(k_i)$ are stripped off for convenience.

\subsection{One-loop renormalization}

As outlined in \hyperlink{cite.Bakx:2025cvu}{Paper~I}, the one-loop contribution can be fixed by considering the single-hard limits of operators in the integrated momenta $q$, or alternatively soft limits of the external momenta $k_i$ as in Eq. \eqref{eq:rencond2}. Now we generalize them to include subleading-gradient operators as
\bea\label{eq:singlehardNLG}
  K_a^{(n+2)}({\bm k}_1,\dots,{\bm k}_{n},{\bm q},-{\bm q})_{\text{av}_{\hat q}}^{q\gg k_i} &=& \sum_b c_{ab}^{(n+2)}K_b^{(n)}({\bm k}_1,\dots,{\bm k}_{n}) \nn\\
  && {} + \frac{1}{q^2}\, \sum_{b_2} c_{ab_2}^{(n+2)} K_{b_2}^{(n)}({\bm k}_1,\dots,{\bm k}_{n})
  +{\cal O}(q^{-4})\,,\nn\\
  K_{a_2}^{(n+2)}({\bm k}_1,\dots,{\bm k}_{n},{\bm q},-{\bm q})_{\text{av}_{\hat q}}^{q\gg k_i} &=& q^2\, \sum_b c_{a_2b}^{(n+2)}K_b^{(n)}({\bm k}_1,\dots,{\bm k}_{n}) \nn\\
  && {} +  \sum_{b_2} c_{a_2b_2}^{(n+2)} K_{b_2}^{(n)}({\bm k}_1,\dots,{\bm k}_{n}) 
  +{\cal O}(q^{-2})\,,
\eea
with coefficient matrices $c^{(n+2)}$ and taking the average over the direction of ${\bm q}$. For all NLG operators we use the minus sign convention, e.g.~$-\nabla^2\delta$, such that the equivalent in Fourier space $k^2\delta$ does not carry the minus sign. 
Using these expansions as well as the renormalization condition in Eq.~\eqref{eq:rencond2} (see Sec.~4 of \hyperlink{cite.Bakx:2025cvu}{Paper~I}), yields
\bea\label{eq:Z1L}
  Z_{ab}^\text{1L} &=& -\sigma_\Lambda^2 s_{ab}^\text{1L},\qquad   Z_{ab_2}^\text{1L} = -\tilde\sigma_{\Lambda}^2 s_{ab_2}^\text{1L},\nn\\
  Z_{a_2b}^\text{1L} &=& -\hat\sigma_\Lambda^2 s_{a_2b}^\text{1L},\qquad   Z_{a_2b_2}^\text{1L} = -\sigma_{\Lambda}^2 s_{a_2b_2}^\text{1L}\,,
\eea
where
\be\label{eq:s1Lab2}
  s_{ab}^\text{1L}\equiv \frac{(n+2)(n+1)}{2}c_{ab}^{(n+2)},\qquad
  s_{ab_2}^\text{1L}\equiv \frac{(n+2)(n+1)}{2}c_{ab_2}^{(n+2)},
\ee
\be\label{eq:s1La2b2}
  s_{a_2b}^\text{1L}\equiv \frac{(n+2)(n+1)}{2}c_{a_2b}^{(n+2)},\qquad
  s_{a_2b_2}^\text{1L}\equiv \frac{(n+2)(n+1)}{2}c_{a_2b_2}^{(n+2)}\,,
\ee
and
\be \label{eq:sigmadef}
  \sigma_\Lambda^2\equiv \int_{p<\Lambda} P^\text{lin}(p),\qquad   \tilde\sigma_{\Lambda}^2\equiv \int_{p<\Lambda} \frac{P^\text{lin}(p)}{p^2},\qquad   \hat\sigma_{\Lambda}^2\equiv \int_{p<\Lambda} p^2 P^\text{lin}(p)\,,
\ee
with $\tilde\sigma$ and $\hat\sigma$ being dimensionful parameters connecting operators of different derivative orders, while the dimensionless parameter $\sigma$ found in \cite{Rubira:2023vzw} and \hyperlink{cite.Bakx:2025cvu}{Paper~I} connects operators at the same order in gradients.
Note that, since the renormalization is an operator relation that holds independently of the correlation function into which it is inserted, the renormalization matrices $s_{AB}^\text{1L}$ need to be independent of $n$, as already outlined in \hyperlink{cite.Bakx:2025cvu}{Paper~I}.
For the leading-gradient operator renormalization matrix $s_{ab}^\text{1L}$ this was checked explicitly up to $n+2=5$ in \hyperlink{cite.Bakx:2025cvu}{Paper~I}, and is reproduced here in \reftab{sab1L} for completeness.
Here, we also obtain results for  $s_{ab_2}^\text{1L}$ (\reftab{sab21L_A} and \reftab{sab21L_B}) , $s_{a_2b}^\text{1L}$ (\reftab{sa2b1L}) and $s_{a_2b_2}^\text{1L}$ (\reftab{sa2b21L_A} and \reftab{sa2b21L_B}), and checked that the entries in the $n+2=5$ matrices that appear also in those for $n+2=3,4$ indeed yield $n$-independent results for $s^\text{1L}$, as expected.  

\subsection{Two-loop renormalization}

\begin{figure}[t]
    \centering
    \includegraphics[width=0.7\columnwidth]{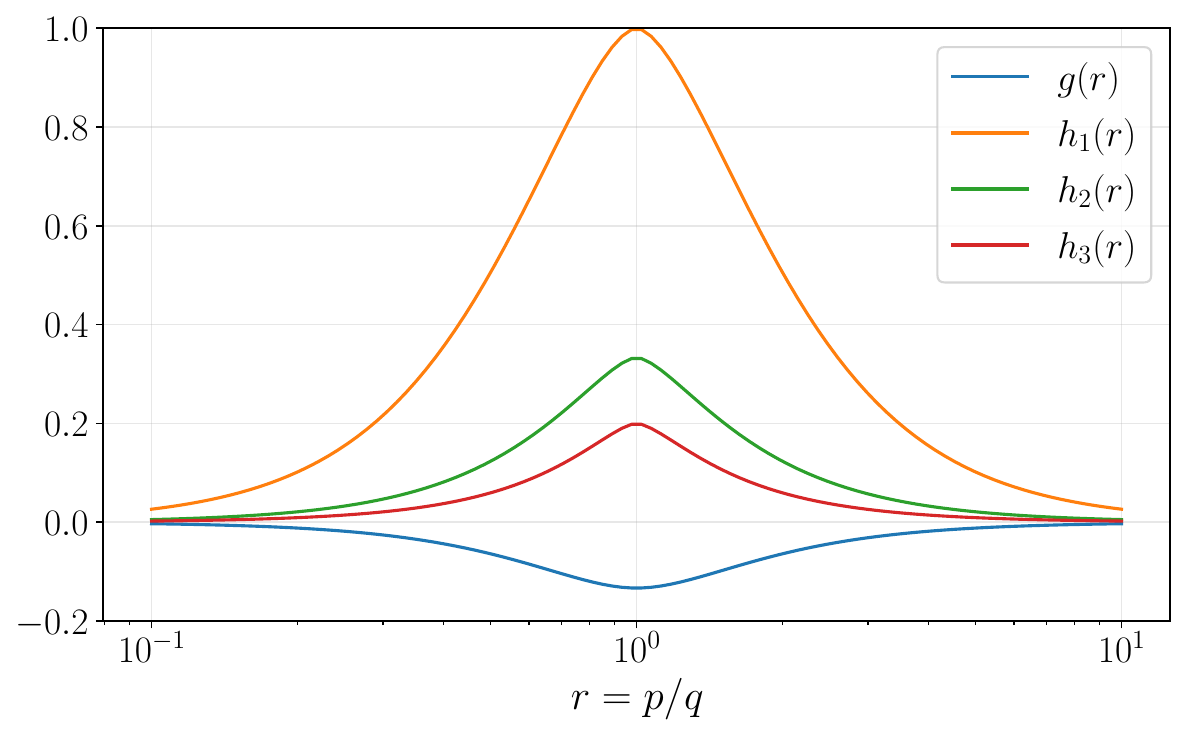}
    \caption{Weight functions $g(r)$ and $h_i(r)$ appearing in the coefficients $d_{a\delta }^{(5)}(r), d_{a\nabla^2\delta }^{(5)}(r)$ in Eqs.~\eqref{eq:doublehardfuncg} and \eqref{eq:doublehard2}, and entering the two-loop kernels $s^{\mathrm{2L}}_{AB}(r)$ defined in Eq.~\eqref{eq:s2LNLG}, shown as functions of the momentum ratio $r=p/q$. Their symmetry under $r\to 1/r$, enhancement near $r=1$, and suppression for $r\to 0,\infty$ encode the selection of the intrinsic two-loop region $p\sim q$.}
    \label{fig:2loopfuncs}
\end{figure}

At two-loop order, we restrict ourselves to renormalizing LG operators ${\cal O}_a$, since as we comment in \refsec{correlators} this is necessary for the renormalization of the two-loop tracer power spectrum.
The corresponding renormalization condition involves double-hard limits, taken at fixed value of the ratio $r=p/q$,
\ba\label{eq:doublehard2}
   K_a^{(n+4)}({\bm k}_1,\dots,{\bm k}_{n},{\bm p},-{\bm p},{\bm q},-{\bm q})_{\text{av}_{\hat p,\hat q}}^{p,q\gg k_i}  = \sum_b d_{ab}^{(n+4)}(r)K_b^{(n)}({\bm k}_1,\dots,{\bm k}_{n}) \nn\\
    + \frac12\left(\frac{1}{p^2}+\frac{1}{q^2}\right)\, \sum_{b_2} d_{ab_2}^{(n+4)}(r) K_{b_2}^{(n)}({\bm k}_1,\dots,{\bm k}_{n})
  +{\cal O}(q^{-4})\,,
\ea
as outlined in \hyperlink{cite.Bakx:2025cvu}{Paper~I} and here also extended to include NLG operators. In the second line, we isolate a $\frac12\left(\frac{1}{p^2}+\frac{1}{q^2}\right)$ factor from the $d_{ab_2}^{(n+4)}(r)$ functions for convenience.
We furthermore consider the lowest possible order $n+4=5$, relevant for the two-loop power spectrum. In this case only the single LG operator ${\cal O}_b=\delta$ contributes to the sum over $b$ on the
right-hand side, and the single NLG operator ${\cal O}_{b_2}=\nabla^2\delta$ to the sum over $b_2$. The coefficients\footnote{Note that we reinstate the subscript $\delta$ on the RHS compared to \hyperlink{cite.Bakx:2025cvu}{Paper~I}, in order to explicitly distinguish it from the double-hard NLG function $d_{a\nabla^2\delta}^{(5)}(r)$ and its associated coefficients below.}
\be
  d_{a\delta}^{(5)}(r)=d_{a\delta}^{(5),0}+\tilde d_{a\delta}^{(5),1}g(r)\,,
\ee
as well as the function
\begin{eqnarray}\label{eq:doublehardfuncg}
  \boxed{g(r) = \frac{15r^8 - 40r^6 + 18r^4- 40r^2 + 15}{240r^4}
  -\frac{(r^2-1)^4(r^2+1)}{64r^5}\ln\left(\frac{1+r}{1-r}\right)^2\,,}
\end{eqnarray} 
were computed in~\cite{Bakx:2025cvu} for all 29 LG bias operators ${\cal O}_a$ up to fifth order. The coefficients $d_{a\delta}^{(5),0}$ and $\tilde d_{a\delta}^{(5),1}$ are reproduced in \reftab{da5} for completeness. At NLG, using $K_{\nabla^2\delta}^{(1)}(k)=k^2$ we find
\be
  d_{a\nabla^2\delta}^{(5)}(r) = d_{a\nabla^2\delta}^{(5),0} + \sum_{i=1}^3 {\tilde d_{a\nabla^2\delta}^{(5),i}}\, h_i(r)\,,
\ee
with coefficients $d_{a\nabla^2\delta}^{(5),0}$ and ${\tilde d_{a\nabla^2\delta}^{(5),i}}$ for all 29 LG bias operators ${\cal O}_a$ computed here up to fifth order and with three distinct $r$-dependent weight functions 
\be
\boxed{
\begin{aligned}
  h_1(r) &= 1 -  \frac{(1 - r^2)^2}{4r (1 + r^2)}\ln\left(\frac{1+r}{1-r}\right)^2\,,\\
  h_2(r) &= \frac{3 - 2 r^2 + 3 r^4}{12r^2} -  \frac{(1 - r^2)^2(1+r^2)}{16r^3}\ln\left(\frac{1+r}{1-r}\right)^2 \,,\\
  h_3(r) &= \frac{15 + 20 r^2 - 22 r^4 + 20 r^6 + 15 r^8}{240 r^4} -  \frac{(1 - r^2)^2(1+r^2)^3}{64r^5}\ln\left(\frac{1+r}{1-r}\right)^2 \,.
\end{aligned}
}
\ee
Those functions are displayed in \reffig{2loopfuncs}, with the corresponding coefficients $d_{a\nabla^2\delta}^{(5),0}$ and ${\tilde d_{a\nabla^2\delta}^{(5),i}}$  presented in \reftab{da5}. Both $g$ and $h_i$ are symmetric under $r\to 1/r$ (i.e. $p\leftrightarrow q$), peak at $r = 1$ and are suppressed as $\propto r^2$ for $r\to 0$ (or $r\to \infty$), i.e. $g(0) = h_1(0)=h_2(0)=h_3(0)=0$. The behavior of $g$ and $h_i$ is a direct consequence of the two-loop renormalization keeping only the `intrinsic two-loop' region $p\sim q$ in the integrand, or equivalently $r\sim 1$, and filtering out the hierarchical regions $p \gg q$ and $q \gg p$, which factorize as two sequential one-loop single-hard contributions (`one-loop-squared').

The renormalization condition Eq.~\eqref{eq:rencond2} together with Eq.~\eqref{eq:fA} yields
\bea\label{eq:Z2L}
  Z_{ab}^\text{2L} &=& Z_{aC}^\text{1L}Z_{Cb}^\text{1L} - \frac12 \int_{pq} s_{ab}^\text{2L}(p/q)P^\text{lin}(p)P^\text{lin}(q)\,,\nn\\
  Z_{ab_2}^\text{2L} &=& Z_{aC}^\text{1L}Z_{Cb_2}^\text{1L} - \frac12 \int_{pq} s_{ab_2}^\text{2L}(p/q)\left(\frac{1}{p^2}+\frac{1}{q^2}\right)P^\text{lin}(p)P^\text{lin}(q)\,,
\eea
where summation over $C$ includes both LG ($c$) and NLG ($c_2$) operators, and
\bea\label{eq:s2LNLG}
  s_{ab}^\text{2L}(r) &=& \frac{(n+4)!}{4n!} d_{ab}^{(n+4)}(r)\,,\nn\\
  s_{ab_2}^\text{2L}(r) &=& \frac12\times\frac{(n+4)!}{4n!} d_{ab_2}^{(n+4)}(r)\,.
\eea
The first line agrees with \hyperlink{cite.Bakx:2025cvu}{Paper~I}, while the second line describes the contribution from second-gradient operators.
Note that compared to Eq.~\eqref{eq:doublehard2}, we find it convenient to absorb a factor $1/2$ in front of the $1/p^2+1/q^2$ factor into the definition of $s_{ab_2}^\text{2L}(r)$.
Further, note that due to the renormalization condition Eq.~\eqref{eq:rencond2}, the renormalization constant $Z_{ab}^\text{2L}$ differs from the one obtained when
imposing the \AB{}, since now it also includes a sum over the NLG terms $Z_{ac_2}^\text{1L}Z_{c_2b}^\text{1L}=\tilde\sigma_\Lambda^2\hat\sigma_\Lambda^2s^\text{1L}_{ac_2} s^\text{1L}_{c_2b}$.
This is expected, as the renormalization constants $Z$ are scheme-dependent. In contrast, the matrices $s^\text{1L}_{AB}$ and $s_{AB}^\text{2L}(r)$ are, for given operators ${\cal O}_A$
and ${\cal O}_B$, scheme-independent quantities, as they are derived from single- and double-hard limits of the (bare) bias operators, which are defined independently of the renormalization scheme.
For the two-loop power spectrum, the relevant contributions are
\bea
  s_{a\delta}^\text{2L}(r) &=& 30 d_{a\delta}^{(5)}(r) = 30\left(d_{a\delta}^{(5),0}+\tilde d_{a\delta}^{(5),1}g(r)\right)\,,\nn\\
  s_{a\nabla^2\delta}^\text{2L}(r) &=& \frac12\times 30 d_{a\nabla^2\delta}^{(5)}(r) \nn\\
  &=& 15\left(d_{a\nabla^2\delta}^{(5),0}+\tilde d_{a\nabla^2\delta}^{(5),1}h_1(r)+\tilde d_{a\nabla^2\delta}^{(5),2}h_2(r)+\tilde d_{a\nabla^2\delta}^{(5),3}h_3(r)\right)\,.
\eea

\subsection{Commutation of limits}\label{subsec:comm_lim}

In \hyperlink{cite.Bakx:2025cvu}{Paper~I}, we derive the commutation-of-limits relation (see Eq.~4.19 therein)
\be\label{eq:s2LdoublesinglePaperI}
  s_{ab}^\text{2L}(0) = s_{ac}^\text{1L}s_{cb}^\text{1L}\,,
\ee
from the two-loop integrals being decomposed into two sequential one-loop integrals in the  hierarchical limit $p\gg q$ or $q\gg p$. Equivalently, the double-hard limit can be written as a product of two single-hard limits when $p\gg q$ or $q\gg p$.  In this section, we discuss additional relations that appear when considering NLG operators. The derivation of the commutation-of-limit relations shown here from sequential one-loop limits is presented in \refapp{consistency}.

By Taylor expanding the $s_{ab}^\text{2L}(r)$ and $s_{ab_2}^\text{2L}(r)$ functions around $r=0$, 
\bea\label{eq:sabTaylor}
  s_{ab}^\text{2L}(r) &=& s_{ab}^\text{2L}(0) + r\times (s_{ab}^\text{2L})'(0) + \frac12 r^2\times (s_{ab}^\text{2L})''(0) + \dots\,,\nn\\
  s_{ab_2}^\text{2L}(r) &=& s_{ab_2}^\text{2L}(0) + r\times (s_{ab_2}^\text{2L})'(0) + \frac12 r^2\times (s_{ab_2}^\text{2L})''(0) + \dots\,,
\eea
where primes denote derivatives with respect to $r=p/q$. We find by considering the $r^0$ term
\be\label{eq:s2Ldoublesingle}
  s_{ab}^\text{2L}(0) = s_{ac}^\text{1L}s_{cb}^\text{1L},\qquad   s_{ab_2}^\text{2L}(0) = s_{ac}^\text{1L}s_{cb_2}^\text{1L}\,,
\ee
with the first relation corresponding to Eq. \eqref{eq:s2LdoublesinglePaperI} and the second one involving NLG operators being new.
Notably, the intermediate index $c$ runs {\it only} over leading-gradient operators, which we explicitly confirmed for $a$ running over all 29 leading-gradient bias operators up to fifth order, and $b$ $(b_2)$ corresponding to $\delta$ $(\nabla^2\delta)$. 
The linear terms in $r$ in Eq. \eqref{eq:sabTaylor} vanish due to analyticity and isotropy, i.e. $(s_{ab}^\text{2L})'(0)=(s_{ab_2}^\text{2L})'(0)=0$.
For the quadratic terms $r^2$, we find the further relations 
\bea\label{eq:s2Ldoublesingle2}
  \frac12 (s_{ab}^\text{2L})''(0) &=& s_{ac_2}^\text{1L}s_{c_2b}^\text{1L}\,,\nn\\
  s_{ab_2}^\text{2L}(0) +\frac12 (s_{ab_2}^\text{2L})''(0) &=& s_{ac_2}^\text{1L}s_{c_2b_2}^\text{1L}\,,
\eea
which we verified for all 29 leading-gradient ${\cal O}_a$ operators, and ${\cal O}_b=\delta$ (${\cal O}_b=\nabla^2\delta$) in the first (second) line, respectively.
In those cases, we use
\be \label{eq:derivhs}
  \frac12 g''(0)=-\frac{32}{105},\quad \frac12 h_1''(0)=\frac{8}{3},\quad \frac12 h_2''(0)=\frac{8}{15},\quad \frac12 h_3''(0)=\frac{8}{35}\,.
\ee

In summary, all the double/single-hard commutation-of-limit relations (see \refapp{consistency} for details) from above then read 
\ba
  30 d_{a\delta}^{(5),0}   &= s_{ac}^\text{1L}s_{c\delta}^\text{1L}\,,\hspace{4cm}\nn\\
  30\tilde d_{a\delta}^{(5),1} \times \frac{-32}{105}   &= s_{ac_2}^\text{1L}s_{c_2\delta}^\text{1L}\,,\nn\\
  30 d_{a\nabla^2\delta}^{(5),0}   &= s_{ac}^\text{1L}s_{c\nabla^2\delta}^\text{1L}\,,\nn\\
  30\left( d_{a\nabla^2\delta}^{(5),0} + \frac{8}{3}\tilde d_{a\nabla^2\delta}^{(5),1}+ \frac{8}{15}\tilde d_{a\nabla^2\delta}^{(5),2}+ \frac{8}{35}\tilde d_{a\nabla^2\delta}^{(5),3}  \right) &= s_{ac_2}^\text{1L}s_{c_2\nabla^2\delta}^\text{1L}\,.
\ea
Notably, the first two lines imply that the double-hard function $s_{a\delta}^\text{2L}(r)$ is completely
fixed by the products of single-hard coefficients, which is a consequence of there only being one $r$-dependent function $g(r)$ in the double-hard LG case, whose  coefficients $\tilde d_{a\delta}^{(5),1}$ are thus entirely fixed by the single-hard limits at NLG. However, the analogous statement fails for the double-hard NLG case, where there are two relations but four sets of coefficients.

\subsection{Renormalized deterministic bias operators and kernels}\label{sec:renkernel}

In summary, for bias operators ${\cal O}_a$ at leading-gradient order, deterministic renormalization as defined by the renormalization conditions Eq.~\eqref{eq:rencond2} amounts to inserting the following renormalized bias operators into one- and two-loop $N$-point functions, respectively, 
\be\label{eq:Oren1L2LNLG}
\boxed{
\begin{array}{lcl}
  \displaystyle   [{\cal O}_a]\Big|_{\text{1L}} &=& \displaystyle {\cal O}_a -\sigma_\Lambda^2 s^\text{1L}_{ab}{\cal O}_b - \tilde\sigma_\Lambda^2 s^\text{1L}_{ab_2}{\cal O}_{b_2} \,,\\[1.5ex]
  \displaystyle {} [{\cal O}_a]  \Big|_{\text{2L}} &=& \displaystyle {\cal O}_a -\sigma_\Lambda^2 s^\text{1L}_{ab}[{\cal O}_b]\Big|_\text{1L} - \tilde\sigma_\Lambda^2 s^\text{1L}_{ab_2}[{\cal O}_{b_2}]\Big|_\text{1L} \\[1.5ex]
  \displaystyle && \displaystyle {} - \frac12 \left(\int_{pq<\Lambda} s_{ab}^\text{2L}(p/q)P^\text{lin}(p)P^\text{lin}(q)\right) {\cal O}_b\\[1.5ex]
  \displaystyle && \displaystyle {} - \frac12 \left(\int_{pq<\Lambda} s_{ab_2}^\text{2L}(p/q)\left(\frac{1}{p^2}+\frac{1}{q^2}\right)P^\text{lin}(p)P^\text{lin}(q)\right) {\cal O}_{b_2}
\end{array}
}\,,
\ee
and
\be\label{eq:O2ren1L}
\boxed{
  [{\cal O}_{a_2}]\Big|_{\text{1L}} = {\cal O}_{a_2} -\hat\sigma_\Lambda^2 s^\text{1L}_{a_2b}{\cal O}_b - \sigma_\Lambda^2 s^\text{1L}_{a_2b_2}{\cal O}_{b_2} 
  }\,.
\ee
This follows from inserting the one- and two-loop renormalization constants Eq.~\eqref{eq:Z1L} and Eq.~\eqref{eq:Z2L} into Eq.~\eqref{eq:Oren}, noting also Eq.~\eqref{eq:fA}.
The case of the LG renormalization scheme [see Eq.~\eqref{eq:rencond} and Eq.~4.20 in~\hyperlink{cite.Bakx:2025cvu}{Paper~I}] is recovered when omitting all contributions from second-gradient operators ${\cal O}_{b_2}$ to the subtraction terms. As noted in~\hyperlink{cite.Bakx:2025cvu}{Paper~I}, the equations above are operator relations that hold when inserted into arbitrary $N$-point functions, or alternatively also at field level.

In practice, within one-loop contributions of the form `${\bm p},-{\bm p}$' to any $N$-point functions, deterministic bias renormalization is equivalent to replacing the bare
bias kernels $K_{a}^{(n+2)}({\bm k}_1,\dots,{\bm k}_n,{\bm p},-{\bm p})$ by renormalized kernels, denoted by $K_{[a]}^{(n+2)}$. 
Specifically, using the first line in Eq.~\eqref{eq:Oren1L2LNLG} as well as Eq.~\eqref{eq:fA}, we find
\bea\label{eq:RenKernel1loop}
  \lefteqn{ K_{[a]}^{(n+2)}({\bm k}_1,\dots,{\bm k}_n,{\bm p},-{\bm p}) = K_{a}^{(n+2)}({\bm k}_1,\dots,{\bm k}_n,{\bm p},-{\bm p}) }\\
  && {} - \frac{2}{(n+2)!}\times n!\Bigg( s^\text{1L}_{ab}K_{[b]}^{(n)}({\bm k}_1,\dots,{\bm k}_n) + \frac{1}{p^2}s^\text{1L}_{ab_2}K_{[b_2]}^{(n)}({\bm k}_1,\dots,{\bm k}_n) \Bigg)\,.\nn
\eea
For practical reasons, to express the renormalized kernel for $a$ without summing over other kernels $b$, we may additionally re-express the subtraction terms in terms of the single-hard limits Eq.~\eqref{eq:singlehardNLG}, using Eq.~\eqref{eq:s1Lab2} as
\bea\label{eq:Karen}
  K_{[a]}^{(n+2)}({\bm k}_1,\dots,{\bm k}_n,{\bm p},-{\bm p}) &=& K_{a}^{(n+2)}({\bm k}_1,\dots,{\bm k}_n,{\bm p},-{\bm p}) \\
  && {} - K_{a}^{(n+2)}({\bm k}_1,\dots,{\bm k}_n,{\bm p},-{\bm p})^{p\gg k_i}_{\text{av}_{\hat p}}\Big|_{(1/p)^0,(1/p)^2}\,,\nn
\eea
where the subscript denotes the sum of leading [$(1/p)^0$] and next-to-leading [$(1/p)^2$] terms in a Taylor expansion for large $p$ with $k_i$ fixed.
Similarly, for second-gradient operators Eq.~\eqref{eq:O2ren1L} as well as Eq.~\eqref{eq:s1La2b2} yield
\ba\label{eq:Ka2ren}
  K_{[a_2]}^{(n+2)}({\bm k}_1,\dots,{\bm k}_n,{\bm p},-{\bm p}) &= K_{a_2}^{(n+2)}({\bm k}_1,\dots,{\bm k}_n,{\bm p},-{\bm p}) \\
  & {} - K_{a_2}^{(n+2)}({\bm k}_1,\dots,{\bm k}_n,{\bm p},-{\bm p})^{p\gg k_i}_{\text{av}_{\hat p}}\Big|_{(1/p)^{-2},(1/p)^0}\,.\nn
\ea
Note that in this case the leading terms start at order $p^2=(1/p)^{-2}$, and next-to-leading contributions are of order $p^0$ here. The renormalized bias therefore removes the UV-sensitive contributions from loop integrals, which for the deterministic bias renormalization at one-loop within the scheme Eq.~\eqref{eq:rencond2} considered in this work, means removing the leading and next-to-leading UV-sensitive contributions in the large-$p$ limit
for loops described by `${\bm p},-{\bm p}$' wavenumber configurations. For comparison, for the \AB{} defined by Eq.~\eqref{eq:rencond}, only the leading $(1/p)^0$-terms for $K_{[a]}$ and $(1/p)^{-2}$ terms for $K_{[a_2]}$ UV contributions are subtracted. Moreover, the renormalized kernels are a convenient form for implementing bias renormalization in numerical loop computations. In particular, it is possible to omit the averaging over the direction $\hat p$ of the loop wavenumber within the subtraction term, since the kernels are inserted in loop integrals for which this averaging is part of the loop integration. This has the advantage that UV sensitivity cancels at the integrand level already~\cite{Lewandowski:2017kes}.

We generalize this concept to deterministic two-loop renormalization in the following. To derive renormalized kernels at two-loop order, we use Eq.~\eqref{eq:Oren1L2LNLG} as well as Eq.~\eqref{eq:fA}, which yields
\bea
  \lefteqn{ K_{[a]}^{(n+4)}({\bm k}_1,\dots,{\bm k}_n,{\bm p},-{\bm p},{\bm q},-{\bm q}) = K_{a}^{(n+4)}({\bm k}_1,\dots,{\bm k}_n,{\bm p},-{\bm p},{\bm q},-{\bm q}) }\\
  && {} - \frac{2^3}{(n+4)!}\Bigg[\frac{(n+2)!}{2}\Bigg( s^\text{1L}_{ab}K_{[b]}^{(n+2)}({\bm k}_1,\dots,{\bm k}_n,{\bm p},-{\bm p}) + \frac{s^\text{1L}_{ab_2}}{q^2}K_{[b_2]}^{(n+2)}({\bm k}_1,\dots,{\bm k}_n,{\bm p},-{\bm p}) \Bigg)\nn\\
  && {} + n!\Bigg( \frac12 s^\text{2L}_{ab}(p/q)K_{b}^{(n)}({\bm k}_1,\dots,{\bm k}_n) + \frac12 s^\text{2L}_{ab_2}(p/q)\left(\frac{1}{p^2}+\frac{1}{q^2}\right)K_{b_2}^{(n)}({\bm k}_1,\dots,{\bm k}_n) \Bigg)\Bigg]\,. \nn
\eea
The subtraction terms in the last line, involving $s^\text{2L}_{ab}(r)$ and
$s^\text{2L}_{ab_2}(r)$, respectively, yield the leading and next-to-leading contributions to the bias kernel expanded in the double-hard limit Eq.~\eqref{eq:doublehard2}
when using Eq.~\eqref{eq:s2LNLG}.
For the one-loop subtraction terms in the middle line, we use Eq.~\eqref{eq:Karen} and Eq.~\eqref{eq:Ka2ren}, together with Eq.~\eqref{eq:s1Lab2} and
Eq.~\eqref{eq:s1La2b2} as well as Eq.~\eqref{eq:singlehardNLG}. Altogether, we obtain the following result for the renormalized two-loop kernel:
\bea\label{eq:Karen2L}
  \lefteqn{ K_{[a]}^{(n+4)}({\bm k}_1,\dots,{\bm k}_n,{\bm p},-{\bm p},{\bm q},-{\bm q}) = K_{a}^{(n+4)}({\bm k}_1,\dots,{\bm k}_n,{\bm p},-{\bm p},{\bm q},-{\bm q}) }\nn\\
  && {} - 2\times\Bigg[ K_{a}^{(n+4)}({\bm k}_1,\dots,{\bm k}_n,{\bm p},-{\bm p},{\bm q},-{\bm q})^{q\gg p,k_i}_{\text{av}_{\hat p,\hat q}}\Big|_{(1/q)^0,(1/q)^2}\nn\\
  && \qquad\quad {} - K_{a}^{(n+4)}({\bm k}_1,\dots,{\bm k}_n,{\bm p},-{\bm p},{\bm q},-{\bm q})^{q\gg p,k_i}_{\text{av}_{\hat p,\hat q}}\Big|_{(1/q)^0}\Big|^{p\gg k_i}_{(1/p)^0,(1/p)^2} \nn\\
  && \qquad\quad {} - K_{a}^{(n+4)}({\bm k}_1,\dots,{\bm k}_n,{\bm p},-{\bm p},{\bm q},-{\bm q})^{q\gg p,k_i}_{\text{av}_{\hat p,\hat q}}\Big|_{(1/q)^2}\Big|^{p\gg k_i}_{(1/p)^{-2},(1/p)^0} \Bigg]\nn\\
  && {} - K_{a}^{(n+4)}({\bm k}_1,\dots,{\bm k}_n,{\bm p},-{\bm p},{\bm q},-{\bm q})^{p,q\gg k_i,\ p/q=\text{fixed}}_{\text{av}_{\hat p,\hat q}}\Big|_{k^0,k^2}\,.
\eea
The last line accounts for the double-hard limit, and the subscript $k^0,k^2$ stands for the sum of the leading and next-to-leading terms in a Taylor expansion
for large $q$ with ratio $p/q$ as well as $k_i$ fixed, see Eq.~\eqref{eq:doublehard2}. Note that this double-hard limit is symmetric in $p\leftrightarrow q$. The terms in the square bracket
account for the single-hard limit $q\gg p,k_i$. The factor of two in front takes care of the identical contribution to the two-loop integral from the region $p\gg q,k_i$.
The second line denotes the sum of the leading and next-to-leading terms in a Taylor series for large $q$ with $p$ and $k_i$ held fixed. The third and fourth lines arise from
the subtraction terms in Eq.~\eqref{eq:Karen} and Eq.~\eqref{eq:Ka2ren}, and account for removing the overlap between single- and double-hard regions. They involve the sequential
single-hard limits, when first expanding for large $q\gg p,k_i$ and afterwards in addition taking the limit $p\gg k_i$. The subscripts indicate the orders in the respective Taylor series
that are kept in each term and for each expansion, respectively.

This result can be promoted to an integrand-level renormalized two-loop kernel by omitting the averages over the directions $\hat p$ and $\hat q$, and replacing the factor of two in front of the square bracket by a sum of two expressions, one being the square bracket itself and the other the square bracket with $p$ and $q$ interchanged.\footnote{In the following, when referring to `renormalized kernels', we refer to the non-angle-averaged renormalized kernels as described here for the two-loop kernel, and as in Eq.~\eqref{eq:Ka2ren} but without angle-average over $\hat p$ for the one-loop case.} This renormalized kernel accounts for subtracting the UV sensitivity arising from either single or double hard regions in the two-loop integration, with a proper removal of overlap contributions at leading and next-to-leading gradient order.
The case of the \AB{}  (i.e.~subtracting only leading-gradient terms) is obtained by omitting all `$1/q^2$' as well as `$k^2$' terms in Eq.~\eqref{eq:Karen2L}.

\newpage
\section{Renormalization of stochastic terms} \label{sec:stochastic}

  In addition to the deterministic bias expansion, biased tracers receive contributions from stochastic fields that encode the effect of small-scale physics on large-scale density perturbations and are uncorrelated with the long-wavelength density field. In the EFT, these contributions appear as contact terms when products of renormalized bias operators are brought to the same spatial point. Formally, they arise from the operator product  of composite operators, and their coefficients require renormalization analogous to the deterministic bias coefficients (see \hyperlink{cite.Bakx:2025cvu}{Paper~I} and~\cite{Carroll:2013oxa,Assassi:2014fva,Rubira:2024tea}). 

  Concretely, the product of two, three, or four renormalized bias operators at coincident points introduces a set of contact-term renormalization tensors $Z_{ab\cdots c}$, one for each combination of operators and each operator $[{\cal O}_c]$ onto which the product can mix. These constants are fixed by requiring that UV-sensitive contributions to the large-scale limit of composite operator correlators are absorbed into the appropriate contact terms. The resulting renormalization conditions take the same form as the deterministic conditions of \refsec{renorm}, but now applied to the correlators of composite operators. Products of two, three, and four operators are needed to renormalize the power spectrum, bispectrum, and trispectrum, respectively, and are discussed in turn below. We note that the application of operator product expansions to stochastic renormalization is standard in the QFT context;  instead, within LSS, a frequently adopted approach consists of introducing \textit{noise fields} $\epsilon_\mathcal{O}({\bm x})$ in the bias expansion, whose $N$-point correlators are then captured by a gradient expansion (see \cite{Desjacques:2016bnm} and references therein). Here, we show that the operator product expansion correctly recovers the known results for LG renormalization for the power spectrum, bispectrum and trispectrum. 
  For a relation between the field formulation through correlators of $\epsilon_\mathcal{O}({\bm x})$ and the contact-operator formulation as presented in \cite{Carroll:2013oxa,Rubira:2024tea}, see \cite{Rubira:2025rqo}. 
  
  Throughout this Section we restrict to stochastic terms at leading gradient order. Hence, throughout this Section, renormalization conditions only include sums over lower-case indices $a,b, \dots$ signifying LG operators. We comment more on this power counting setup and the relation to the size of deterministic contributions in \refsec{correlators}. 

\subsection{Product of two operators}

The renormalization of a product of two operators at distinct spatial points ${\bm x}$ and ${\bm y}$ takes the form~\cite{Bakx:2025cvu}
\be\label{eq:stochxy}
   [{\cal O}_a({\bm x}){\cal O}_b({\bm y})] = [{\cal O}_a({\bm
x})][{\cal O}_b({\bm y})]+Z_{abc}[{\cal O}_c({\bm
X})](2\pi)^3\delta_D({\bm x}-{\bm y})\,,
\ee
where ${\bm X}=({\bm x}+{\bm y})/2$ and the sum over $c$ includes the unit operator ${\bm 1}$. The contact-term coefficients $Z_{abc}$ are fixed by the renormalization condition
\be\label{eq:stochxyrencond}
   \langle[{\cal O}_a({\bm k}){\cal O}_b({\bm k}')]\delta_L({\bm
k}_1)\cdots\delta_L({\bm k_n})\rangle\Big|_{{\bm k},{\bm k}',{\bm k}_i\to 0}
   = \langle[{\cal O}_a({\bm k})][{\cal O}_b({\bm k}')]\delta_L({\bm
k}_1)\cdots\delta_L({\bm k_n})\rangle\Big|^\text{tree}_{{\bm k},{\bm
k}',{\bm k}_i\to 0}\,,
\ee
which requires that the large-scale limit of the composite-operator correlator is reproduced by the tree-level result, in direct analogy with the deterministic renormalization conditions of \refsec{renorm}.

For the renormalization of the power spectrum only the unit-operator component is needed. Defining
\be
   \Delta N_{ab}\equiv Z_{ab{\bm 1}}\,,
\ee
the condition Eq.~\eqref{eq:stochxyrencond} evaluated at $n=0$ gives, to any loop order (using $P^\text{lin}(0)=0$),
\be
   \Delta N_{ab}=-P_{[a][b]}(0)\,.
\ee
At one-loop order this yields
\be\label{eq:Nab1L}
   \Delta
N_{ab}^\text{1L}=-2n_{ab}^{(22)}\int_{p<\Lambda}P^\text{lin}(p)^2\,,
\ee
where
\be
   n_{ab}^{(22)}=\left(K^{(2)}_a({\bm k}-{\bm p},{\bm p})K^{(2)}_b({\bm
k}-{\bm p},{\bm p})\right)^{p\to\infty}_{\text{av}_{\hat p}} =
n_a^{(2)}n_b^{(2)}\,,
\ee
with $n_a^{(2)}=K^{(2)}_a(-{\bm p},{\bm p})$ equal to $0$ for $a=\delta$ and $1$ for $a=\delta^2,\,\text{tr}[(\Pi^{[1]})^2]$. In \hyperlink{cite.Bakx:2025cvu}{Paper~I} (see Sections 3.4 and 4.4 therein), we also derived the corresponding double-hard limits for the stochastic contributions at two-loop order in the power spectrum. We do not use their explicit form here, but only note that no new parameters need to be introduced at leading gradient order relative to the one-loop case.

The contact terms with $c\neq{\bm 1}$ are required to renormalize
mode-coupling loop contributions to higher-point functions.
Concretely, the $B^{(321),I}$ bispectrum contribution (in the notation
of~\cite{Scoccimarro:1997st}) featuring a mode-coupling loop reads
\bea
   B^{(321),I}_{abc}(k,k',k'') &=& 6\int_p P^\text{lin}(p)\Big( K^{(3)}_{a}({\bm
k}+{\bm k}',{\bm p}-{\bm k}',-{\bm p})K^{(2)}_{b}({\bm k}'-{\bm p},{\bm
p})P^\text{lin}(|{\bm k}'-{\bm p}|)\nn\\
   && {} + K^{(2)}_{a}({\bm k}+{\bm p},-{\bm p})K^{(3)}_{b}({\bm k}+{\bm
k}',-{\bm p}-{\bm k},{\bm p})P^\text{lin}(|{\bm k}+{\bm p}|) \Big) \nn\\
   && {} \qquad {} \times K^{(1)}_{c}(k'')P^\text{lin}(k'')  + 3\
\text{permutations}\,.
\eea
Renormalizing this contribution requires $Z_{ab\delta}^\text{1L}$, which is
fixed by evaluating Eq.~\eqref{eq:stochxyrencond} at $n=1$ and one-loop order.
Evaluating at $n=2$ similarly yields $Z_{abc}^\text{1L}$ for operators
$[{\cal O}_c^{[2]}]$ starting at second perturbative order, which
enters the mode-coupling loops of the one-loop trispectrum in
$T^{(4211)}+T^{(3311)}$ (see below for details). More generally, the two-operator contact
terms in Eq.~\eqref{eq:stochxy} at one-loop order renormalize the mode-coupling
loops of any $N$-point function of the form $G^{(n_1 n_2 1\cdots 1)}$ with
$n_1+n_2=N+2$ and $n_1,n_2>1$, with the corresponding $Z_{abc}^\text{1L}$
fixed by Eq.~\eqref{eq:stochxyrencond} at $n=N-2$.

For general $n$, the one-loop renormalization condition is obtained by plugging Eq. \eqref{eq:stochxy} into Eq. \eqref{eq:stochxyrencond} and reads
\be\label{eq:stochxyrencond1L}
   \langle [{\cal O}_a({\bm k})][{\cal O}_b({\bm k}')]\delta_L({\bm
k}_1)\cdots\delta_L({\bm k_n})\rangle^\text{1L}
   + Z_{abc}^\text{1L}\langle[{\cal O}_c({\bm k}+{\bm k}')]\delta_L({\bm
k}_1)\cdots\delta_L({\bm k_n})\rangle^\text{tree}\Big|_{{\bm k},{\bm
k}',{\bm k}_i\to 0}=0\,.
\ee
Since the deterministic renormalization of individual operators already
absorbs all `deterministic' loop contributions\footnote{As an example, for the bispectrum at one-loop the deterministic renormalization already renormalizes $B^{(411)}$ as well as the `factorizable' diagram $B^{(321),II}$.}, only the mode-coupling
(`stochastic') loops remain on the left-hand side; we may therefore replace
$[{\cal O}]$ by ${\cal O}$ when isolating these terms. Stripping the
overall momentum-conserving Dirac delta and external power spectra
$P^\text{lin}(k_i)$ (indicated by the prime), the one-loop mode-coupling
contribution is
\bea\label{eq:stochxyoneloop}
   &&  {\langle [{\cal O}_a({\bm k})][{\cal O}_b({\bm
k}')]\delta_L({\bm k}_1)\cdots\delta_L({\bm
k_n})\rangle'}^\text{1L}_\text{stoch.}  
   = \int_p \, \sum_{m=0}^n \ \sum_{\{{\bm q}_1,\dots,{\bm q}_m\}
\atop  \subset\{{\bm k}_1,\dots,{\bm k}_n\}} 2\frac{(m+2)!}{2}\frac{(n-m+2)!}{2} \vs
&& \quad \times P^\text{lin}(|{\bm
p}+{\bm k}+{\bm q}_{12\cdots m}|)P^\text{lin}(p) \,\,K_a^{(m+2)}(-{\bm q}_1,\dots,-{\bm
q}_m,{\bm p}+{\bm k}+{\bm q}_{12\cdots m},-{\bm p})\nn\\
   && \qquad  \times K_b^{(n-m+2)}(-{\bm q}_{m+1},\dots,-{\bm q}_n,{\bm
k}'+{\bm q}_{(m+1)(m+2)\cdots n}-{\bm p},{\bm p}) \,,
\eea
where the inner sum runs over all $\binom{n}{m}$ ways of partitioning
$\{{\bm k}_1,\dots,{\bm k}_n\}$ into the subset $\{{\bm q}_1,\dots,{\bm q}_m\}$
assigned to operator ${\cal O}_a$ and the complementary set assigned to ${\cal O}_b$,
with the momentum-conservation constraint ${\bm k}'+{\bm
q}_{(m+1)\cdots n}=-({\bm k}+{\bm q}_{12\cdots m})$ implied by the stripped
Dirac delta. Taking the limit ${\bm k},{\bm k}',{\bm k}_i\to 0$ amounts to
expanding the integrand for large $p$ and averaging over the direction
of ${\bm p}$. At leading gradient order, both linear power spectra in the
integrand may be evaluated at $p$. One then expands the UV limit as a
linear combination of bias kernels, isolating a $1/n!$ factor for convenience
\ba\label{eq:sabc1Ldef}
&  \frac{1}{n!}\Bigg[ \sum_{m=0}^n \ \sum_{\{{\bm q}_1,\dots,{\bm
q}_m\} \atop  \subset\{{\bm k}_1,\dots,{\bm k}_n\}} \
2\frac{(m+2)!}{2}\frac{(n-m+2)!}{2}\, K_a^{(m+2)}(-{\bm q}_1,\dots,-{\bm
q}_m,{\bm p}+{\bm k}+{\bm q}_{12\cdots m},-{\bm p})  \nn\\
& \times K_b^{(n-m+2)}(-{\bm q}_{m+1},\dots,-{\bm q}_n,-{\bm
k}-{\bm q}_{12\cdots m}-{\bm p},{\bm p})
\Bigg]^{p\to\infty}_{\text{av}_{\hat p}}
   = s^\text{1L}_{abc} K_c^{(n)}({\bm k}_1,\dots,{\bm k}_n)\,,
\ea
which defines the numerical coefficients $s^\text{1L}_{abc}$; the same coefficients were already considered in \cite{Rubira:2024tea} (see \refapp{tables}). Since Eq.~\eqref{eq:sabc1Ldef} is an
operator relation, the $s^\text{1L}_{abc}$ must be independent of $n$,
providing a non-trivial consistency check analogous to the $n$-independence
of $s^\text{1L}_{ab}$ in the deterministic sector. Inserting this
expansion into Eq.~\eqref{eq:stochxyrencond1L} gives
\be\label{eq:Zabc1L}
   Z_{abc}^\text{1L} =
-s^\text{1L}_{abc}\int_{p<\Lambda}P^\text{lin}(p)^2\,.
\ee
Therefore, while $\sigma^2_\Lambda = \int_{p<\Lambda}P^\text{lin}(p)$ controls the renormalization of the deterministic part, integrals of the type $\int_{p<\Lambda}\left[P^\text{lin}(p)\right]^n$ parametrize the renormalization of stochastic parameters (see \cite{Rubira:2024tea}).   

For $c={\bm 1}$, Eq. \eqref{eq:sabc1Ldef} reproduces Eq.~\eqref{eq:Nab1L}, with
\be\label{eq:sab11L}
   s^\text{1L}_{ab{\bm 1}} = 2n_{ab}^{(22)} = 2 n_a^{(2)}n_b^{(2)}\,,
\ee
while the $c\neq{\bm 1}$ coefficients are obtained from Eq.~\eqref{eq:sabc1Ldef}
at $n>0$.
For $n=1$ the sum over subsets in Eq.~\eqref{eq:sabc1Ldef} is trivial, and
the two terms $m=1$ and $m=0$ give
\ba\label{eq:stoch_n1}
 &  \Bigg[  6 K_a^{(3)}(-{\bm k}_1,{\bm p}+{\bm k}+{\bm k}_{1},-{\bm
p})K_b^{(2)}(-{\bm p}-{\bm k}-{\bm k}_{1},{\bm p})\\
  & \quad + 6 K_a^{(2)}({\bm p}+{\bm k},-{\bm p})K_b^{(3)}(-{\bm
k}_1,-{\bm p}-{\bm k},{\bm p})\Bigg]^{p\to\infty}_{\text{av}_{\hat p}}
   = s^\text{1L}_{abc} K_c^{(1)}({\bm k}_1) = s^\text{1L}_{ab\delta}\,. \nn
\ea
Explicit computation confirms that the left-hand side is independent of
${\bm k}$ and ${\bm k}_1$, as required by the operator relation. Note that
when both ${\cal O}_a$ and ${\cal O}_b$ are second-order operators,
terms scaling as ${\bm k}_1\cdot{\bm k}/{\bm k}_1^2$ cancel between the
two summands, as can be verified by expanding in the limit $k_1\ll k\ll p$
and using the Galilean-invariance relations (see e.g. App.~C of \cite{Garny:2022kbk})
\be
  K_a^{(m)}({\bm q}_1,{\bm q}_2,\dots,{\bm q}_m)\to
  \frac{1}{m}\frac{{\bm q}_1\cdot{\bm q}}{{\bm q}_1^2}K_a^{(m-1)}({\bm q}_2,\dots,{\bm q}_m)+{\cal O}(q_1^0)\quad \mbox{for}\ q_1\to 0
\ee
among the bias kernels, where ${\bm q}=\sum_i {\bm q}_i$. 

The $n=1$ case in Eq.~\eqref{eq:sabc1Ldef} corresponds precisely to the $B^{(321),I}_{abc}$ bispectrum (with
${\bm k}_1=-{\bm k}-{\bm k}'$), confirming that its renormalization is
governed by the $s^\text{1L}_{ab\delta}$ coefficients.
For $n=2$ the sum over subsets produces three groups of terms,
\ba\label{eq:stoch_n2}
   &\Bigg[  12 K_a^{(4)}(-{\bm k}_1,-{\bm k}_2,{\bm p}+{\bm
k}+{\bm k}_{12},-{\bm p})K_b^{(2)}(-{\bm p}-{\bm k}-{\bm k}_{12},{\bm
p})\\
   &\quad + 9 \left( K_a^{(3)}(-{\bm k}_1,{\bm p}+{\bm k}+{\bm
k}_{1},-{\bm p})K_b^{(3)}(-{\bm k}_2,-{\bm p}-{\bm k}-{\bm k}_{1},{\bm
p}) + ({\bm k}_1\leftrightarrow {\bm k}_2)
\right)\nn\\
   &\quad + 12 K_a^{(2)}({\bm p}+{\bm k},-{\bm p})K_b^{(4)}(-{\bm
k}_1,-{\bm k}_2,-{\bm p}-{\bm k},{\bm p})\Bigg]^{p\to\infty}_{\text{av}_{\hat p}}
   = s^\text{1L}_{abc} K_c^{(2)}({\bm k}_1,{\bm k}_2) \,.\nn
\ea
These are the stochastic loop contributions from the product of two
operators that enter the renormalization of the one-loop trispectrum
contributions\footnote{Here the superscript $I$ refers to the topology with the mode-coupling loop with two vertices, as for the bispectrum loop. In the notation of \cite{Steele:2021lnz} these are $T_{3311b}$ and $T_{4211b}$, see e.g. their Fig. 4.} $T^{(4211),I}$ and $T^{(3311),I}$.
We obtain $s^\text{1L}_{abc}$ for operators ${\cal O}_c$ up to second
perturbative order, and operator pairs $({\cal O}_a,{\cal O}_b)$ at orders
$(3,3)$ or $(4,2)$, respectively.
The $n$-independence of the $s^\text{1L}_{ab\delta}$ entries obtained from
the $n=2$ limits was verified to be consistent with the $n=1$ results
above, as expected.
The full set of coefficients\footnote{Note that this set of coefficients is symmetric in $a,b$, as can be checked from Eqs. \eqref{eq:stoch_n1},\eqref{eq:stoch_n2} in the $p \to \infty$ limit.} is presented in
Tables~\ref{tab:sabtrPi11L}--\ref{tab:sabtrPi1trPi11L}.

\subsection{Product of three operators}

For three operators at locations ${\bm x}$, ${\bm y}$, and ${\bm z}$, we generalize Eq.~\eqref{eq:stochxy} to
\ba\label{eq:stochxyz}
   & [{\cal O}_a({\bm x}){\cal O}_b({\bm y}){\cal O}_c({\bm z})] =
[{\cal O}_a({\bm x})][{\cal O}_b({\bm y})][{\cal O}_c({\bm z})] + Z_{abd}[{\cal O}_d({\bm X}_{xy}){\cal O}_c({\bm
z})](2\pi)^3\delta_D({\bm x}-{\bm y}) \nn\\
   & \quad + Z_{bcd}[{\cal O}_d({\bm X}_{yz}){\cal O}_a({\bm
x})](2\pi)^3\delta_D({\bm y}-{\bm z})  + Z_{acd}[{\cal O}_d({\bm X}_{xz}){\cal O}_b({\bm
y})](2\pi)^3\delta_D({\bm x}-{\bm z})  \vs 
& \quad + Z_{abcd}[{\cal O}_d({\bm X}_{xyz})](2\pi)^3\delta_D({\bm
x}-{\bm y})(2\pi)^3\delta_D({\bm x}-{\bm z})\,,
\ea
where ${\bm X}_{xy}=({\bm x}+{\bm y})/2$ and ${\bm X}_{xyz}=({\bm x}+{\bm y}+{\bm z})/3$. The terms involving a single Dirac delta contain renormalization constants already fixed by the two-operator product renormalization of the previous subsection. The last line introduces new coefficients $Z_{abcd}$ that capture mode-coupling loops involving all three bias operators simultaneously. Once more, the sum over $d$ includes the unit operator and we assume ${\cal O}_a$, ${\cal O}_b$, and ${\cal O}_c$ are all at least linear in the density field. The Fourier-space form of Eq.~\eqref{eq:stochxyz} reads
\ba\label{eq:stochxyzFourier}
   &[{\cal O}_a({\bm k}){\cal O}_b({\bm k}'){\cal O}_c({\bm k}'')] =
[{\cal O}_a({\bm k})][{\cal O}_b({\bm k}')][{\cal O}_c({\bm k}'')] + Z_{abd}[{\cal O}_d({\bm k}+{\bm k}'){\cal O}_c({\bm k}'')] \\
 &\quad+ Z_{bcd}[{\cal O}_d({\bm k}'+{\bm k}''){\cal O}_a({\bm k})] + Z_{acd}[{\cal O}_d({\bm k}+{\bm k}''){\cal O}_b({\bm k}')]+ Z_{abcd}[{\cal O}_d({\bm k}+{\bm k}'+{\bm k}'')]\,,\nn
\ea
and the new coefficients $Z_{abcd}$ are fixed by the renormalization condition that generalizes Eq.~\eqref{eq:stochxyrencond}:
\bea\label{eq:stochxyzrencond}
   \lefteqn{\langle[{\cal O}_a({\bm k}){\cal O}_b({\bm k}'){\cal
O}_c({\bm k}'')]\delta_L({\bm k}_1)\cdots\delta_L({\bm
k_n})\rangle\Big|_{{\bm k},{\bm k}',{\bm k}'',{\bm k}_i\to 0}}\nn\\
   &=& \langle[{\cal O}_a({\bm k})][{\cal O}_b({\bm k}')][{\cal
O}_c({\bm k}'')]\delta_L({\bm k}_1)\cdots\delta_L({\bm
k_n})\rangle\Big|^\text{tree}_{{\bm k},{\bm k}',{\bm k}'',{\bm k}_i\to 0}\,.
\eea

The case $n=0$ determines the unit-operator contact term, which encodes the stochastic bispectrum noise amplitude. Next, we define
\be\label{eq:dn_abc}
   \Delta N_{abc} \equiv Z_{abc{\bm 1}}\,,
\ee
and introduce the renormalized bispectrum
\be
   \langle[{\cal O}_a({\bm k}){\cal O}_b({\bm k}'){\cal O}_c({\bm
k}'')]\rangle = B_{[abc]}(k,k',k'')(2\pi)^3\delta_D({\bm k}+{\bm
k}'+{\bm k}'')\,,
\ee
as well as the partially renormalized bispectrum with only deterministic renormalization applied,
\be
   \langle[{\cal O}_a({\bm k})][{\cal O}_b({\bm k}')][{\cal O}_c({\bm
k}'')]\rangle = B_{[a][b][c]}(k,k',k'')(2\pi)^3\delta_D({\bm k}+{\bm
k}'+{\bm k}'')\,.
\ee
Applying the condition Eq.~\eqref{eq:stochxyzrencond} at $n=0$ and using $P^\text{lin}(0)=0$ then gives
\be
   \Delta N_{abc} = - B_{[a][b][c]}(0,0,0)\,.
\ee
At one-loop order only the $B^{(222)}$ contribution survives for vanishing external wavenumbers, giving
\be\label{eq:Nabc1L}
   \Delta N_{abc}^\text{1L} =
-8n_{abc}^{(222)}\int_{p<\Lambda}P^\text{lin}(p)^3\,,
\ee
where
\be
   n_{abc}^{(222)}=\left(K^{(2)}_a({\bm k}-{\bm p},{\bm
p})K^{(2)}_b({\bm k}'+{\bm k}-{\bm p},{\bm p}-{\bm k})K^{(2)}_c({\bm
k}''+{\bm p},-{\bm p})\right)^{p\to\infty}_{\text{av}_{\hat p}} =
n_a^{(2)}n_b^{(2)}n_c^{(2)}\,.
\ee

At one-loop order, the renormalization condition Eq.~\eqref{eq:stochxyzrencond} reads
\ba\label{eq:stochxyzrencond1L}
    & \langle [{\cal O}_a({\bm k})][{\cal O}_b({\bm k}')][{\cal
O}_c({\bm k}'')]\delta_L({\bm k}_1)\cdots\delta_L({\bm
k_n})\rangle^\text{1L}  \nn\\
     &\quad + \left(Z_{abd}^\text{1L}\langle[{\cal O}_d({\bm k}+{\bm k}'){\cal
O}_c({\bm k}'')]\delta_L({\bm k}_1)\cdots\delta_L({\bm
k_n})\rangle^\text{tree} \ + 2\ \text{permutations}\right)\nn\\
     &\quad + Z_{abcd}^\text{1L}\langle[{\cal O}_d({\bm k}+{\bm k}'+{\bm
k}'')]\delta_L({\bm k}_1)\cdots\delta_L({\bm
k_n})\rangle^\text{tree}\Big|_{{\bm k},{\bm k}',{\bm k}'',{\bm k}_i\to
0}=0\,.
\ea
At tree level, square brackets may be dropped since renormalized and bare operators agree. The second line absorbs stochastic loops connecting
two of the three operators (already fixed in the previous subsection), while deterministic loops are accounted for by the individual-operator
renormalization in the first line. Only the genuinely three-operator stochastic loop remains, which is cancelled by the last line, thereby
fixing $Z_{abcd}^\text{1L}$. Expanding the mode-coupling loop in the UV limit gives
\ba\label{eq:sabcd1Ldef}
  &\frac{1}{n!}\Bigg[ \sum_{m_1=0}^n\sum_{m_2=0}^{n-m_1} \
\sum_{\{{\bm q}_1,\dots,{\bm q}_{m_1+m_2}\} \atop  \subset\{{\bm
k}_1,\dots,{\bm k}_n\}}\ \sum_{\{{\bm q}_1,\dots,{\bm q}_{m_1}\} \atop
\subset\{{\bm q}_1,\dots,{\bm q}_{m_1+m_2}\}} \hspace{-0.8cm} 4\times
2\frac{(m_1+2)!}{2}\frac{(m_2+2)!}{2}\frac{(n-m_1-m_2+2)!}{2}\nn\\
& \quad   \times K_c^{(n-m_1-m_2+2)}(-{\bm q}_{m_1+m_2+1},\dots,-{\bm
q}_n,-{\bm p}-{\bm k}-{\bm k}'-{\bm q}_{12\cdots {m_1+m_2}},{\bm p})
\\
& \quad   \times K_b^{(m_2+2)}(-{\bm q}_{m_1+1},\dots,-{\bm
q}_{m_1+m_2},-{\bm p}-{\bm k}-{\bm q}_{12\cdots {m_1}},{\bm p}+{\bm
k}+{\bm k}'+{\bm q}_{12\cdots {m_1+m_2}})  \nn\\
& \quad   \times K_a^{(m_1+2)}(-{\bm q}_1,\dots,-{\bm q}_{m_1},{\bm
p}+{\bm k}+{\bm q}_{12\cdots {m_1}},-{\bm p})\Bigg]^{p\to\infty}_{\text{av}_{\hat p}}    = s^\text{1L}_{abcd} K_d^{(n)}({\bm k}_1,\dots,{\bm k}_n)\,,\nn
\ea
which defines the coefficients $s^\text{1L}_{abcd}$. Inserting this
into Eq.~\eqref{eq:stochxyzrencond1L} gives
\be \label{eq:Zabcd}
   Z_{abcd}^\text{1L} =
-s^\text{1L}_{abcd}\int_{p<\Lambda}P^\text{lin}(p)^3\,.
\ee
For $d={\bm 1}$ this reproduces Eq.~\eqref{eq:Nabc1L}, with
\be\label{eq:sabc11L}
   s^\text{1L}_{abc{\bm 1}} = 8n_{abc}^{(222)} = 8
n_a^{(2)}n_b^{(2)}n_c^{(2)}\,.
\ee
The $n=1$ case is relevant for renormalizing the trispectrum contribution\footnote{Again in the notation of \cite{Steele:2021lnz}, the relevant diagram is $T_{3221c}$ in their Fig. 4. The other two topologies $T_{3221a}, T_{3221b}$ are already covered by deterministic renormalization and the two-operator contact terms, respectively. In what follows, we will omit the explicit labeling to avoid clutter. }
$T^{(3221),I}$. Evaluating Eq.~\eqref{eq:sabcd1Ldef} at $n=1$ yields
\ba
&4\times 2\times 3 \Bigg[ K_a^{(2)}({\bm p}+{\bm k},-{\bm p})K_b^{(2)}(-{\bm p}-{\bm k},{\bm p}+{\bm k}+{\bm k}')  K_c^{(3)}(-{\bm k}_1,-{\bm p}-{\bm k}-{\bm k}',{\bm p})+\nn\\   
&    
K_a^{(3)}(-{\bm k}_1,{\bm p}+{\bm k}+{\bm k}_{1},-{\bm p})
    K_b^{(2)}(-{\bm p}-{\bm k}-{\bm k}_{1},{\bm
p}+{\bm k}+{\bm k}'+{\bm k}_{1}) K_c^{(2)}(-{\bm p}-{\bm k}-{\bm k}'-{\bm
k}_{1},{\bm p}) \vs
&+ K_a^{(2)}({\bm p}+{\bm k},-{\bm p})    K_b^{(3)}(-{\bm k}_1,-{\bm p}-{\bm k},{\bm
p}+{\bm k}+{\bm k}'+{\bm k}_{1}) K_c^{(2)}(-{\bm p}-{\bm k}-{\bm k}'-{\bm k}_{1},{\bm p})\Bigg]^{p\to\infty}_{\text{av}_{\hat p}}   \vs
 & \hspace{7cm}= s^\text{1L}_{abcd} K_d^{(1)}({\bm k}_1) = s^\text{1L}_{abc\delta}\,,
\ea
We find $s^\text{1L}_{ab\delta\delta}=0$; the non-trivial coefficients for $c=\text{tr}(\Pi^{[1]})^2$ and $c=(\text{tr}\Pi^{[1]})^2$ are listed in Tables~\ref{tab:sabtrPi1Pi1delta1L} and~\ref{tab:sabtrPi1trPi1delta1L}.

\subsection{Product of four operators}

The four-operator product is required to renormalize the trispectrum contribution $T^{(2222)}$ and, more generally, any mode-coupling loop involving four bias operators. The generalization of Eq.~\eqref{eq:stochxyz} reads
\bea\label{eq:stochxyzw}
   \lefteqn{[{\cal O}_a({\bm x}){\cal O}_b({\bm y}){\cal O}_c({\bm
z}){\cal O}_d({\bm w})] = [{\cal O}_a({\bm x})][{\cal O}_b({\bm
y})][{\cal O}_c({\bm z})][{\cal O}_d({\bm w})] } \nn\\
   && {} + Z_{abe}[{\cal O}_e({\bm X}_{xy}){\cal O}_c({\bm z}){\cal
O}_d({\bm w})](2\pi)^3\delta_D({\bm x}-{\bm y}) + 5\
\text{permutations}\nn\\
   && {} + Z_{abce}[{\cal O}_e({\bm X}_{xyz}){\cal O}_d({\bm
w})](2\pi)^3\delta_D({\bm x}-{\bm y})(2\pi)^3\delta_D({\bm x}-{\bm z}) +
3\ \text{permutations}\nn\\
   && {} + Z_{abe}Z_{cdf}[{\cal O}_e({\bm X}_{xy}){\cal O}_f({\bm
X}_{zw})](2\pi)^3\delta_D({\bm x}-{\bm y})(2\pi)^3\delta_D({\bm z}-{\bm
w}) + 2\ \text{permutations}\nn\\
   && {} + Z_{abcde}[{\cal O}_e({\bm X}_{xyzw})](2\pi)^3\delta_D({\bm
x}-{\bm y})(2\pi)^3\delta_D({\bm x}-{\bm z})(2\pi)^3\delta_D({\bm
x}-{\bm w})\,,
\eea
where the last line introduces the new coefficient $Z_{abcde}$. The second and third lines account for contact terms in which two or three of the four arguments coincide. The fourth line is a `two-loop term' in the sense that it is only required to renormalize two-loop diagrams (since $Z_{abc}$ starts at one-loop order)\footnote{This contribution would renormalize e.g. the $T^{(3322)}$ diagram in the two-loop trispectrum where two $P_{22}$-like diagrams are joined along a vertex. The resulting diagram is renormalized by the squares of the same coefficients that render $B^{(321),I}$ finite. }. However, looking ahead, we include it here since the (leading-order) renormalized contribution is expected to be of order $P_L(k)/\bar{n}^2\sim P_L(k)^3$. Thus, it is as important as a \textit{tree-level} term; we expand on this in Section \ref{subsec:trisp}. All of the coefficients from lines two to four are already determined by the two- and three-operator renormalization of the previous subsections. For the trispectrum, only the unit-operator component of the last line is needed, defining
\be\label{eq:dn_abcd}
   \Delta N_{abcd} \equiv Z_{abcd{\bm 1}}\,.
\ee
The new coefficient $Z_{abcde}$ is fixed by the renormalization
condition that generalizes Eq.~\eqref{eq:stochxyzrencond}:
\ba\label{eq:stochxyzwrencond}
   &\langle[{\cal O}_a({\bm k}){\cal O}_b({\bm k}'){\cal
O}_c({\bm k}''){\cal O}_d({\bm k}''')]\delta_L({\bm
k}_1)\cdots\delta_L({\bm k_n})\rangle\Big|_{{\bm k},{\bm k}',{\bm
k}'',{\bm k}''',{\bm k}_i\to 0}\nn\\
   &\quad = \langle[{\cal O}_a({\bm k})][{\cal O}_b({\bm k}')][{\cal
O}_c({\bm k}'')][{\cal O}_d({\bm k}''')]\delta_L({\bm
k}_1)\cdots\delta_L({\bm k_n})\rangle\Big|^\text{tree}_{{\bm k},{\bm
k}',{\bm k}'',{\bm k}''',{\bm k}_i\to 0}\,.
\ea
Following the same steps as in the previous subsections, we introduce
the renormalized trispectrum
\bea
   \langle[{\cal O}_a({\bm k}){\cal O}_b({\bm k}'){\cal O}_c({\bm
k}''){\cal O}_d({\bm k}''')]\rangle' &=& T_{[abcd]}({\bm k},{\bm k}',{\bm
k}'') ,
\eea
and the partially renormalized trispectrum with only deterministic
renormalization applied,
\bea
   \langle[{\cal O}_a({\bm k})][{\cal O}_b({\bm k}')][{\cal O}_c({\bm
k}'')][{\cal O}_d({\bm k}''')]\rangle' &=& T_{[a][b][c][d]}({\bm k},{\bm
k}',{\bm k}'')\,.
\eea
Using $P^\text{lin}(0)=0$, Eq.~\eqref{eq:stochxyzwrencond} then gives
\be
   \Delta N_{abcd} = - T_{[a][b][c][d]}(0,0,0)\,.
\ee
At one-loop order, only the $T^{(2222)}$ contribution survives for
vanishing external wavenumbers, giving
\be\label{eq:Nabcd1L}
   \Delta N_{abcd}^\text{1L} = -(6\times 4\times
2)n_{abcd}^{(2222)}\int_{p<\Lambda}P^\text{lin}(p)^4\,,
\ee
where
\be
   n_{abcd}^{(2222)}= n_a^{(2)}n_b^{(2)}n_c^{(2)}n_d^{(2)}\,.
\ee

\newpage
\section{From renormalized operators to correlators}\label{sec:correlators}

Up to this point we have established the renormalization of the deterministic bias operators (\refsec{renorm}) and of the stochastic contact terms (\refsec{stochastic}). In this section we translate those operator-level results into explicit prescriptions for the renormalized $N$-point functions of biased tracers. We first discuss the general structure of renormalized correlators and the way in which deterministic and stochastic renormalization enter at each loop order, and then focus on the two-loop power spectrum, the one-loop bispectrum, and the one-loop trispectrum. In each case we identify which operators yield distinct, non-degenerate contributions.
We close the section with a unified counting (\reftab{operator_counting}) of the LG, NLG, and stochastic operators required for the renormalization of different combinations of $N$-point functions of practical relevance, ranging from a stand-alone one-loop power spectrum to the full set $\{P^\text{2L},B^\text{1L},T^\text{1L}\}$.

In all cases, bias coefficients that appear in several $N$-point
functions are renormalized consistently, i.e.~have identical values in
the respective
spectra. This is ensured by the operator-level renormalization
procedure, which ensures cancellation of UV-sensitive contributions to loop
integrals when inserted into any correlators. We start with outlining the general structure (which is applicable \textit{at any loop
order and to any correlator}). For conciseness, we use indices with capital letters to denote
both LG and NLG bias operators. 

\subsection{General structure}

For any $N$-point function of the biased-tracer field, the bias expansion $\delta_g = b_A {\cal O}_A$ together with the renormalization relations of Eqs.~\eqref{eq:Oren1L2LNLG}, \eqref{eq:O2ren1L}, and the stochastic contact terms of Eqs.~\eqref{eq:stochxy}, \eqref{eq:stochxyz}, \eqref{eq:stochxyzw} yields a finite, $\Lambda$-independent prediction provided that all bias coefficients are reinterpreted as renormalized parameters.
Schematically, at a given loop order $\ell$ we have
\ba\label{eq:Ngen}
   \langle \delta_g({\bm k}_1)\cdots\delta_g({\bm k}_N)\rangle^{\text{$\ell$-loop}}_{\text{ren.}}
   = \sum_{A_1,\dots,A_N} b_{[A_1]}\cdots b_{[A_N]}\,\langle [{\cal O}_{A_1} \cdots {\cal O}_{A_N}]\rangle^\text{$\ell$-loop} + \text{ ren. stoch.}
\ea
We comment on the second `renormalized stochastic' piece shortly. The \emph{deterministic loops} (first term above) arise from Wick contractions among the fields inside the individual renormalized operators $[{\cal O}_{A_i}]$. The second term contains the \emph{stochastic (contact-term) loops}, arising from diagrams in which subsets of the external operators are evaluated at coincident spatial points. These are the mode-coupling loops absorbed by the contact-term renormalization constants of \refsec{stochastic}. Indeed, to evaluate the renormalized product at $\ell$ loops in Fourier space, one takes the Fourier transform of the real-space expression, e.g. Eq. \eqref{eq:stochxyzFourier} and truncates it at the appropriate loop order by expanding the operators (using also $Z = Z^{1\text{L}} + \dots$). In general, the stochastic contributions have rich structure, where individual lower $N'$-point  functions ($N' < N$) can contribute at $\ell'$-loop order for any $\ell' < \ell$. These are evaluated at specific linear combinations of the external momenta ${\bm k_1}, \dots {\bm k_N}$. 

The renormalization constants $Z$ occurring in the operator product expansion, whose values need to be fixed by imposing renormalization conditions, are $\Lambda$-dependent. Their purpose is precisely to cancel the UV dependence in products of operators (evaluated beyond tree level) that is not already accounted for via deterministic renormalization. We saw that these constants could be fixed by evaluating the appropriate $N$-point function renormalization condition. For example, for $Z_{ab{\bm 1}}$ this is the power spectrum; for $Z_{ab\delta}$ it is the bispectrum; for $Z_{abc}$ with $c$ starting at second order it is the trispectrum and so on. Evaluating these renormalization conditions at a given loop order then yields a new contribution to said $N$-point function; in the examples above it corresponds to a constant term, a term proportional to $P^\text{lin}(k_1) + P^\text{lin}(k_2) + P^\text{lin}(k_3)$ (after symmetrizing) and a term proportional to a tracer bispectrum, respectively. 

In the spirit of the EFT, such contributions need to be included in predictions for the corresponding $N$-point functions \textit{with free coefficients}. This is completely analogous to e.g. the NLG renormalization of deterministic operators: at second order in gradients, additional UV limits of bias kernels arise. Then, each of these UV limits is promoted to a new bias operator with a computable `bare' component, which cancels the UV limit, and a `finite' renormalized part, which cannot be predicted and needs to be determined by matching to data. Note however that since the operator product expansion is valid at the operator level, a new parameter only needs to be introduced for the \textit{lowest} order operator product to which a renormalization constant $Z$ contributes. Thus, for every renormalization constant $Z_{A_1A_2\dots A_n}$, a \textit{single} new stochastic amplitude corresponding to the $A_n$ operator is introduced (as one would also do in the $n=2$ case in the NLG example above).

Let us now explain our notation for the EFT parameters related to stochastic contributions.
In this work, we  need $n=3,4,5$. The corresponding free EFT parameters associated with these cases will be indicated by the letters $X = c,d,e$ (in analogy to $X = b$ for $n=2$, which corresponds to deterministic renormalization). In each case, a subscript $A$ denotes the operator ${\cal O}_A$ associated with renormalizing the contact terms for the product of $n-1$ operators at coincident locations, i.e.~related to the corresponding  stochastic renormalization constants $Z_{A_1\dots A_{n-1}A}$. In this way, we also recover the notation $[c_\mathbf{1}] \equiv [c]$ from \hyperlink{cite.Bakx:2025cvu}{Paper~I} for the stochastic renormalization of the power spectrum. We thus denote the renormalized coefficients by $[c_A], [d_A], [e_A]$ for $n=3,4,5$, respectively\footnote{This is slightly different from the $b_{[A]}$ notation we had for $n=2$; the reason for this distinction will become clear below.}.   

All in all, this structure implies that the renormalized stochastic piece in Eq. \eqref{eq:Ngen} contains combinations of \textit{renormalized} lower-order, lower-loop $N$-point functions multiplied by free coefficients. Schematically, this amounts to 
\begin{eqnarray}\label{eq:ren_stoch_ctr}
    \text{ren. stoch.} = \sum_{m=1}^{N-1}[X_{A_1}]\cdots [X_{A_m}]\,\langle [{\cal O}_{A_1} \cdots {\cal O}_{A_m}]\rangle^\text{$\ell'$-loop}
\end{eqnarray}
where the renormalized coefficient $[X_{A_i}]$ equals $b_{[A_i]}$ (a deterministic bias) or a stochastic amplitude $[\{c,d,e,\cdots \}_{A_i}]$. These amplitudes are renormalized by mode-coupling loops with the corresponding number $n-1=2,3,4$ of vertices, and thus acquire a $\Lambda$-dependence for which we will derive a stochastic RG equation. 

In this work we restrict to stochastic renormalization at LG order. Since the LG amplitudes $[\{c,d,e,\cdots \}_{A_i}]$ are dimensionful (they have units of $n-1$ powers of volume for $n=3,4,\dots$), this implies that there is a scale $k_\epsilon$ associated to them\footnote{Note that this refers to the scale of the \textit{renormalized} parameter, while the running scale is governed by the dimensionful constants $Z_{a_1a_2\cdots a_n}(\Lambda) \sim \int^\Lambda P_L(p)^{n-1}$, as we saw in \refsec{stochastic}. Unlike $\bar{n}^{-1}$, these constants do not depend on sample properties, but rather only on the initial conditions.}. This scale is expected to be of order $k_\epsilon^3 = \bar{n}$ where $\bar{n}$ is the number density of the tracer in question. Typical spectroscopic surveys roughly reach scales where $\bar{n} P^\text{lin}(k) \approx 1$, which implies that $k_\epsilon \approx k_{\text{NL}}$. Thus, each occurrence of a renormalized stochastic amplitude is counted as $2(n-2)$ powers of $\delta$. In this setup, the tree-level renormalized stochastic contribution from Eq. \eqref{eq:ren_stoch_ctr} is  \textit{of the same order as deterministic tree-level contributions}. We will therefore group these two contributions together in what follows. Consequently, it is not strictly consistent to compute the deterministic expansion for e.g. the power spectrum to fifth order (and NLG terms to third order) without also including stochastic corrections at higher orders in gradients and higher loops. However, for simplicity,
we will ignore this issue and only perform stochastic renormalization at LG order and evaluate these contributions at tree-level. A generalization to stochastic renormalization beyond LG order is left for future work.

We are now ready to specialize to the $N$-point functions of interest. For each of the power spectrum, bispectrum and trispectrum, we  first write out explicitly the bare and renormalized expressions in terms of bias parameters and operator spectra. Then, we specialize further to the loop order relevant for this work (two for the power spectrum, one for the bispectrum and trispectrum).
Keeping in mind
that second-gradient operators are counted equally to a loop
suppression, only LG operators contribute in many of the summations in
practice, in particular
within the noise terms, for which we include LG renormalization only at
present. 

\subsection{Power Spectrum}

Let us start with the power spectrum, generalizing results from \hyperlink{cite.Bakx:2025cvu}{Paper~I}.
For the power spectrum, the structure of the (LG) renormalized stochastic contribution is trivial: we have 
\be
\text{ren. stoch. }(P) = [c_{\bm 1}].
\ee
The tracer power spectrum can be written in terms of bare or renormalized quantities as (using summation convention for repeated
indices) 
\bea\label{eq:pren}
P(k) &=& b_{A} b_{B} \, P_{AB}(k)
+
c_{\bm 1} \nn \\
&=& b_{[A]} b_{[B]} \, P_{[AB]}(k)
+
[c_{\bm 1}]\,.
\eea
The $\Lambda$-independent coefficient $[c_{\bm 1}] \equiv [c]$ (whose definition is the above identity) is renormalized by the UV-limit of the product of two operators: indeed, taking the Fourier transform of Eq. \eqref{eq:stochxy}, contracting with $b_{[A]} b_{[B]}$ and taking the ensemble average we get
\bea \label{eq:prenorm}
b_{[A]} b_{[B]} \, P_{[AB]} (k) &=& b_{[A]} b_{[B]} P_{[A][B]} (k)+ b_{[A]}b_{[B]}Z_{AB\bm 1} \nn\\
&=& b_A b_B P_{AB} (k)+ b_{[A]}b_{[B]}Z_{AB\bm 1}
\eea
so that $[c]$ can be regarded as the difference between a bare coefficient $c$ and the UV-dependent part of the operator product, which is the second term in Eq. \eqref{eq:prenorm}.

Thus, the bare and renormalized coefficients are related via 
\be
   b_{[A]} = b_{C}([1+Z]^{-1})_{CA},\qquad  [c_{\bm 1}]=c_{\bm 1}-b_{[A]}b_{[B]}\Delta
N_{AB}\,,
\ee
where the matrices $Z_{AB}$ and $\Delta N_{AB} \equiv Z_{AB\bm 1}$ are determined by the
respective deterministic and stochastic operator renormalization conditions. 
We give concrete expressions for the one- and two-loop approximation below.

At up to one-loop order, the power spectrum  expressed in terms of renormalized bias and noise terms reads\footnote{Here and in what follows, we reserve the superscript `$n$-loop' for contributions that are specifically of that order, in contrast to the `$n$L' superscript, which includes all terms up to a given loop order $n$.}
\bea\label{eq:P1L}
P^\text{tree}(k)
&=&
b_{[\delta]}^2P^\text{lin}(k) +  [c_{\bm 1}]\,,\nn\\
P^\text{1L}(k)
&=&
P^\text{tree}(k) + \sum_{ab} b_{[a]} b_{[b]} \,
P_{[ab]}^\text{1-loop}(k)
+ 2b_{[\nabla^2\delta]}b_{[\delta]}\, k^2 P^\text{lin}(k)
\, ,
\eea
and at two-loop order 
\bea\label{eq:P2L}
P^\text{2L}(k)
&=& P^\text{1L}(k) + \sum_{ab} b_{[a]} b_{[b]} \, P_{[ab]}^\text{2-loop}(k) \nn\\
&& {} +
2\sum_{a_2b} b_{[a_2]} b_{[b]} \, P_{[a_2b]}^\text{1-loop}(k) + \left((b_{[\nabla^2\delta]})^2+2b_{[\nabla^4\delta]}b_{[\delta]}\right)\, k^4 P^\text{lin}(k)
  \, .
\eea
We now summarize the operator content in the one- and two-loop power spectrum. Before turning to two-loop order, we briefly recall the operator content of the renormalized one-loop power spectrum. The one-loop power spectrum receives two types of contributions $P^{(22)}_{[ab]}$ and $P^{(13)}_{[ab]}$. Thus, the sum over $ab$ in the second line of Eq. \eqref{eq:P1L} receives contributions from up to third order LG operators. Restricting the third-order operators to the $K_{[b]}^{(3)}({\bm k},{\bm q},-{\bm q})$ configuration and taking into account that deterministic bias renormalization amounts to replacing the naive kernel $K_{b}^{(3)}({\bm k},{\bm q},-{\bm q})$ of the bare operator by the UV-subtracted renormalized kernel from Eq.~\eqref{eq:RenKernel1loop},\footnote{This can be done in two ways: (i) using the renormalized kernel as stated in Eq.~\eqref{eq:RenKernel1loop} and taking into account that the direction of ${\bm q}$ is integrated over, or (ii) omitting the angle-average in the subtraction term in Eq.~\eqref{eq:RenKernel1loop}. We find that both methods yield identical results, and therefore adopt the more convenient option (ii). We furthermore note that the operator redundancy analysis yields equivalent results when adopting either the LG or the NLG renormalization scheme, which amounts to subtracting either only the leading UV-limit or also the $1/q^2$-part (as in Eq.~\ref{eq:RenKernel1loop}) in the renormalized kernel. The same applies analogously at two-loop order.} reduces the number of operators to 4. The single NLG operator that does enter is $\nabla^2\delta$. In the stochastic sector, the one-loop power spectrum  requires a single stochastic coefficient $[c_{\bm 1}]$.

The two-loop power spectrum requires bias kernels up to perturbative order five, involving the fourth and fifth-order kernels $K_{[a]}^{(4)}({\bm k},{\bm k} - {\bm q},{\bm p},-{\bm p})$ and $K_{[a]}^{(5)}({\bm k},{\bm p},-{\bm p},{\bm q},-{\bm q})$ only at these fixed momenta configurations, and evaluated for the renormalized bias operators, amounting to using the UV subtracted kernels  from Eq.~\eqref{eq:RenKernel1loop} and Eq.~\eqref{eq:Ka2ren} at the one- and two-loop level for deterministic loops for the fourth- and fifth-order kernels, respectively. Out of the 29 LG operators, we only need 17, as per the result from \hyperlink{cite.Bakx:2025cvu}{Paper~I} (see also~\cite{Eggemeier:2018qae}). These operators are 
\bea
    \label{eq:bias_basis}
{\rm (1)} \qquad 
    &{\rm tr}\big[ \Pi^{[1]} \big], \\
{\rm (2)} \qquad 
    &{\rm tr}\big[ \big( \Pi^{[1]} \big)^2 \big], 
    ~~ \Big( {\rm tr}\big[ \Pi^{[1]} \big] \Big)^2, \non\\
{\rm (3)} \qquad 
   &{\rm tr}\big[ \big( \Pi^{[1]} \big)^3 \big],
   ~~{\rm tr}\big[ \big( \Pi^{[1]} \big)^2 \big] {\rm tr}\big[ \Pi^{[1]} \big],
   ~~ \Big( {\rm tr}\big[ \Pi^{[1]} \big] \Big)^3,
   ~~{\rm tr}\big[\Pi^{[1]} \Pi^{[2]} \big], \non\\
{\rm (4)}  \qquad 
   &{\rm tr}\big[\Pi^{[1]} \Pi^{[3]} \big], 
   ~~{\rm tr}\big[\Pi^{[1]} \Pi^{[1]} \Pi^{[2]} \big],    
   ~~{\rm tr}\big[\Pi^{[2]} \Pi^{[2]} \big],   
   ~~{\rm tr}\big[ \Pi^{[1]} \big] {\rm tr}\big[\Pi^{[1]} \Pi^{[2]} \big], \non\\
 {\rm (5)} \qquad 
 & {\rm tr}\big[ \Pi^{[1]} \big] {\rm tr}\big[\Pi^{[1]} \Pi^{[3]} \big],
 ~~{\rm tr}\big[\Pi^{[1]} \Pi^{[4]} \big],   
 ~~{\rm tr}\big[\Pi^{[1]} \Pi^{[1]} \Pi^{[3]} \big], \non\\  
& {\rm tr}\big[\Pi^{[1]} \Pi^{[2]} \Pi^{[2]} \big],    
~~{\rm tr}\big[\Pi^{[2]} \Pi^{[3]} \big],   
~~{\rm tr}\big[ \Pi^{[1]} \big] {\rm tr}\big[\Pi^{[2]} \Pi^{[2]} \big],\non
\eea

At second gradient order, the relevant new contributions are the mixed LG$\times$NLG one-loop diagrams, which take three forms depending on which operator carries the loop. First, $P^{(13)}_{a_2b}$, where the NLG operator contributes its tree-level kernel $K_{a_2}^{(1)}(k)=k^2$ and the LG operator carries the one-loop kernel $K_{[b]}^{(3)}$ (renormalized via Eq.~\eqref{eq:Karen}). 
Secondly, we have $P^{(31)}_{a_2b}$, where the NLG operator carries the one-loop kernel $K_{[a_2]}^{(3)}$ (renormalized via Eq.~\eqref{eq:Karen2L}), and finally $P^{(22)}_{a_2b}$, where both operators contribute second-order kernels. These require both LG and NLG operators to be up to second-order operators. 
For example, the well-known counterterm $2b_{[\nabla^2\delta]}b_{[\delta]}\,k^2 P_{\delta\delta}^\text{1-loop}(k)$ is recovered from $2b_{[a_2]}b_{[b]}P_{[a_2b]}^\text{1-loop}(k)$ for $a_2=\nabla^2\delta$ and $b=\delta$.
Note that the {\it renormalized} $b_{[\nabla^2\delta]}$ appearing there has the same value as that in $2b_{[\nabla^2\delta]}b_{[\delta]}\,k^2 P^\text{lin}(k)$, consistent with the fact that $\Lambda$-dependent UV contributions are already removed within the renormalized power spectra, and the corresponding renormalized counterterms capture only the physical, `finite' part which is independent of the loop order of the power spectrum that it is multiplied with.
Further, note that when counting gradient-suppression equally to insertions of $\delta_L$, 
contributions $P^{\text{1L}}_{a_2b_2}$ involving two NLG operators inside a one-loop term are formally counted equally to three-loop order, i.e.~beyond what we consider here. The only contributions suppressed by $(kR_L)^4$ that occur in the two-loop power spectrum are thus the $k^4P^\text{lin}$ terms from the last line in Eq.~\eqref{eq:P2L}.
Identifying which NLG operators give rise to \emph{distinct, non-degenerate} contributions when inserted into the one-loop power spectrum requires evaluating the corresponding kernels on the wavenumber configurations $K_{[a_2]}^{(2)}({\bm q},{\bm k}-{\bm q})$ and $K_{[a_2]}^{(3)}({\bm k},{\bm q},-{\bm q})$, and checking for linear dependencies after taking the UV-subtraction terms from deterministic bias renormalization according to Eq.~\eqref{eq:Ka2ren} into account, i.e.~using the renormalized instead of the bare kernel $K_{a_2}^{(3)}({\bm k},{\bm q},-{\bm q})$. We find that only the following nine NLG operators yield independent contributions: 
\bea\label{eq:NLGpklist}
   && \nabla^2\mbox{tr}(\Pi^{[1]}),\quad
      \nabla^2\mbox{tr}[(\Pi^{[1]})^2],\quad
      \nabla^2[\mbox{tr}(\Pi^{[1]})]^2,\quad
      \nabla_k\nabla_l[\Pi^{[1]}_{kl}\mbox{tr}(\Pi^{[1]})],\nn\\
   && \nabla^2\mbox{tr}[\Pi^{[1]}\Pi^{[2]}],\quad
      \nabla_k\nabla_l[(\Pi^{[1]}\Pi^{[2]})_{kl}],\quad
      \hat\Pi^{[3]},\quad \Pi_{B1},\nn\\
   && \mbox{tr}(\Pi^{[1]})\nabla^2\mbox{tr}(\Pi^{[1]})\,.
\eea
Of these, eight carry two overall gradients and a single one, $\mbox{tr}(\Pi^{[1]})\nabla^2\mbox{tr}(\Pi^{[1]})=\delta\nabla^2\delta$, does not. Momentum conservation guarantees that this last one is not needed for renormalizing the matter density two-loop power spectrum. It is required for a biased tracer, however. Thus,  we find that generalizing renormalization of the two-loop power spectrum from the matter density field to a general biased tracer requires introducing only a {\it single} extra NLG operator (i.e.~a single additional `counterterm' parameter). Furthermore, note that this same NLG operator already contributes to the one-loop galaxy bispectrum~\cite{Eggemeier:2018qae}. Notably, our results imply that for the two-loop galaxy power spectrum there are no extra operators without two overall derivatives at third order in perturbations.

The remaining $20-9=11$ NLG operators of \refsec{basis} that are not contained in Eq.~\eqref{eq:NLGpklist} are either trivially degenerate with this set when evaluated on the one-loop wavenumber configuration (the case of $\mbox{tr}[\Pi^{[2]}\nabla^2\Pi^{[1]}]$ and $\Pi_{B2}$, which reduce to combinations of operators in Eq.~\eqref{eq:NLGpklist} on $K^{(3)}({\bm k},{\bm q},-{\bm q})$), or have vanishing renormalized $K^{(3)}({\bm k},{\bm q},-{\bm q})$ kernels (all operators involving three or more factors of $\Pi^{[1]}$, analogously to the corresponding LG operators at one-loop). Moreover, we find that even though $\Pi_{B1}$ is independent of the other operators in Eq. \eqref{eq:NLGpklist} when evaluated on $({\bm k},{\bm p},-{\bm p})$, it is not required to absorb the single-hard UV limit of any of the kernels $K_a^{(5)}({\bm k},{\bm p},-{\bm p},{\bm q},-{\bm q})$ for all 29 LG bias operators ${\cal O}_a$ (in contrast to the case of a general configuration $K_a^{(5)}({\bm k}_1,{\bm k}_2,{\bm k}_3,{\bm q},-{\bm q})$ relevant for the trispectrum, see below). In that sense, one could omit this operator from the renormalized two-loop power spectrum\footnote{\label{eq:footnoteCompAnastasiou:2025jsy} This is consistent with the fact that the $K^{(2)}({\bm k}-{\bm q},{\bm q})$ and $K^{(3)}({\bm k},{\bm q},-{\bm q})$ kernels of the seven NLG counterterms with two overall derivatives reported in Eqs.~B.2 and B.6 of \cite{Anastasiou:2025jsy} can be written as linear combinations of the first seven operators in Eq.~\eqref{eq:NLGpklist} (after projecting out contributions that are degenerate with the leading EFT correction proportional to $k^2P^\text{lin}$, i.e.~considering the UV-subtracted kernels $K^{(3)}({\bm k},{\bm q},-{\bm q})-K^{(3)}({\bm k},{\bm q},-{\bm q})^{q\to\infty}$ that correspond to the kernels of renormalized operators as defined in Eq.~\ref{eq:Ka2ren}).}, but we do not do this here.  
We checked that, apart from $\Pi_{B1}$, all other NLG bias operators from Eq.~\eqref{eq:NLGpklist} are required with independent coefficients in order to be able to absorb the UV-dependence from the single-hard limit in $P^{(24)}$ as well as $P^{(15)}$ for a biased tracer described by a general linear combination of the 29 LG bias operators.\footnote{In other words, the $1/q^2$ contributions to the single-hard limits of linear combinations of kernels $\sum_a b_a K_a^{(5)}({\bm k},{\bm p},-{\bm p},{\bm q},-{\bm q})$ and $\sum_a b_a K_a^{(4)}({\bm k}-{\bm p},{\bm p},{\bm q},-{\bm q})$ with coefficients $b_a$ span the entire space of arbitrary linear combinations of the $K_{a_2}^{(3)}({\bm k},{\bm p},-{\bm p})$ and $K_{a_2}^{(2)}({\bm k}-{\bm p},{\bm p})$ kernels of the operators given in  Eq.~\eqref{eq:NLGpklist}, with the single exception of $\Pi_{B1}$.} This implies that in particular the extra operator $\hat\Pi^{[3]}$ is needed.

Adding the stochastic sector, the renormalization of the two-loop power spectrum generates a constant shot-noise renormalization $[c_{\bm 1}]$. We retain only its leading-gradient version here. As noted in \hyperlink{cite.Bakx:2025cvu}{Paper~I}, this means no additional free stochastic LG EFT parameters are needed at two- as compared to one-loop order.

\subsection{Bispectrum}

Here we demonstrate that applying the NLG bias renormalization approach to the one-loop galaxy bispectrum yields results consistent with previous literature~\cite{Eggemeier:2018qae} (see also \cite{Philcox:2022frc,DAmico:2022ukl}), before turning to the one-loop trispectrum.
The structure for the tracer bispectrum can be read off from the Fourier-space form of the triple operator product expansion Eq. \eqref{eq:stochxyzFourier}: to find the renormalized stochastic counterterms (cf.~Eq.~\eqref{eq:ren_stoch_ctr}), one contracts Eq. \eqref{eq:stochxyzFourier} with $b_{[A]}b_{[B]}b_{[C]}$, takes the expectation value and replaces all instances of $b_{[A]}b_{[B]}Z_{ABD}$ and $b_{[A]}b_{[B]}b_{[C]}Z_{ABCD}$ with renormalized coefficients $[c_D]$ and $[d_D]$, respectively. Concretely, 
\be
\text{ren. stoch. }(B) = \left(  [c_{D}] b_{[C]} P_{[DC]}(k_3) +2\,\text{permutations}\right)
+ [d_{\bm 1}].
\ee
Thus, we get
\bea
\lefteqn{ B(k_1,k_2,k_3) }\\
&=&
  b_{A} b_{B} b_{C} \, B_{ABC}(k_1,k_2,k_3)
+\left(  c_{D}b_C P_{DC}(k_3) +2\,\text{permutations}\right)
+ d_{\bm 1}\nn\\
&=&
  b_{[A]} b_{[B]} b_{[C]} \, B_{[ABC]}(k_1,k_2,k_3)
+\left(  [c_{D}] b_{[C]} P_{[DC]}(k_3) +2\,\text{permutations}\right)
+ [d_{\bm 1}]\,,\nn
\eea
noting again that only $D = \bm 1$ contributes to $d$ for the bispectrum\footnote{This pattern persists at all loop orders, since ${\bm 1}$ is the only operator with a non-vanishing average, after applying deterministic operator renormalization.}. The bare coefficients $c_D$,$d_D$ are again defined via the above expression. To relate the bare and renormalized stochastic coefficients, one simply plugs in expressions for the renormalized operator products $B_{[ABC]}$, $P_{[DC]}$ and equates coefficients, using also  $b_A {\cal O}_{A}=b_{[A]} [{\cal O}_{A}]$. The result is 
\begin{eqnarray}\label{eq:bisp_cd}
    [c_{D}] = c_{[D]} - b_{[A]}b_{[B]}Z_{ABD} \quad \text{where} \quad c_{[D]} = c_{C}([1+Z]^{-1})_{CD}.
\end{eqnarray}
and 
\be
   [d_{\bm 1}] =   d_{\bm 1} - b_{[A]} b_{[B]} b_{[C]}\Delta N_{ABC} - 3 [c_{D}]
b_{[C]} \Delta N_{DC} \,,
\ee
where $\Delta N_{ABC} = Z_{ABC \bm 1}$ and $\Delta N_{AB} = Z_{AB \bm 1}$. This also explains our slightly different notation $[c_D]$ for fully renormalized stochastic coefficients; we choose to reserve the notation $c_{[D]}$ for the second expression in Eq. \eqref{eq:bisp_cd}, which transforms like deterministic bias.   
At one-loop order, the $[c_{D}]$ coefficients are only needed for ${\cal
O}_D={\cal O}_C=\delta$ being the linear bias, and renormalize the
stochastic contribution to $B^{(321)}$, see below\footnote{Recently, \cite{Bakx:2025pop} also included the power spectrum up to one-loop order in the stochastic renormalization.}.
The three-operator contact term renormalization is accounted for by the
overall noise term $d_{\bm 1}$, and its renormalized counterpart $[d_{\bm 1}]$ is given by the above identity.
This accounts for renormalizing $B^{(222)}$ at one-loop order.

The bispectrum expressed in terms of renormalized bias and
noise terms at tree level reads 
\bea\label{eq:Btree}
   B^\text{tree}(k_1,k_2,k_3)
  &=& \sum_a b_{[a]}b_{[\delta]}^2\left(
B^\text{tree}_{a\delta\delta}(k_1,k_2,k_3)+
2\,\text{permutations}\right) \nn\\
  &&
+
[c_{\delta}]b_{[\delta]}
\Bigl(
P^{\rm lin}(k_1)
+
P^{\rm lin}(k_2)
+
P^{\rm lin}(k_3)
\Bigr) +
[d_{\bm 1}]\, 
\eea
while at one-loop order we have
\bea\label{eq:B1L}
   B^\text{1L}(k_1,k_2,k_3)
  &=& B^\text{tree}(k_1,k_2,k_3) \\
  && {} + \sum_{abc} b_{[a]}b_{[b]} b_{[c]}
\,B_{[abc]}^\text{1-loop}(k_1,k_2,k_3)\nn\\
  && {} + \sum_{a_2} b_{[a_2]}b_{[\delta]}^2\left(
B^\text{(211)}_{a_2\delta\delta}(k_1,k_2,k_3)+
2\,\text{permutations}\right) \nn\\
  && {} + \sum_{b} b_{[\nabla^2\delta]}b_{[b]}b_{[\delta]}\left(
B^\text{(121)}_{(\nabla^2\delta)b\delta}(k_1,k_2,k_3)+
5\,\text{permutations}\right) \,.\nn
  \eea
Here the sum in the first line of Eq.~\eqref{eq:Btree} is over up to second order LG
bias operators, as is the last line of Eq. \eqref{eq:B1L}.
The sum in the line containing $B_{[abc]}^\text{1-loop}(k_1,k_2,k_3)$ is over up to fourth order LG bias operators,
excluding those composed of four $\Pi^{[1]}$ fields ~\cite{Eggemeier:2018qae} (their renormalized
one-loop
kernels Eq.~\eqref{eq:RenKernel1loop} entering $B^{(411)}$ are identically zero). In the next-to last line, the sum is over up to second order NLG bias
operators (note that $b_{[\nabla^2\delta]}$ is included here).

The origin of the $[c_{\delta}]$ term in Eq. \eqref{eq:Btree} is the stochastic
renormalization of the product of the (up to) third- and second-order bias
contributions in $B^{(321)}$ by a linear field via $s^{\text{1L}}_{ab\delta}$.
This is then correlated with the remaining third $\delta$ in
$B^{(321)}$, giving $P^{\rm lin}(k_3)$.
After symmetrizing, this yields a stochastic term as above.
The constant $[d_{\bm 1}]$ accounts for noise terms related to the UV limit of
$B^{(222)}$. We emphasize again that under the assumption that $\bar{n}^{-1} \sim P_L$, these terms should all be counted as tree-level contributions.

We now turn to the explicit operator content. The one-loop bispectrum involves bias kernels up to perturbative order four. Out of the total 15 operators, since the fourth-order kernel only appears at $K_a^{(4)}({\bm k},{\bm k}',{\bm q},-{\bm q})$ configuration, the total number of non-degenerate operators reduces to 11, consistent with previous literature~\cite{Eggemeier:2018qae}  (see also \cite{Philcox:2022frc,DAmico:2022ukl}). These are the operators in the first four rows of Eq. \eqref{eq:bias_basis}.

At second gradient order, the relevant NLG operators comprise the full list defined in \refsec{basis}, since there are no redundancies for wavenumber configurations $K_{a_2}^{(2)}({\bm q}, {\bm k}-{\bm q})$. Thus, there are five NLG operators for biased tracers and four for matter. 

In the stochastic sector, the renormalization of $B^{(321),I}$ via the mode-coupling loops between two operators requires the single coefficient $[c_{\delta}]$, multiplying the sum of three linear power spectra. Finally, the $[d_{\bm 1}]$ coefficient is required to renormalize the product of three operators, occurring in $B^{(222)}$. This is consistent with the set of parameters introduced in \cite{McDonald:2009dh}.

\subsection{Trispectrum}\label{subsec:trisp}

In this subsection, we present the general structure as well as the specific one-loop expression for the renormalized, connected galaxy trispectrum. Renormalizing the deterministic part involves the UV limits of bias kernels $K^{(5)}_a({\bm k}_1,{\bm k}_2,{\bm k}_3,{\bm q},-{\bm q})$, which requires the full NLG operator basis at up to third order as derived in~\refsec{basis} in this work. We furthermore discuss stochastic renormalization and compare the latter to~\cite{Spezzati:2025zsb}.
The structure of the renormalized stochastic counterterms follows from the four-operator product expansion from Eq.~\eqref{eq:stochxyzw}. Proceeding in analogy to the bispectrum, we obtain for the LG stochastic EFT corrections to the galaxy trispectrum
\bea\label{eq:bren}
\lefteqn{ \text{ren. stoch. }(T) }\nn\\
&=&
     \left(  [d_{E}]b_{[D]} P_{[ED]}(k_4) +3\,\text{perm.}\right)+ \left([c_{E}]b_{[C]}b_{[D]} B_{[ECD]}(k_{12},k_3,k_4) +5\,\text{perm.}\right) \nn\\
&& {}
+ \left(  [c_{E}][c_{F}] P_{[EF]}(k_{12})
+2\,\text{perm.}\right)
+ [e_{\bm 1}]\,.
\eea
The structure of the two terms in the middle line matches those discussed in \cite{Spezzati:2025zsb} when evaluated at lowest non-zero order in perturbation theory.
This time, we also encounter a contribution proportional to the \textit{second} power of a stochastic coefficient (first term on the bottom line). 
We now expand on the origin of this term in more detail. 

One can now see explicitly that if this term is evaluated at tree-level, only $E = F = \delta$ contributes\footnote{The contribution with $E=F=\bm 1$ contributes precisely to the \textit{disconnected} trispectrum $\sim P^2$, which is thus renormalized by the same $Z_{ab\bm 1}$ coefficients as $P$, in line with expectation.} and we get a contribution $\propto [c_\delta]^2 P^\text{lin}(k_{12})\, + $ 2 perms. In the noise-field formulation, which is more common in EFT literature, this term arises from a contraction of the form 
\be
\langle (\epsilon_\delta \delta) (\bm k_1) \epsilon (\bm k_2)(\epsilon_\delta \delta) (\bm k_3)  \epsilon (\bm k_4)\rangle = \int_{\bm p_1 \bm p_2} \langle \delta (\bm p_1) \delta (\bm p_2)\rangle  \times \langle \epsilon_\delta (\bm k_1-\bm p_1) \epsilon (\bm k_2 ) \epsilon_\delta(\bm k_3 -\bm p_2) \epsilon (\bm k_4)\rangle\,.
\ee
The four-point stochastic correlator can be split into a disconnected and a connected piece, the former of which contains the two terms
\begin{eqnarray}
     \left(\delta^D(\bm k_{12}-\bm p_{1}) \delta^D(\bm k_{34}-\bm p_2)\, + \delta^D(\bm k_{14}-\bm p_{1}) \delta^D(\bm k_{23}-\bm p_2)\right) \langle \epsilon_\delta \epsilon \rangle'(\bm k_2)\langle \epsilon_\delta \epsilon \rangle'(\bm k_4)  
\end{eqnarray}
which upon integration gives the desired structure. In other words, in this way we recover the statement that the coefficient of the $P^\text{lin}(k_1) + P^\text{lin}(k_2) + P^\text{lin}(k_3)$ counterterm piece of the bispectrum should precisely equal the renormalized bias $b_{[\delta]}$ times the square root of the coefficient that enters the $P^\text{lin}(k_{12}) + P^\text{lin}(k_{13}) + P^\text{lin}(k_{14})$ counterterm for the (connected) trispectrum. 

The disconnected part of the four-point stochastic correlator also contains a contribution proportional to $\langle \epsilon_\delta \epsilon_\delta \rangle \cdot \langle \epsilon \epsilon \rangle  $. However, for this piece the momentum constraints yield $\bm k_{13} = \bm k_{24} = 0$, so that this does not contribute to the connected trispectrum. We defer a more in-depth discussion of the connection between the noise-field formulation and the operator product expansion to future work.

All in all, the fully general renormalized trispectrum then reads
\bea\label{eq:tren}
\lefteqn{ T({\bm k}_1,{\bm k}_2,{\bm k}_3,{\bm k}_4) }\nn\\
&=&
  b_{A} b_{B} b_{C} b_{D} \, T_{ABCD}({\bm k}_1,{\bm k}_2,{\bm k}_3,{\bm k}_4)
+\left(  c_{E}b_Cb_D B_{ECD}(k_{12},k_3,k_4)
+5\,\text{perm.}\right)\\
&& {}
+ \left(  d_{E}b_D P_{ED}(k_4) +3\,\text{perm.}\right)
+ \left(  c_{E}c_F P_{EF}(k_{12})
+2\,\text{perm.}\right)
+ e_{\bm 1}\nn\\
&=&
  b_{[A]} b_{[B]} b_{[C]} b_{[D]} \, T_{[ABCD]}({\bm k}_1,{\bm k}_2,{\bm k}_3,{\bm k}_4)
+\left(  [c_{E}]b_{[C]}b_{[D]} B_{[ECD]}(k_{12},k_3,k_4) +5\,\text{perm.}\right)\nn\\
&& {}
+ \left(  [d_{E}]b_{[D]} P_{[ED]}(k_4) +3\,\text{perm.}\right)
+ \left(  [c_{E}][c_{F}] P_{[EF]}(k_{12})
+2\,\text{perm.}\right)
+ [e_{\bm 1}]\nn
\eea
where $k_{12} = |{\bm k}_1+{\bm
k}_2|$. The $c_E$, $d_E$ and $e$ coefficients account for renormalization of
stochastic contributions to $T^{(4211)}+T^{(3311)}$, $T^{(3221)}$ and
$T^{(2222)}$ at one-loop order, respectively.
The corresponding bare and renormalized coefficients $[c_E]$ and $c_E$ are
related as in Eq. \eqref{eq:bisp_cd}, and correspond to the generalization of contact
terms for which the arguments
of two operators coincide. For the trispectrum, we also require contact
terms of products of three and four operators. After going through the algebra once more, the relation between bare and renormalized counterparts reads 
\be\label{eq:dren}
   [d_{E}] = d_{[E]} - b_{[A]}b_{[B]}b_{[C]}Z_{ABCE} -
3[c_{F}]b_{[C]}Z_{FCE}\, \quad \text{ where}\quad   d_{[E]} = d_{C}([1+Z]^{-1})_{CE}
\ee
and 
\ba \label{eq:e_def}
   [e_{\bm 1}] =&   e_{\bm 1} - b_{[A]} b_{[B]} b_{[C]} b_{[D]}\Delta N_{ABCD}
   - 6 [c_{E}]b_{[C]}b_{[D]} \Delta N_{ECD} \nn \\
   &- 4 [d_{E}]b_{[D]} \Delta N_{ED} 
   - 3 [c_{E}][c_{F}] \Delta N_{EF}\,.
\ea
We highlight that symmetry factors work out as needed, including in
higher-order terms such as those contributing to renormalization only beyond one-loop
order (the last term), which serves
as a consistency check. 
The trispectrum expressed in terms of fully renormalized
quantities at tree level reads
\bea\label{eq:Ttree}
T^\text{tree}({\bm k}_1,{\bm k}_2,{\bm k}_3,{\bm k}_4)
  &=& \sum_{ab} b_{[a]}b_{[b]}b_{[\delta]}^2\left(
T^\text{tree}_{ab\delta\delta}({\bm k}_1,{\bm k}_2,{\bm k}_3,{\bm k}_4)+
11\,\text{perm.}\right) \\
  && {}
+
\sum_{a}
[c_{a}]b_{[\delta]}^2
\Bigl(
B^{\rm tree}_{a\delta\delta}(|{\bm k}_1+{\bm k}_2|,k_3,k_4)
+
5~{\rm perm.}
\Bigr) \nn\\
&& {}
+
[c_{\delta}]^2
\Bigl(
P^\text{lin}(|{\bm k}_1+{\bm k}_2|)
+
2~{\rm perm.}
\Bigr) \nn\\
&& {}
+
[d_{\delta}]b_{[\delta]}
\Bigl(
P^{\rm lin}(k_1)
+
P^{\rm lin}(k_2)
+
P^{\rm lin}(k_3)
+
P^{\rm lin}(k_4)
\Bigr)
  +
[e_{\bm 1}]. \nn
\eea
while at one-loop order we have 
\bea\label{eq:T1L}
   T^\text{1L}({\bm k}_1,{\bm k}_2,{\bm k}_3,{\bm k}_4)
  &=& 
T^\text{tree} + \sum_{abcd} b_{[a]}b_{[b]} b_{[c]}
b_{[d]}\,T_{[abcd]}^\text{1-loop}({\bm k}_1,{\bm k}_2,{\bm k}_3,{\bm k}_4)\\
  && {} + \sum_{a_2b} b_{[a_2]}b_{[b]}b_{[\delta]}^2\left(
T^{(3111+2211)}_{a_2b\delta\delta}({\bm k}_1,{\bm k}_2,{\bm k}_3,{\bm k}_4)+
11\,\text{perm.}\right) \nn\\
  && {} + \sum_{b} b_{[\nabla^2\delta]}b_{[b]}b_{[\delta]}^2\left(
T^{(1311)}_{(\nabla^2\delta)b\delta\delta}({\bm k}_1,{\bm k}_2,{\bm k}_3,{\bm k}_4)+
11\,\text{perm.}\right) \nn
\eea
Summation in the first (second) line of Eq. \eqref{eq:Ttree} is over up to third-order (second-order) LG bias
operators, respectively. In the first line of Eq. \eqref{eq:T1L}, the sum is over deterministic operators up to fifth order; we specify the list below. In the second line of Eq. \eqref{eq:T1L}, which renormalizes $T^{(5111)}+T^{(4211)}$
up to third-order NLG and up to second-order LG bias operators  are summed over. Finally, the last line sums over up to third-order LG bias operators and renormalizes $T^{(3311)}$.

The $[c_{a}]$ terms come from the
renormalization of the $T^{(4211)}$ and $T^{(3311)}$ loops by up to 
second-order operators $\mathcal{O}_a^{[2]}$.
Correlating them with the remaining two linear fields yields the
tree-level bispectrum of these $\mathcal{O}_a^{[2]}$ operators with two linear
$\delta$'s, and a free coefficient
$[c_{a}]$ for each of these. Note that for $a = \delta$ we already encountered this contribution in the bispectrum. 
The $[d_{\delta}]$ coefficient accounts for renormalizing $T^{(3221)}$,
and the constant $[e_{\bm 1}]$ for $T^{(2222)}$. This is in agreement with \cite{Spezzati:2025zsb}, up to the missing term proportional to $[c_\delta]^2$ mentioned above. 

Finally, we turn to the explicit operator content in the one-loop trispectrum. The renormalized one-loop trispectrum involves bias kernels up to perturbative order five. Since the diagram $T^{(5111)}_{abcd}$ contains $K_a^{(5)}({\bm k_1},{\bm k_2},{\bm k_3},{\bm q},-{\bm q})$, out of the 29 LG terms, only 24 are needed in that specific configuration. These are 7 more than the combinations that only involve the two-loop power spectrum, due to the more general momentum configuration. Relative to Eq. \eqref{eq:bias_basis} there are the extra $4 + 3 = 7$ operators, 
\bea \label{eq:bias_basis_trisp}
{\rm (4)} 
   &{\rm tr}\big[ \big( \Pi^{[1]} \big)^2 \big] \big({\rm tr}\big[ \Pi^{[1]} \big]\big)^2, 
   ~{\rm tr}\big[ \big( \Pi^{[1]} \big)^3 \big] {\rm tr}\big[ \Pi^{[1]} \big],
   ~\Big( {\rm tr}\big[ \big( \Pi^{[1]} \big)^2 \big] \Big)^2,   
   ~\Big( {\rm tr}\big[ \Pi^{[1]} \big] \Big)^4, \\
 {\rm (5)}  
    & {\rm tr}\big[ \Pi^{[1]} \big] {\rm tr}\big[\Pi^{[1]} \Pi^{[1]} \Pi^{[2]}\big]\,,
~~{\rm tr}\big[ \Pi^{[1]}  \Pi^{[1]} \big] {\rm tr}\big[\Pi^{[1]} \Pi^{[2]}\big]\,,
~~\left({\rm tr}\big[ \Pi^{[1]}\big]\right)^2 {\rm tr}\big[\Pi^{[1]} \Pi^{[2]} \big]. \non
\eea

NLG operators at third perturbative order with two overall derivatives enter the renormalization of the one-loop trispectrum, contributing through diagrams analogous to $T^{(3111)}$. Here, the momentum dependence is fully general and hence we end up with all 20 NLG operators from \refsec{basis}. However, here it turns out that only the combination $\Pi_{B1} - \Pi_{B2}$ is needed to absorb the single-hard limit of general fifth-order kernels. Hence, in principle 19 coefficients would suffice. We checked that for a general biased tracer, the UV limits of the corresponding linear combinations of LG bias operators up to fifth order span the full space of linear combinations of those 19 NLG operators. That is, all of them are in principle required for renormalizing the one-loop trispectrum of a completely generic biased tracer. This includes in particular $\hat\Pi^{[3]}$, along with the combination $\Pi_{B1} - \Pi_{B2}$ as stated above.
\bigskip

\subsection{Summary: operator counting}\label{sec:counting}
The operator content required to renormalize a given combination of $N$-point functions is set by the maximal perturbative order of the bias kernels involved, which fixes the LG and NLG sectors, together with the specific momenta configuration involved. For the stochastic sector, the maximal number of bias operators meeting at coincident points determines the number of operators. \reftab{operator_counting} collects this information for the combinations of practical interest, ranging from a stand-alone one-loop power spectrum to the full $\{P^\text{2L},B^\text{1L},T^\text{1L}\}$ set. Note that these counts only pertain to \textit{auto-correlations} of the same tracer. We defer a more in-depth discussion of the combinations relevant for e.g. cross-correlations between matter and biased tracers, as probed by CMB lensing, to future work. In order to specialize to auto-correlations of only the matter field, one should omit the $n=2,3,4,5$ contributions from the deterministic LG sector, as well as all stochastic parameters and the two non-$\nabla^2$ rows. Specifically, the one-loop matter power spectrum and bispectrum combination contains a total of $1 + 4 + 0 = 5$ free parameters. The two-loop matter power spectrum contains $8$ free parameters, which is the same as the combination with the one-loop matter bispectrum. The one-loop matter trispectrum requires $15$ free parameters. 

The counting for each of the individual instances of $P^{\text{1L}},P^{\text{2L}},B^{\text{1L}},T^{\text{1L}}$ has been addressed above. To combine them, note that in the deterministic sector $B^{\text{1L}}$ consists of a subset of $P^\text{2L}$, and only six new fifth-order LG operators and four overall-derivative NLG operators (second line of Eq. \eqref{eq:NLGpklist}) need to be added. In the stochastic sector, the power spectrum and the bispectrum do not share any free coefficients and hence they need to be added together, yielding $1+2 = 3$ free coefficients. 

When combining all three statistics up to fifth order in perturbation theory, we simply note that $T^\text{1L}$ encompasses both $B^\text{1L}$ and $P^\text{2L}$ in the deterministic sector. In the stochastic sector, the same parameter $[c_{\delta}]$ that renormalizes $B^{(321)}$ in the bispectrum also occurs here, so that the only new parameters in addition to those of $T^\text{1L}$ itself are the `pure-stochastic' contributions $[c_{\bm 1}]$ and $[d_{\bm 1}]$. Thus, there are $5 + 1 + 1 = 7$ stochastic parameters at LG order in the combination $P^\text{2L} + B^\text{1L} + T^\text{1L}$. 

Finally, we comment on how to consistently implement operator redundancies in joint analyses of different $N$-point functions. First, note that the one-loop renormalized kernels of products of three $\Pi^{[1]}$ operators are identically zero. Thus, these operators simply give zero contribution to $P^{\text{1L}}$ even in a combined analysis with e.g. $B^\text{1L}$ or $T^{\text{1L}}$ (or even $T^\text{tree}$), where they yield non-trivial contributions from general momentum configurations. The same reasoning also applies for the two-loop power spectrum, where all four third-order operators contribute to $P^{(33)}$. The $P^{(13)}$ contribution from LG deterministic operators in our two-loop calculation is identical to this contribution in our one-loop calculation. As a consequence, there is no ambiguity in writing e.g. Eq. \eqref{eq:P2L} as a sum of the previous one-loop calculation and a new two-loop piece. 

Beyond third-order, similar remarks also apply to other classes of operators that are non-redundant in $T^\text{1L}$ but redundant in $B^\text{1L}$ and/or $P^\text{2L}$. Specifically, at fourth order, one-loop renormalized kernels of four $\Pi^{[1]}$ operators are zero. Hence, these operators yield zero contribution to $P^{(24)}$ and $B^{(411)}$ and these loops are, after renormalization, unchanged when combining with $T^\text{1L}$ (where they yield distinct, non-zero contributions to $T^{(4211)}$). At fifth order, one-loop renormalized kernels of products of five $\Pi^{[1]}$ operators are zero and hence they do not contribute to any statistic. Moreover, when combining $P^\text{2L}$ and $T^\text{1L}$, we find that \textit{two-loop} renormalized $K_a^{(5)}$ kernels are zero for all operators that are not needed for $P^\text{2L}$. These are {\it (i)} products of four or five $\Pi^{[1]}$ operators, and {\it (ii)} products of three $\Pi^{[1]}$ operators and one $\Pi^{[2]}$. Hence, the (renormalized) $P^{(15)}$ calculation remains unchanged between a $P^\text{2L}$ and $P^\text{2L} + T^\text{1L}$ analysis.

We emphasize once more that this is a consequence of performing renormalization at the operator level, so that the identically vanishing renormalized kernel yields a vanishing contribution when plugged into any statistic at any loop order. Additionally, this implies that \textit{renormalized} bias parameters that are inferred when fitting a lower-loop calculation to data do not need to be modified when adding higher-loop corrections to the model. This is in stark contrast to the completely bare case, where e.g. the addition of several $P^{(13)}$ diagrams from products of three $\Pi^{[1]}$ operators would yield contributions that are exactly degenerate with $P^\text{lin}$, thus requiring careful interpretation of the linear bias coefficient when fitting the bare one-loop model to the same data. Similar problems would arise when different $N$-point functions are combined; for example, fitting the bare one-loop power spectrum in combination with the tree-level bispectrum (which is trivially a renormalized quantity) would require using different linear bias coefficients for the two statistics (even though this difference is of course still computable and dictated by the RG).

In the NLG sector, we also encounter redundancies where the one-loop renormalized kernel of an operator is \textit{not} identically zero for our choice of operator basis. This happens specifically for $\mbox{tr}[\Pi^{[2]}\nabla^2\Pi^{[1]}]$ and $\Pi_{B2}$ in the two-loop power spectrum; see the discussion below Eq. \eqref{eq:NLGpklist}. These two operators are non-redundant in $T^\text{1L}$. When evaluated on $\bm k, \bm q, -\bm q$ they are redundant with operators from Eq. \eqref{eq:NLGpklist}, but their one-loop renormalized kernels are not zero\footnote{Of course, given the redundancy for the bare kernels, it is always possible to redefine the set of chosen operators such that the bare (and hence the renormalized) kernels vanish. This is similar to the construction of the minimal NLG basis needed for $T^\text{1L}$, which is discussed in App. \ref{subsec:redundancy}. }. In other words, given a (hypothetical and very precise) measurement of all NLG operator biases from $T^\text{1L}$, the contribution from e.g. $\Pi_{B2}$ needs to be added to $P^\text{2L}$ as well, since it does not vanish. Thus, our operator counting summarizes the \textit{number of independent free parameters} that need to be fitted to data at a given loop order for a specific combination of $N$-point functions. However, care must be taken to associate these to specific operators when degeneracies can be broken by combining multiple $N$-point functions.  

To summarize, bias renormalization ensures that when complementing $P^\text{1L}$ with $B^\text{1L}$, the coefficients of the extra $7$ third-order renormalized LG bias operators and the extra $4$ second-order NLG operators contribute only to $B^\text{1L}$, while the implementation of $P^\text{1L}$ from Eq.~\eqref{eq:P1L} remains completely unchanged, including in particular the sets of operators that are summed over. Furthermore, the coefficients of operators that appear both in $P^\text{1L}$ and $B^\text{1L}$ are identical. The same applies when going from  $P^\text{1L}$ to $P^\text{2L}$, and analogously when going from a $P^\text{1L}+B^\text{1L}$ analysis to $P^\text{2L}+B^\text{1L}$. When going from $P^\text{2L}+B^\text{1L}$ to $P^\text{2L}+B^\text{1L}+T^\text{1L}$, the extra $4$ ($3$) renormalized LG fourth (fifth) order bias operators contribute only to $T^\text{1L}$. This means the summation range over $a,b$ in the first line of Eq.~\eqref{eq:P2L} for $P^\text{2L}$ remains unchanged. For the $7+4=11$ extra NLG operators, the $9$ out of them composed of a product of three $\Pi^{[1]}$'s do not enter $P^\text{2L}$, while the remaining two (being $\mbox{tr}[\Pi^{[2]}\nabla^2\Pi^{[1]}]$ and $\Pi_{B2}$) need to be taken into account in the summation over $a_2$ in the second line of Eq.~\eqref{eq:P2L} for $P^\text{2L}$.

\begin{table}[!ht]
\centering
\renewcommand{\arraystretch}{1.25}
\setlength{\tabcolsep}{4pt}
\begin{tabular}{l|ccccc}
\hline\hline
 & $P^\text{1L}$ & $P^\text{1L}\!+\!B^\text{1L}$ & $P^\text{2L}$ & $P^\text{2L}\!+\!B^\text{1L}$ & \,$P^\text{2L}\!+\!B^\text{1L}\!+\!T^\text{1L}$\, \\
\hline
\multicolumn{6}{l}{\emph{Maximal kernel order needed}}\\
\hline
$n_\text{max}^{\rm LG}, n_\text{max}^{\rm NLG}+2$  & 3 & 4 & 5 & 5 & 5 \\
\hline
\multicolumn{6}{l}{\emph{Deterministic LG operators (per perturbative order $n$)}} \\
\hline
$n=1$  & 1 & 1 & 1 & 1 & 1 \\
$n=2$  & 2 & 2 & 2 & 2 & 2 \\
$n=3$  & 1 & 4 & 4 & 4 & 4 \\
$n=4$  & --              & 4 & 4 & 4 & 8 \\
$n=5$  & --              & --              & 6 & 6 & 9 \\
\textbf{LG subtotal} & $4$ & $11$ & $17$ & $17$ & $24$ \\
\hline
\multicolumn{6}{l}{\emph{Deterministic NLG operators}} \\
\hline
$\nabla^2(\cdots)$, $n=1$       & 1 & 1 & 1 & 1 & 1 \\
$\nabla^2(\cdots)$, $n=2$       & --        & 3 & 3 & 3 & 3 \\
$\nabla^2(\cdots)$, $n=3$       & --        & --        & 4 & 4 & 11 \\
non-$\nabla^2$, $n=2$           & --        & 1        & 1 & 1 & 1 \\
non-$\nabla^2$, $n=3$           & --        & --        & 0        & 0        & 4 \\
\textbf{NLG subtotal} & 1 & $5$ & $9$ & $9$ & $20$ \\
\hline
\multicolumn{6}{l}{\emph{Stochastic parameters and mode-coupling coefficients}} \\
\hline
$[c_{\bm 1}]$                & $1$ & $1$ & $1$ & $1$ & $1$ \\
$[c_{\delta}]$         & --        & $1$ & --        & $1$ & $1$ \\
$[c_{a}], a \in O^{[2]}$             & --        & --        & --        & --        & $2$ \\
$[d_{\bm 1}]$                 & --        & $1$ & --        & $1$ & $1$ \\
$[d_{\delta}]$        & --        & --        & --        & --        & $1$ \\
$[e_{\bm 1}]$                & --        & --        & --        & --        & $1$ \\
\textbf{Stoch.\ subtotal} & $1$ & $3$ & $1$ & $3$ & $7$ \\
\hline\hline
\textbf{Total (no NNLG)} & $6$ & $19$ & $27$ & $29$ & $51$ \\
\hline\hline
\end{tabular}
\caption{Operator counting for the renormalization of various combinations of renormalized $N$-point functions. The structure of the deterministic LG and NLG operator content is organized by perturbative order. For NLG, the rows distinguish operators with two overall gradients [$\nabla^2(\cdots)$], which are the operators ${\cal O}^{\nabla^2}_{a_2}$ introduced in \refsec{basis}, from operators without [$\text{non-}\nabla^2$]. We also list the independent stochastic parameters, with the number $n$ of the external $\delta_L$ fields at which the renormalization condition is imposed. We omitted the $\nabla^4\delta$ operator which occurs in $P^\text{2L}$ only.
}
\label{tab:operator_counting}
\end{table}

\newpage
\section{The renormalization group}\label{sec:RGE}

In this section, we derive the one- and two-loop renormalization group equations (RGEs) governing the cutoff-dependence of the bare bias coefficients in the presence of both leading-gradient and second-gradient operators. In practice, these equations could be used to check whether the values of EFT coefficients obtained from fitting a set of $N$-point functions, with loops computed with a given UV cutoff $\Lambda$, to simulation or data, feature the theoretically expected `running' when varying $\Lambda$. Deviations could be a sign of unresolved degeneracies given a certain accuracy, or over-fitting. Another potential application is the resummation of higher-order loop contributions, akin to the use in particle physics. While these applications go beyond the scope of this work, we derive the RGEs and discuss their properties at one- and two-loop order, serving as a non-trivial consistency check of the bias renormalization approach.

The derivation of the one-loop RGEs for LG deterministic bias coefficients was discussed in \cite{Rubira:2023vzw} (see also~\cite{McDonald:2006hf,Carroll:2013oxa}),  being generalized to two-loop order in \hyperlink{cite.Bakx:2025cvu}{Paper~I}, one-loop stochastic parameters in \cite{Rubira:2024tea} and to include primordial non-Gaussianities in \cite{Nikolis:2024kbx}. Here we extend this program to include NLG operators, which also allows us to consider the case of the matter density field itself as a special case (see also~\cite{Peron:2025lgh}), for which LG renormalization is absent due to momentum conservation. For biased tracers, on the other hand, mixing of LG and NLG operators occurs.

Since the renormalized bias coefficients are, by construction, physical parameters, they must be independent of the arbitrary UV cutoff $\Lambda$. This requirement implies a set of RGEs, generalizing the discussion of \hyperlink{cite.Bakx:2025cvu}{Paper~I} to the case where the operator basis contains both leading-gradient and second-gradient operators. Writing the renormalized operators in terms of bare ones through the mixing matrix $Z_{AB}$, the RG flow of the bare coefficients $b_A(\Lambda)$ is given by (see Sections 5 and 6 of \hyperlink{cite.Bakx:2025cvu}{Paper~I} for a derivation)
\be \label{eq:generalRG}
  \frac{d}{d\Lambda}b_A = b_B\gamma_{BA},\qquad \gamma_{BA}\equiv[(1+Z)^{-1}]_{BC}\frac{dZ_{CA}}{d\Lambda}\,,
\ee
with $A,B,C$  running over all deterministic bias operators in the basis, including both leading-gradient operators ${\cal O}_a$ and second-gradient operators ${\cal O}_{a_2}$. The matrix $\gamma_{BA}$ is the anomalous-dimension matrix associated with this operator basis and encodes how renormalization induces mixing among operators as the cutoff is varied such that any change in $\Lambda$ reshuffles the bare coefficients with the premise that the renormalized expansion remains invariant.

With the decomposition of the operator basis into leading-gradient and second-gradient contributions, the RGEs take the block form
\bea
  \frac{d}{d\Lambda}b_a &=&  b_b\gamma_{ba} + b_{b_2}\gamma_{b_2a}\,,\nn\\
  \frac{d}{d\Lambda}b_{a_2} &=&  b_b\gamma_{ba_2} + b_{b_2}\gamma_{b_2a_2}\,,
\eea
with the first equation describing the running of the leading-gradient bias coefficients $b_a$, which receive contributions both from mixing among leading-gradient operators themselves and from mixing with second-gradient operators. 
The second equation analogously sets the running of the second-gradient coefficients $b_{a_2}$. The RG structure will be worked out explicitly at one- and two-loop order below, expanding the anomalous-dimension matrix as $\gamma_{AB}=\gamma_{AB}^\text{1L}+\gamma_{AB}^\text{2L}+\cdots$, with a vanishing tree-level contribution, $\gamma_{AB}^\text{tree}=0$.

The RGEs derived here apply to general biased tracers. A particular subcase is that of the matter density field, for which $b_\delta=1$ while all other leading-gradient bias coefficients vanish, $b_a=0$. We will discuss this specific case in more detail in \refsec{matter}.

\subsection{One-loop RG}

Since the renormalization matrix starts at $O(Z^\text{1L})$, we can expand $(1+Z)^{-1}=1+O(Z)$ and set $1+Z\mapsto 1$ in Eq. \eqref{eq:generalRG} at one-loop order. The anomalous dimension is then directly given by the derivative of the one-loop renormalization constants,
\be
  \gamma_{AB}^\text{1L} = \frac{dZ^\text{1L}_{AB}}{d\Lambda}\,.
\ee
Using the explicit form of the one-loop counterterms in Eq.~\eqref{eq:Z1L}, we obtain
\bea\label{eq:gammaab1L}
  \gamma_{ab}^\text{1L} &=& -\frac{d\sigma_\Lambda^2}{d\Lambda} s_{ab}^\text{1L},\qquad 
  \gamma_{ab_2}^\text{1L} = -\frac{d\tilde\sigma_\Lambda^2}{d\Lambda} s_{ab_2}^\text{1L},\nn\\
  \gamma_{a_2b}^\text{1L} &=& -\frac{d\hat\sigma_\Lambda^2}{d\Lambda} s_{a_2b}^\text{1L},\qquad 
  \gamma_{a_2b_2}^\text{1L} = -\frac{d\sigma_\Lambda^2}{d\Lambda} s_{a_2b_2}^\text{1L}\,,
\eea
where $\sigma,\, \tilde\sigma$ and $\hat\sigma$ are defined in Eq. \eqref{eq:sigmadef}, such that
\be
  \frac{d\sigma_\Lambda^2}{d\Lambda}=\frac{4\pi\Lambda^2}{(2\pi)^3}P^\text{lin}(\Lambda),\quad
  \frac{d\tilde\sigma_\Lambda^2}{d\Lambda}=\frac{1}{\Lambda^2}\frac{d\sigma_\Lambda^2}{d\Lambda},\quad
  \frac{d\hat\sigma_\Lambda^2}{d\Lambda}=\Lambda^2\frac{d\sigma_\Lambda^2}{d\Lambda}\,.
\ee
The different powers of $\Lambda$ reflect the fact that mixing between operators of different derivative order is accompanied by dimensionful factors.

Inserting these expressions into the general RGEs, we obtain the one-loop running of the bias coefficients,
\be
\boxed{
\begin{array}{lcl}
  \displaystyle \frac{d}{d\Lambda}b_a &=&\displaystyle   -\frac{d\sigma_\Lambda^2}{d\Lambda} \left[ b_b s^\text{1L}_{ba} + \Lambda^2 b_{b_2}s^\text{1L}_{b_2a}\right]\,,\\[1.5ex]
  \displaystyle \frac{d}{d\Lambda}b_{a_2} &=&\displaystyle   -\frac{d\sigma_\Lambda^2}{d\Lambda} \left[ \frac{1}{\Lambda^2} b_b s^\text{1L}_{ba_2} + b_{b_2}s^\text{1L}_{b_2a_2}\right]\,,
\end{array}
}
\ee
which generalize the LG case of \cite{Rubira:2023vzw}.
The structure of these equations makes explicit that the RG flow induces mixing between operators of different derivative order, since leading-gradient coefficients are sourced not only by themselves but also by second-gradient coefficients, and vice versa. 
The matrices $s^\text{1L}_{AB}$ encode this mixing and are determined by the single-hard limits of the corresponding bias kernels, as discussed in \refsec{renorm}.

\subsection{Matter} 
\label{sec:matter}

A particular example is the case of matter density renormalization. In that case $b_\delta=1$, and all other $b_a=0$. From the structure of \reftab{sab1L}, the operator $\delta$ does not source any other $\O_a$. From \reftab{sa2b1L}, the only higher-in-derivative operators that could source leading-in-derivative operators are $b_{a_2}^{\text{non-}\nabla^2}$, as defined in Eq. \eqref{eq:ctbar}, but since those operators do not contain two {\it overall} gradients, they cannot appear as counterterms for the matter density due to momentum conservation. Also note that $b_{a_2}^{\text{non-}\nabla^2}$, in turn, cannot also be sourced by $\delta$ (\reftab{sab21L_B}) or any other overall-gradient bias coefficient $b_{a_2}^{\nabla^2}$ (\reftab{sa2b21L_B}). 

Therefore, the choice $b_{\delta}=1$ together with $b_{a_2}^{\text{non-}\nabla^2}=0$ is stable, since both $\frac{d}{d\Lambda}b_{a} = \frac{d}{d\Lambda}b_{a_2}^{\text{non-}\nabla^2} = 0$, and the running of the counterterm coefficients renormalizing the matter density field is given by
\be \label{eq:matterRG}
\boxed{
\begin{array}{lcl}
    \frac{d}{d\Lambda}b_{a_2}^{\nabla^2} =  -\frac{d\sigma_\Lambda^2}{d\Lambda} \left[ \frac{1}{\Lambda^2}  s^\text{1L}_{\delta a_2} + b_{b_2}^{\nabla^2}s^\text{1L}_{b_2a_2}\right]\,.\qquad (\text{counterterm RG})\\  
\end{array}
}
\ee
The running is sourced by the matter density, given by the term $s^\text{1L}_{\delta a_2}$. For instance, a well-known example is
$s^\text{1L}_{\delta \nabla^2\delta}=-\frac{61}{630}$ which is related to the hard limit of the $F_3({\bm k},{\bm q},-{\bm q})$ kernel. In this work, we provide explicit results for all coefficients for NLG operators  ${\cal O}_{a_2}$ up to third order, see (\reftab{sab21L_B}) and  (\reftab{sa2b21L_B}).
We discuss solutions for the matter RGE in \refsec{RGsol}.

\subsection{Two-loop RG}

The structure of the RG equation beyond its leading (one-loop) approximation provides a non-trivial consistency check of the bias renormalization procedure. Moreover, including the two-loop contribution $\gamma_{AB}^\text{2L}$ to the RG equation allows us to quantify corrections to the running of bias parameters. In particular, the success of the RG technique in terms of resumming the leading UV contributions to loop integrals can only be assessed when comparing solutions to the RG equations with one- and two-loop running.

The two-loop running is described by the anomalous-dimension matrix
\be
  \gamma_{AB}^\text{2L} = \frac{dZ^\text{2L}_{AB}}{d\Lambda} - Z^\text{1L}_{AC}\frac{dZ^\text{1L}_{CB}}{d\Lambda} \,,
\ee
which follows directly from Eq. \eqref{eq:generalRG}. 
Using the renormalization constants from Eq.~\eqref{eq:Z2L}, we find for the case of leading-gradient operator mixing (i.e. $A=a, B=b$) as well as for the generation of second-gradient bias (i.e. $A=a, B=b_2$)
\bea
  \gamma_{ab}^\text{2L} &=& \frac{dZ^\text{1L}_{aC}}{d\Lambda}Z^\text{1L}_{Cb} - \frac{d\sigma_\Lambda^2}{d\Lambda}\int_{p<\Lambda} s_{ab}^\text{2L}(p/\Lambda)P^\text{lin}(p),\nn\\
  \gamma_{ab_2}^\text{2L} &=& \frac{dZ^\text{1L}_{aC}}{d\Lambda}Z^\text{1L}_{Cb_2} - \frac{d\sigma_\Lambda^2}{d\Lambda}\int_{p<\Lambda} s_{ab_2}^\text{2L}(p/\Lambda)\left(\frac{1}{p^2}+\frac{1}{\Lambda^2}\right)P^\text{lin}(p)\,,
\eea
where the index $C$ runs over both leading-gradient (LG, $C=c$) and second-gradient (NLG, $C=c_2$) operators. The last terms in each equation arise from the double-hard limits, involving the $s_{AB}^\text{2L}(r)$ functions.
The first terms are given by a product of two one-loop terms. Their role is to subtract those two-loop contributions that are already taken into account by iteratively solving the one-loop RGEs, such that $\gamma^\text{2L}$ captures only the `intrinsically' two-loop part of the running.\footnote{For common applications in particle physics, the two-loop contribution generated by iteratively solving the one-loop RG corresponds to the so-called `leading logarithms', while the `intrinsic' two-loop part captured by $\gamma^\text{2L}$ corresponds to `next-to leading logarithmic' terms. For further discussion of analogies in the present context, we refer to \hyperlink{cite.Bakx:2025cvu}{Paper~I}.}

Next, using the one-loop renormalization constants from Eq. ~\eqref{eq:Z1L}, together with the double/single-hard commutation of limits relations Eq.~\eqref{eq:s2Ldoublesingle} and Eq.~\eqref{eq:s2Ldoublesingle2}, the contributions from the squared one-loop terms can be written explicitly as
\bea
  \frac{dZ^\text{1L}_{ac}}{d\Lambda}Z^\text{1L}_{cb} 
  &=& s^\text{1L}_{ac}s^\text{1L}_{cb} \frac{d\sigma_\Lambda^2}{d\Lambda}\sigma_\Lambda^2
   = s_{ab}^\text{2L}(0)  \frac{d\sigma_\Lambda^2}{d\Lambda}\int_{p<\Lambda} P^\text{lin}(p)   \,, \\
  \frac{dZ^\text{1L}_{ac_2}}{d\Lambda}Z^\text{1L}_{c_2b} 
  &=& s^\text{1L}_{ac_2}s^\text{1L}_{c_2b}  \frac{d\tilde\sigma_\Lambda^2}{d\Lambda}\hat\sigma_\Lambda^2
  = \frac12 (s_{ab}^\text{2L})''(0) \frac{1}{\Lambda^2}\frac{d\sigma_\Lambda^2}{d\Lambda}\int_{p<\Lambda} p^2  P^\text{lin}(p)  \,,\nn\\
  \frac{dZ^\text{1L}_{ac}}{d\Lambda}Z^\text{1L}_{cb_2} 
  &=& s^\text{1L}_{ac}s^\text{1L}_{cb_2} \frac{d\sigma_\Lambda^2}{d\Lambda}\tilde\sigma_\Lambda^2
  = s_{ab_2}^\text{2L}(0)  \frac{d\sigma_\Lambda^2}{d\Lambda}\int_{p<\Lambda} \frac{P^\text{lin}(p)}{p^2}   \,,\nn\\
  \frac{dZ^\text{1L}_{ac_2}}{d\Lambda}Z^\text{1L}_{c_2b_2} 
  &=& s^\text{1L}_{ac_2}s^\text{1L}_{c_2b_2} \frac{d\tilde\sigma_\Lambda^2}{d\Lambda}\sigma_\Lambda^2
  = \left( s_{ab_2}^\text{2L}(0) +\frac12 (s_{ab_2}^\text{2L})''(0) \right) \frac{d\sigma_\Lambda^2}{d\Lambda}\frac{1}{\Lambda^2} \int_{p<\Lambda}  P^\text{lin}(p) \nn\\
  &=& s_{ab_2}^\text{2L}(0)\frac{d\sigma_\Lambda^2}{d\Lambda}\frac{1}{\Lambda^2} \int_{p<\Lambda}  P^\text{lin}(p) 
      +\frac12 (s_{ab_2}^\text{2L})''(0) \frac{d\sigma_\Lambda^2}{d\Lambda}\int_{p<\Lambda} \frac{p^2}{\Lambda^2}  \frac{P^\text{lin}(p) }{p^2}\,. \nn
\eea
These expressions make clear how the products of one-loop coefficients reproduce the low-$r$ (i.e.~hierarchical) limits of the two-loop kernels, as given by the commutation-of-limit relations. 
From the expressions after the last equality sign for each of the four contributions, respectively, we see that these one-loop-squared terms provide precisely the subtraction terms that account for the `overlap' with the two-loop contribution, including both leading-gradient and gradient-suppressed pieces, in direct analogy with \hyperlink{cite.Bakx:2025cvu}{Paper~I}. In particular, the terms proportional to $s_{AB}^\text{2L}(0)$ reproduce the leading contributions in the hierarchical limit, while those involving $(s_{AB}^\text{2L})''(0)$ capture the next-to-leading, second-gradient corrections. Note in particular how the second-gradient contribution from the last line in Eq.~\eqref{eq:s2Ldoublesingle2} is required to correctly reproduce the two separate terms appearing in the final line above.

Thus, combining the intrinsic two-loop contributions with the subtraction terms arising from iterated one-loop renormalization, we find that the two-loop RG coefficients isolate precisely the `intrinsic' two-loop part of the running. Importantly, this separation consistently includes contributions from both leading-gradient and second-gradient operators. The resulting expressions can be written as
\be\label{eq:gamma2Labab2}
\boxed{
\begin{array}{lcl}
  \displaystyle \gamma_{ab}^\text{2L} &=& \displaystyle - \frac{d\sigma_\Lambda^2}{d\Lambda}\int_{p<\Lambda} s_{ab}^\text{2L}(p/\Lambda)\big|_{\text{sub}_{0+2}} P^\text{lin}(p),\\
  \displaystyle \gamma_{ab_2}^\text{2L} &=& \displaystyle - \frac{d\sigma_\Lambda^2}{d\Lambda}\int_{p<\Lambda} \left(\frac{s_{ab_2}^\text{2L}(p/\Lambda)\big|_{\text{sub}_{0+2}}}{p^2}+\frac{s_{ab_2}^\text{2L}(p/\Lambda)\big|_{\text{sub}_{0}}}{\Lambda^2}\right)P^\text{lin}(p)\,,
\end{array}
}
\ee
where we introduced subtracted versions of the two-loop double-hard kernels,
\bea \label{eq:ssubs}
  s_{aB}^\text{2L}(p/\Lambda)\big|_{\text{sub}_{0}} &\equiv& s_{aB}^\text{2L}(p/\Lambda) - s_{aB}^\text{2L}(0) \,,\nn\\
  s_{aB}^\text{2L}(p/\Lambda)\big|_{\text{sub}_{0+2}} &\equiv& s_{aB}^\text{2L}(p/\Lambda) - s_{aB}^\text{2L}(0) - \frac12 (s_{aB}^\text{2L})''(0) \frac{p^2}{\Lambda^2}\,,
\eea
valid for both $B=b$ and $B=b_2$. Here derivatives are taken with respect to $r=p/\Lambda$. The subtraction terms remove the contributions that are already accounted for by lower-order renormalization. In particular, the $\text{sub}_{0}$ subtraction removes the leading constant piece in the hierarchical limit $p\ll\Lambda$, while $\text{sub}_{0+2}$ additionally removes the subleading $p^2/\Lambda^2$ contribution. Since odd powers vanish by isotropy and analyticity, the $\text{sub}_{0+2}$ functions start at order $p^4/\Lambda^4$. This makes explicit that all contributions scaling as $p^0$ and $p^2$ are already absorbed by one-loop renormalization (including second-gradient operators), and that the two-loop RG isolates only the intrinsic higher-order UV-sensitive terms.

The different subtraction structures appearing in Eq.~\eqref{eq:gamma2Labab2} reflect the operator content retained in the renormalization scheme. In particular, for the $1/\Lambda^2$ term in $\gamma_{ab_2}^\text{2L}$ only the $\text{sub}_{0}$ subtraction appears, since removing also the $p^2/\Lambda^2$ piece would require including operators with four gradients in the basis, which lies beyond the scope of this work. Similarly, the absence of further subtractions in $\gamma_{ab}^\text{2L}$ indicates that contributions of order $p^4/\Lambda^4$ are not captured within the present operator basis. Finally, the structure of the $1/p^2$ term in $\gamma_{ab_2}^\text{2L}$ ensures that no residual contributions proportional to $P^\text{lin}(p)/p^2$ remain, consistent with the absence of additional lower-derivative counterterms.

This result can be directly compared to the two-loop RG obtained in \hyperlink{cite.Bakx:2025cvu}{Paper~I}, which was based on renormalization of leading-gradient operators only, within the \AB{}  rather than the generalized NLG scheme defined by Eq.~\eqref{eq:rencond2} adopted here. In that case, while the one-loop anomalous dimension $\gamma_{ab}^\text{1L}$ is identical, the two-loop contribution reads
\be\label{eq:gamma2LAssassiBaumann}
  \gamma_{ab}^\text{2L}\Big|_\text{\AB} = - \frac{d\sigma_\Lambda^2}{d\Lambda}\int_{p<\Lambda} s_{ab}^\text{2L}(p/\Lambda)\big|_{\text{sub}_{0}} P^\text{lin}(p)\,.
\ee
Thus, only the leading-gradient contribution is subtracted from the overlap between double-hard and sequential single-hard configurations, while subleading gradient contributions are not removed. The generalized scheme considered here extends this subtraction to include the $p^2/\Lambda^2$ terms, also incorporating second-gradient operators into the RG flow.

\begin{figure}[t]
    \centering
    \includegraphics[width=0.5\columnwidth]{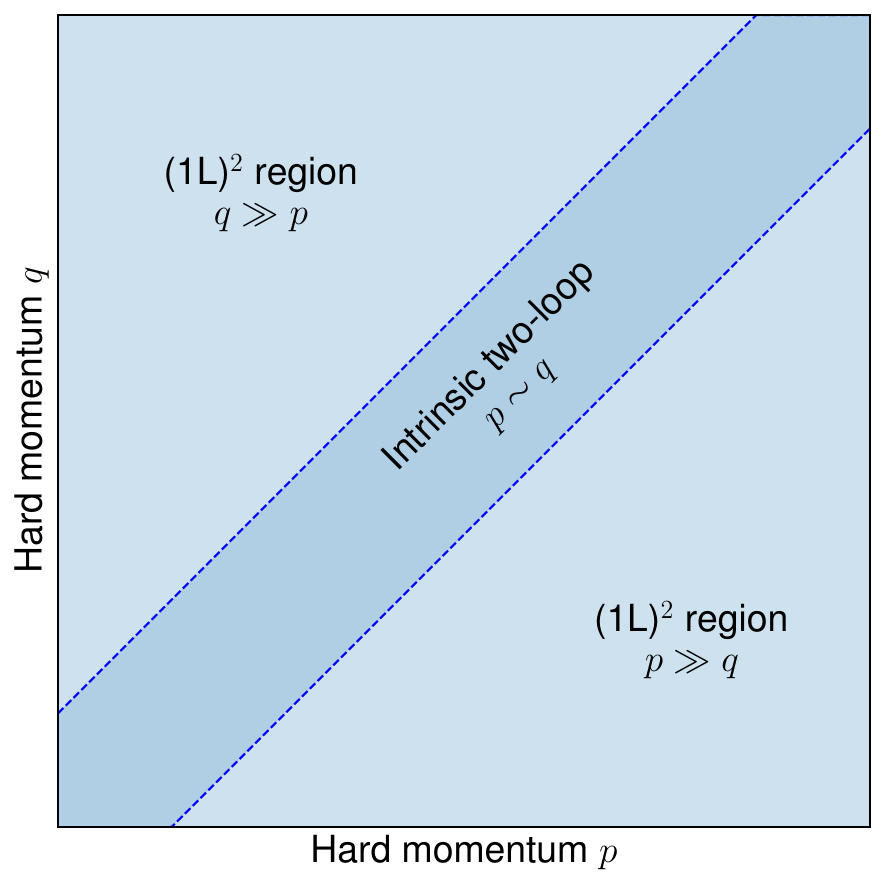}
    \caption{Schematic decomposition of the two-loop momentum plane in terms of regions in the hard loop momenta $(p,q)$. The diagonal band with $p\sim q$ corresponds to the intrinsic two-loop region, encoded by the double-hard kernels $s_{aB}^\text{2L}(r)$. The off-diagonal regions with $p\gg q$ and $q\gg p$ correspond to hierarchical configurations that factorize into two sequential single-hard limits, or $(\mathrm{1L})^2$, contributions. }
    \label{fig:methodOfRegionsPQ}
\end{figure}

We emphasize the relation between the single double-hard limits and commutation-of-limit relations from Eqs.~\eqref{eq:s2Ldoublesingle} and~\eqref{eq:s2Ldoublesingle2} to the one and two-loop RG. These relations encode the factorization properties of the bias kernels in hierarchical momentum configurations, where the double-hard limit decomposes into sequential single-hard limits. As a result, they ensure that the contributions from iterated one-loop renormalization exactly reproduce the leading and subleading terms in the small-$r$ expansion of the two-loop kernels. This guarantees a precise cancellation between the one-loop-squared terms and the corresponding parts of the two-loop integrals. This structure can be understood from the method of regions~\cite{Beneke:1997zp}, where we can decompose the two-loop momentum plane $(p,q)$ into a diagonal region ($p\sim q$, filtered out by the functions described in \reffig{2loopfuncs}) and two off-diagonal hierarchical regions ($p\gg q$ and $q\gg p$) as in Fig.~\ref{fig:methodOfRegionsPQ}. The diagonal region corresponds to the intrinsic two-loop contribution and is absorbed by the intrinsic two-loop double-hard kernels $s_{aB}^\text{2L}(r)$, while the off-diagonal regions correspond to hierarchical configurations in which one hard loop is integrated out first and the second loop acts subsequently, so that the contribution factorizes into two sequential single-hard limits. These are precisely the $(\mathrm{1L})^2$ regions reproduced by products of one-loop mixing coefficients.

In line with our findings from \hyperlink{cite.Bakx:2025cvu}{Paper~I}, we will see in \refsec{RGsol} that also in the NLG case, most of the UV sensitivity from loop integrals is already captured by the one-loop RG evolution, which resums the effects associated with sequential single-hard configurations.

\subsection{Stochastic RG equations}

The correlators can equivalently be expressed in terms of bare loops and
bare bias and noise terms.
In that case, bare loops depend on the UV cutoff, and the
cutoff-dependence needs to be cancelled by
choosing the bias and noise terms to depend on $\Lambda$. For the bias
this is described by the deterministic RG equations \cite{Rubira:2023vzw}.
Analogously, RG equations for the bare noise coefficients can be derived as in \cite{Rubira:2024tea}, using the relation to the renormalized noise coefficients derived above together with the fact that the latter are, by definition, $\Lambda$-independent.

For the noise terms related to the contact terms from products of {\it
two} operators this yields, using Eq. \eqref{eq:bisp_cd}, 
\be \label{eq:dLcd}
   \frac{d}{d\Lambda}c_{[d]} = b_{[a]}b_{[b]} \frac{d}{d\Lambda} Z_{abd}\,,
\ee
using that $\frac{d}{d\Lambda} c_{[D]} =  ([1+Z]^{-1})_{CD}\frac{d}{d\Lambda}c_{C} + c_{C}\frac{d}{d\Lambda}([1+Z]^{-1})_{CD}$ from Eq. \eqref{eq:bisp_cd} and also that
\be \label{eq:Zderiv}
\frac{d}{d\Lambda} Z_{abd}^\text{1L} = -2 s^\text{1L}_{abd}P^\text{lin}(\Lambda) \frac{d\sigma_\Lambda^2}{d\Lambda} \,,
\ee
 from Eq. \eqref{eq:Zabc1L}, we end up after expanding both sides of Eq.~\eqref{eq:dLcd} at one loop
\be \label{eq:dLcdfinal}
   \frac{d}{d\Lambda}c_{d}\Big|_{\rm 1L} = -\frac{d\sigma_\Lambda^2}{d\Lambda} \left[s^\text{1L}_{ad}c_{a}  -2 s^\text{1L}_{abd} b_{a}b_{b} P^\text{lin}(\Lambda) \right]\,.
\ee
This equation matches the $m=2$ case of Eq.~3.36 of \cite{Rubira:2024tea} (see also Eqs.~4.7 and 4.8 therein). Note that the different dimensions of $c$ and $b$ coefficients are compensated by a factor $P^\text{lin}(\Lambda)$.
The leading stochastic coefficients therefore receive two contributions to their running; one linear in the other stochastic parameters and another quadratic via the deterministic bias. See Fig.~4 of \cite{Rubira:2024tea} for a schematic representation of the stochastic RGE structure.

For the noise terms related to the contact terms from products of {\it
three} operators this yields, using Eq. \eqref{eq:dren}, 
\be
   \frac{d}{d\Lambda} d_{[e]} = b_{[a]}b_{[b]}b_{[c]}\frac{d}{d\Lambda}
Z_{abce} + 3[c_{f}]b_{[c]}\frac{d}{d\Lambda} Z_{fce}\,,
\ee
corresponding to the $m=3$ case of Eq.~3.36 of~\cite{Rubira:2024tea} (see also Eqs.~4.11 and 4.12 therein).
Truncating once more at one-loop and computing the derivative of $Z_{abce}$ from Eq.~\eqref{eq:Zabcd}, similar to Eq.~\eqref{eq:Zderiv}, we find
\be
   \frac{d}{d\Lambda} d_{e}\Big|_{\rm 1L} = - \frac{d\sigma_\Lambda^2}{d\Lambda} \left[s^\text{1L}_{ae} d_{a}  + 6s^\text{1L}_{abe}c_{a}b_{b}P^\text{lin}(\Lambda) +  3s^\text{1L}_{abce}b_{a}b_{b}b_{c}\left(P^\text{lin}(\Lambda)\right)^2 \right]\,.
\ee
The RGE for the product of four operators proceeds similarly and corresponds to the $m=4$ case of Eq.~3.36 of \cite{Rubira:2024tea}.

\subsection{RG solutions} \label{sec:RGsol}

We now turn to explicit solutions of the RG equations derived in the previous subsections. The RGEs of the deterministic bias coefficients can be solved independently of those for stochastic contributions, while the latter are coupled to the former. We discuss RG solutions involving NLG deterministic operators in the following, in line with the focus of this work.

Naively, the expected size of the dimensionful NLG bias coefficients is set by either $R_L^2$ or $1/k_\text{NL}^2$. Given the RG running, this may however be realized only within a certain range of scales. To make this point more clear, consider first the well-known case of the matter density power spectrum. In the large-scale limit, the leading non-linear correction scales as $k^2P^\text{lin}(k)$. Within SPT, its coefficient receives contributions from all loop orders, starting at $P^{(13)}$, then $P^{(15)}$ at two-loop order, and so on, with each perturbative order becoming increasingly cutoff-dependent. In addition, the leading EFT correction contributes to this term, captured by $b_{\nabla^2\delta}(\Lambda)$. On the other hand, when expanding the power spectrum on large scales, only the overall size of the $k^2P^\text{lin}(k)$ correction is an (in principle) measurable quantity. The NLG renormalization scheme precisely imposes the condition that the running $b_{\nabla^2\delta}(\Lambda)$ approaches this overall size, i.e.
$b_{\nabla^2\delta}(\Lambda=0)=b_{[\nabla^2\delta]}$  such that
\be
  P_{\delta\delta}(k) = P^\text{lin}(k)+2b_{[\nabla^2\delta]} k^2P^\text{lin}(k)+{\cal O}(k^4P^\text{lin}, k^4)\quad \text{for}\ k\to 0\,.
\ee
Thus, the value of $b_{[\nabla^2\delta]}$ corresponds to the sum of all loop and EFT corrections that yield contributions scaling as $k^2P^\text{lin}(k)$, while the latter are subtracted out from the renormalized loop integrals. When running to higher $\Lambda$ starting from $\Lambda=0$, the loops are gradually `switched on' again, and the absolute size of $b_{\nabla^2\delta}(\Lambda)$ is correspondingly reduced, {\it i.e.} schematically $b_{[\nabla^2\delta]}=b_{\nabla^2\delta}(\Lambda)\,+\,$sum of contributions  from all loops, evaluated with cutoff $\Lambda$. 
This may be compared to the frequently used $c_{s,\infty}^2$ EFT correction to the matter density power spectrum, which is in practice conventionally determined by requiring that the sum of $P^{(13)}$ (computed with infinite cutoff) as well as a term of the form $-2c_{s,\infty}^2k^2P^\text{lin}$ matches simulated or real data, while either ignoring or subtracting additional contributions such as from $P^{(15)}$. This yields the translation 
\be\label{eq:cssqdef}
  c_{s,\infty}^2 \equiv \left(- b_{\nabla^2\delta}(\Lambda=0)\right)\Big|_\text{matter} + \frac{-61}{630}\int_p^\infty \frac{P^\text{lin}(p)}{p^2} \,,
\ee
where the first term
is related to the separate universe measurement~\cite{Lazeyras:2019dcx} and the last term on the RHS comes from the low-$k$ limit of $P^{(13)}$,
also given by the $s^\text{1L}_{\delta \nabla^2\delta}$ entry in \reftab{sab21L_A}. Typical values found in fits to simulations are of
order $c_{s,\infty}^2 \approx 1 \, {\rm Mpc}^2/h^2$ (see e.g.~\cite{Lazeyras:2019dcx}), while 
\be \label{eq:sigmatilde_limit}
\tilde{\sigma}^2_\infty = \int_p^\infty \frac{P^\text{lin}(p)}{p^2} \approx 110 \, {\rm Mpc}^2/h^2\,,
\ee 
for $\Lambda$CDM, which yields values for $b_{\nabla^2\delta}(\Lambda=0)$ of order $10 \, {\rm Mpc}^2/h^2$. When increasing $\Lambda$, part of the contribution to $k^2P^\text{lin}(k)$ is provided by the loop corrections, such that the absolute size of $b_{\nabla^2\delta}(\Lambda)$ is expected to decrease. The RG approach allows us to analyze this scale-dependence in detail, starting from a physically meaningful initial value at $\Lambda=0$.

We note that the properties discussed above can be generalized. All initial values of LG and NLG bias coefficients at $\Lambda=0$ match their renormalized values, 
\be
  b_A(\Lambda=0)=b_{[A]}\qquad \mbox{for}\ A=a,a_2\,,
\ee
which in turn capture the complete physical contribution to all $N$-point functions in an expansion around the low-$k_i$ limit to which they contribute. Thus the initial values could in principle be fixed by measuring these contributions,  e.g.~via the separate-universe technique. Then, the RG predicts how the coefficients evolve with scale when increasing $\Lambda$.

We begin with the `counterterm RG' for the matter density field, for which
\be
  b_\delta=1,\qquad b_a=0\quad (a\neq \delta),
\ee
and the leading-gradient sector is not sourced by the counterterms. Therefore the RG equation for the higher-derivative (NLG) coefficients is a closed system. Starting from Eq.~\eqref{eq:gammaab1L}, and using \(d\tilde\sigma_\Lambda^2/d\Lambda=\Lambda^{-2}d\sigma_\Lambda^2/d\Lambda\), we write at one-loop
\be
  \frac{d b_{a_2}}{d\sigma_\Lambda^2}
  =
  - \bar{s}^{\rm 1L}_{a_2b_2}\,b_{b_2}
  - J_{a_2}(\Lambda)\,, \quad \textrm{with} \quad J_{a_2}(\Lambda) = \frac{1}{\Lambda^2}\,s^{\rm 1L}_{\delta a_2} \,,
  \label{eq:RGE_a2_affine}
\ee
which decomposes into a homogeneous part and an inhomogeneous ODE sourced by $J_{a_2}(\Lambda)$. Here we defined $\bar{s}^{\rm 1L}_{a_2b_2} \equiv {s}^{\rm 1L}_{b_2a_2}$ to be able to write the homogeneous part in usual matrix-vector form.

For the solution of the homogeneous ODE, diagonalize the second-gradient mixing matrix,
\be
  \bar{s}^{\rm 1L}_{a_2b_2}p_{b_2 i_2}=p_{a_2 i_2}\lambda_{i_2},
\ee
and define the diagonal operators and coefficients by
\be \label{eq:RGmattersol}
  {\cal O}^{\rm diag}_{i_2}\equiv {\cal O}_{a_2}p_{a_2 i_2},
  \qquad
  b^{\rm diag}_{i_2}\equiv (p^{-1})_{i_2a_2}b_{a_2},
  \qquad
  s^{\rm 1L,diag}_{\delta i_2}\equiv (p^{-1})_{i_2a_2}s^{\rm 1L}_{\delta a_2}.
\ee
In this basis the RG equations decouple, up to the fixed matter source,
\be
  \frac{d b^{\rm diag}_{i_2}}{d\sigma_\Lambda^2}
  =
  -\lambda_{i_2} b^{\rm diag}_{i_2}
  -\frac{1}{\Lambda^2}s^{\rm 1L,diag}_{\delta i_2}
  \qquad\text{(no sum over \(i_2\)).}
  \label{eq:RGE_a2_diag}
\ee

When truncating the NLG sector to second order, the spectrum is given by
\be
\lambda_{i_2}=\{0,0,-1.66,-3.69\}.
\ee
In this reduced basis there are only relevant and marginal directions.
The eigenvalues (ordered by decreasing modulus) of the third-order $-\bar s^{\rm 1L}_{a_2b_2}$ mixing matrix are given by
\be
\begin{aligned}\label{eq:eigs_s_a2b2}
  \lambda_{i_2}=\{&
  -16.4649,\; -12.6144,\; -10.1159,\; -8.02957,\;
  -3.4373,\; -3.33714,\; -3.28188,\; \\
  &3.19651,\; -2.01088,\; -1.69142,\; -1.69142,\; -1.69142,\;
  -1.58244,\; 1.52725,\;
\\
  &  -0.632969-0.407688\,i,\; -0.632969+0.407688\,i,\;
  0.219837,\; 0,\;0,\;0\}.
\end{aligned}
\ee
Most eigenvalues are relevant (decaying in the UV), but now with three positive irrelevant operators, two of those larger than unity.  
The spectrum also contains marginal directions and a complex-conjugate pair gives a pair of scaling directions with the same real part and an additional rotational component in operator space, whose relevance is determined by the real part of the eigenvalue. 
An interesting feature in the eigensolution is the presence of three eigenvectors that are aligned with the operators $\hat{\Pi}_3$, $\Pi_{B1}$ and $\Pi_{B2}$, all sharing the same eigenvalue $\lambda \simeq -1.69$. This reflects the existence of a three-dimensional degenerate eigenspace of the mixing matrix. In such a situation, the eigenvectors are not uniquely defined, since any orthogonal linear combination of these three directions is also an eigenvector with the same eigenvalue.

Thus the homogeneous evolution of each eigenmode is controlled by
$\exp[-\lambda_{i_2}\Delta\sigma^2]$, while the matter density provides an inhomogeneous forcing term.
The exact solution between two cutoffs $\Lambda_*$ and $\Lambda$ is
\be
  \boxed{
  b^{\rm diag}_{i_2}(\Lambda)
  =
  e^{-\lambda_{i_2}[\sigma_\Lambda^2-\sigma_{\Lambda_*}^2]}
  b^{\rm diag}_{i_2}(\Lambda_*)
  -
  s^{\rm 1L,diag}_{\delta i_2}
  \int_{\Lambda_*}^{\Lambda}d\Lambda'\,
  \frac{1}{\Lambda'^2}
  \frac{d\sigma_{\Lambda'}^2}{d\Lambda'}
  e^{-\lambda_{i_2}[\sigma_\Lambda^2-\sigma_{\Lambda'}^2]}
  }\,.
  \label{eq:RGE_a2_diag_solution}
\ee
Expanding for small running interval gives
\be
\begin{aligned}
  b^{\rm diag}_{i_2}(\Lambda)
  =
  &\,b^{\rm diag}_{i_2}(\Lambda_*)
  -s^{\rm 1L,diag}_{\delta i_2}
  \left(\tilde\sigma_\Lambda^2-\tilde\sigma_{\Lambda_*}^2\right)
  -\lambda_{i_2}b^{\rm diag}_{i_2}(\Lambda_*)
  \left(\sigma_\Lambda^2-\sigma_{\Lambda_*}^2\right)
  \\
  &+
  \frac{\lambda_{i_2}^2}{2}
  b^{\rm diag}_{i_2}(\Lambda_*)
  \left(\sigma_\Lambda^2-\sigma_{\Lambda_*}^2\right)^2
  +
  \lambda_{i_2}s^{\rm 1L,diag}_{\delta i_2}
  \int_{\Lambda_*}^{\Lambda}
  d\tilde\sigma_{\Lambda'}^2
  \left(\sigma_\Lambda^2-\sigma_{\Lambda'}^2\right)
  +\cdots .
\end{aligned}
\ee
The first correction is the direct generation of higher-derivative counterterms by the matter density, while the second is the homogeneous mixing among second-gradient operators. The diagonal basis therefore separates the two effects: a fixed matter source controlled by \(s^{\rm 1L}_{\delta a_2}\), and eigenmode-by-eigenmode amplification or damping controlled by the spectrum of \(s^{\rm 1L}_{a_2b_2}\).

\begin{figure}[t]
    \centering
    \includegraphics[width=0.47\textwidth]{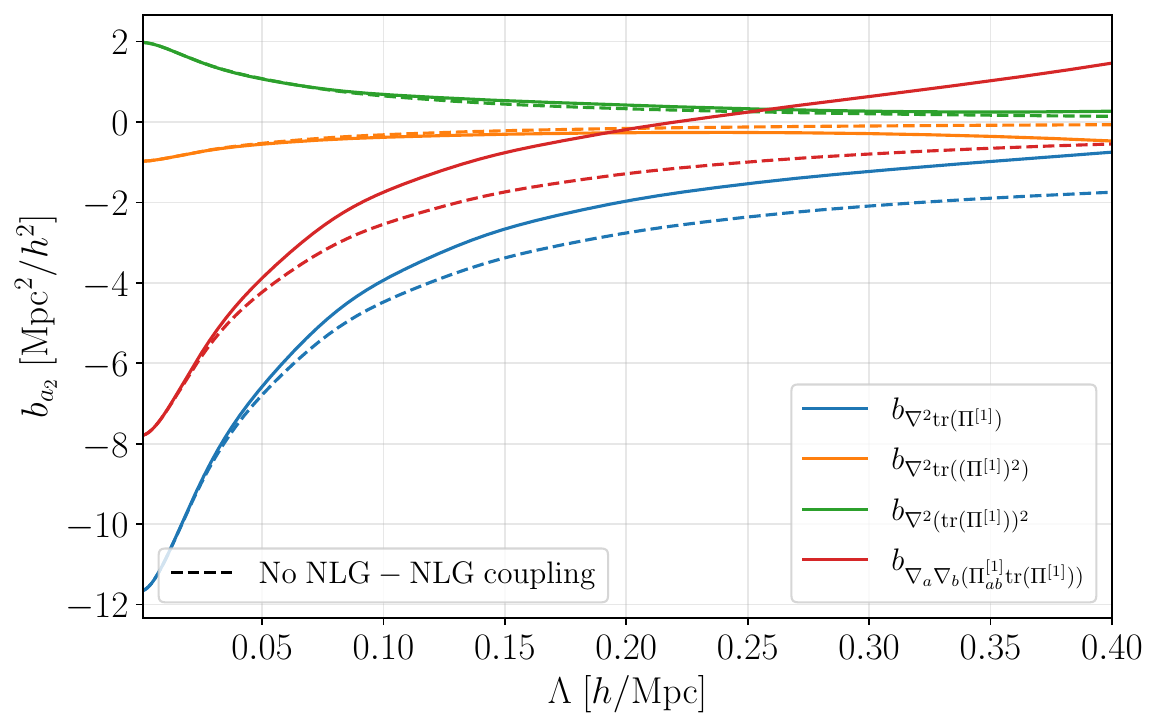}
    \includegraphics[width=0.47\columnwidth]{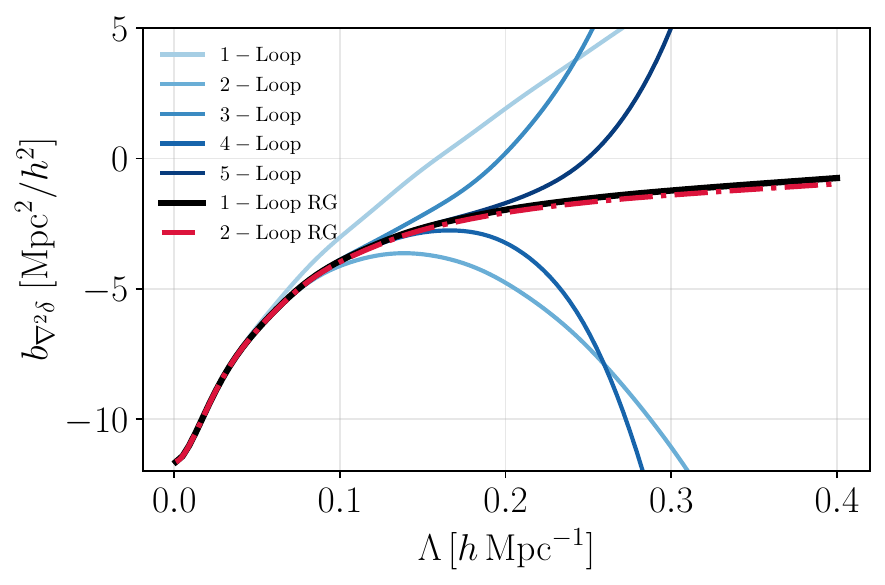}
    \caption{
    \textit{Left:} One-loop RG running of the first four NLG bias parameters. All curves are evolved from initial conditions at $\Lambda \to 0\,h/$Mpc 
    as explained below Eq.~\eqref{eq:sigmatilde_limit}, and interpolate between an initial value capturing the entire physical scale-dependent non-linear corrections to the density field from the respective NLG operators at $\Lambda\to 0$, and cutoff-dependent values of order $1/k_\text{NL}^2$ on mildly non-linear scales.
      Solid lines of each colour include the full NLG--NLG mixing ($s^{1{\rm L}}_{a_2 b_2}$
      coupling) while dashed lines of matching colour show the result with this coupling
      set to zero. 
      \textit{Right:} Convergence of the iterative solution for the NLG bias parameter $b_{\nabla^2\delta}$ as a function of the cutoff scale $\Lambda$ for the same initial conditions.
  Blue lines of increasing shade show successive total contributions from different loop terms, while
  the solid black line gives the exact one-loop RG result obtained by numerical integration of the flow
  equation.
  The red dash-dotted line shows the two-loop RG result. 
    }
    \label{fig:1looprunningmatter}
\end{figure}

The left panel of \reffig{1looprunningmatter} shows the one-loop RG evolution of the NLG bias coefficients in the matter case. To choose physically motivated initial conditions, we 
relate the coefficient $b_{\nabla^2\delta}$ to the conventional EFT counterterm $c_{s,\infty}^2$ using Eq.~\eqref{eq:cssqdef}, and set
\be
b_{\nabla^2\delta}(\Lambda=0)=s^\text{1L}_{\delta \nabla^2\delta}\tilde{\sigma}^2_\infty +(-c_{s,\infty}^2) \simeq -11.65\,{\rm Mpc}^2/h^2 \,,
\ee
as initial conditions, where we used Eq.~\eqref{eq:sigmatilde_limit}
and the estimate $c_{s,\infty}^2 \approx 1 \, {\rm Mpc}^2/h^2$ (see e.g.~\cite{Lazeyras:2019dcx}) for illustration.
For the remaining NLG counterterms,  adopting the analogous parameterization yields
\be \label{eq:ba2IC}
b_{a_2}(\Lambda=0) = s^\text{1L}_{\delta a_2} \tilde{\sigma}^2_\infty +\Delta b_{a_2}\,.
\ee
Here $\Delta b_{a_2}$, which is defined through Eq.~\eqref{eq:ba2IC}, plays the analog role of $c_{s,\infty}^2$ for the higher-order NLG counterterms, i.e. is {\it defined} as the mismatch of the full physical value and the SPT one-loop prediction computed with infinite cutoff. By virtue of this definition, it implicitly absorbs residual higher-loop contributions along with EFT corrections. The optimal choice for the initial conditions $b_{a_2}(\Lambda=0)$  would be  separate-universe measurements of the renormalized coefficients of the respective NLG operators. In absence of such values, in \reffig{1looprunningmatter}, for illustration, we instead assume that the physical value is not parametrically different from the SPT one-loop prediction and thus set $\Delta b_{a_2}=0$. We checked that our results are qualitatively insensitive to this choice as long as the assumption from above is satisfied.
By starting from Eq.~\eqref{eq:ba2IC}, the part of the NLG coefficients that is fixed by the one-loop renormalization of $\delta$ is partially removed when running towards increasing $\Lambda$ and the counterterms are order one (in units of $1/k_\text{NL}^2$) around mildly nonlinear scales. For comparison,  the dashed curves show the evolution obtained after switching off the NLG--NLG mixing block, $s^{1{\rm L}}_{a_2b_2}=0$. In that case, note that $b_{\nabla^2\delta}$ asymptotes to $-1\,{\rm Mpc}^2/h^2$ as $\Lambda \to \infty$, as set by the physical estimate of $c_{s,\infty}^2$ we use as input.  Comparing the dashed and solid curves, we find order-one shifts in the NLG coefficients at scales approaching the nonlinear regime. This indicates that priors on higher-derivative bias parameters based only on the natural size inferred from the $b_\delta$-sourced one-loop counterterms can miss an important part of the RG evolution, and should therefore be used with caution.

We also highlight an interesting feature related to the NLG-NLG mixing pattern arising from $s^{1{\rm L}}_{a_2b_2}$. The associated large negative eigenvalues, see Eq.~\eqref{eq:eigs_s_a2b2}, make the solution of the RGE rather robust when running from $\Lambda=0$ towards larger values, meaning that the qualitative behaviour does not depend very sensitively on the precise details of the initial conditions. This suggests that initializing LG and NLG bias coefficients with separate-universe values (or deriving priors based on those) can be a viable option in practice. Conversely, when attempting to initialize the RG at $\Lambda\sim k_\text{NL}$ and running towards $\Lambda\to 0$ the large negative eigenvalues of $s^{1{\rm L}}_{a_2b_2}$ imply an exponential sensitivity of the IR solution to initial conditions set in the UV. Overall, the NLG-NLG mixing may imprint non-trivial features on the pattern of NLG bias coefficients, leaving a more detailed investigation to future work.

The right panel of \reffig{1looprunningmatter} illustrates how the RG evolution reorganizes the perturbative expansion for the higher-derivative bias coefficient $b_{\nabla^2\delta}$.
The successive lines, resulting from the series expansion in Eq.~\eqref{eq:RGE_a2_diag_solution}, approach the RG solution. This illustrates that solving the RG equation resums the tower of higher-loop contributions generated by repeated one-loop insertions, as indicated in Eq.~\eqref{eq:RGE_a2_diag_solution}. In the method-of-regions language, these terms correspond to hierarchical hard-momentum configurations, where the multi-loop integral factorizes into a sequence of single-hard subdiagrams. This is the NLG analogue of the result found in \hyperlink{cite.Bakx:2025cvu}{Paper~I} (see Fig.~2 and Sec.~8 therein) for leading-in-derivative operators, where the one-loop RG similarly captured the iterated contributions from ordered hard regions.
The red dash-dotted curve shows that including the intrinsic two-loop RG contribution $\gamma^\text{2L}$ from Eq.~\eqref{eq:gamma2Labab2} only mildly changes the one-loop RG evolution over the range of cutoffs shown. This suggests that the intrinsic two-loop contribution associated with the diagonal region ($p\sim q\gg k$) is numerically subdominant compared to the hierarchical regions already resummed by the one-loop RG. Equivalently, most of the cutoff dependence is captured by successive one-loop mixings, while the additional non-factorizable two-loop hard region gives only a small correction. 

\begin{figure}[t]
    \centering
    \includegraphics[width=0.47\textwidth]{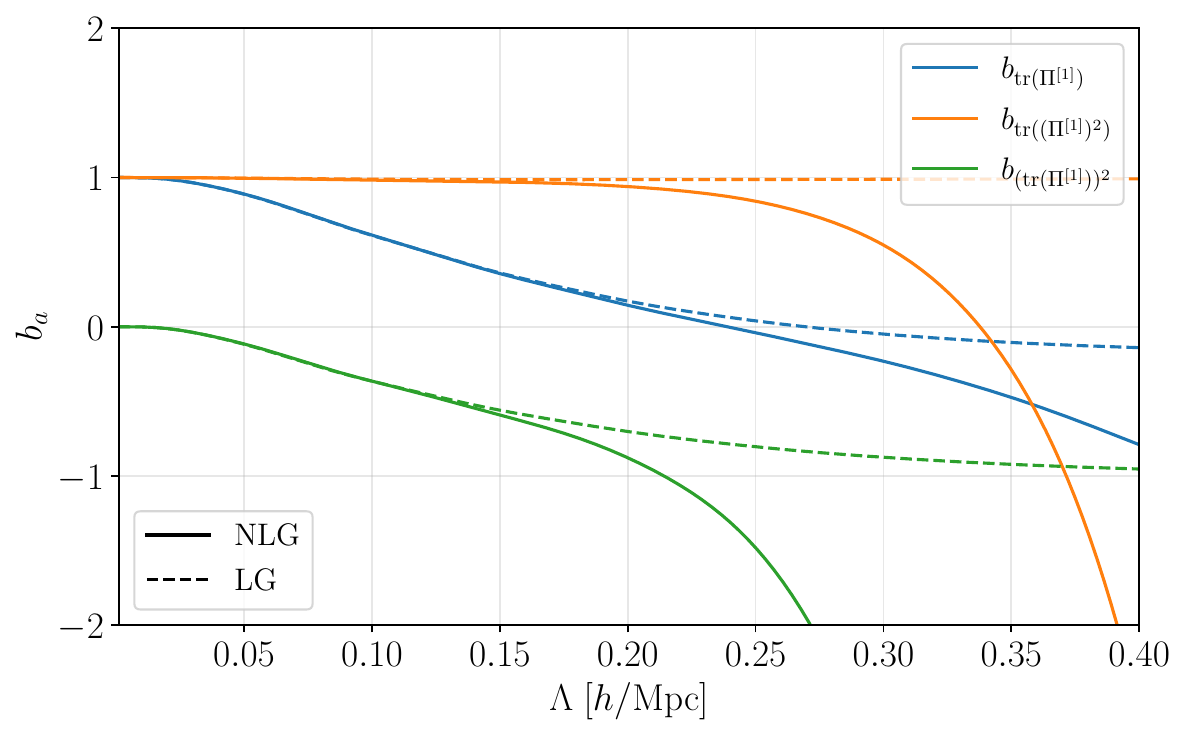}
    \includegraphics[width=0.47\textwidth]{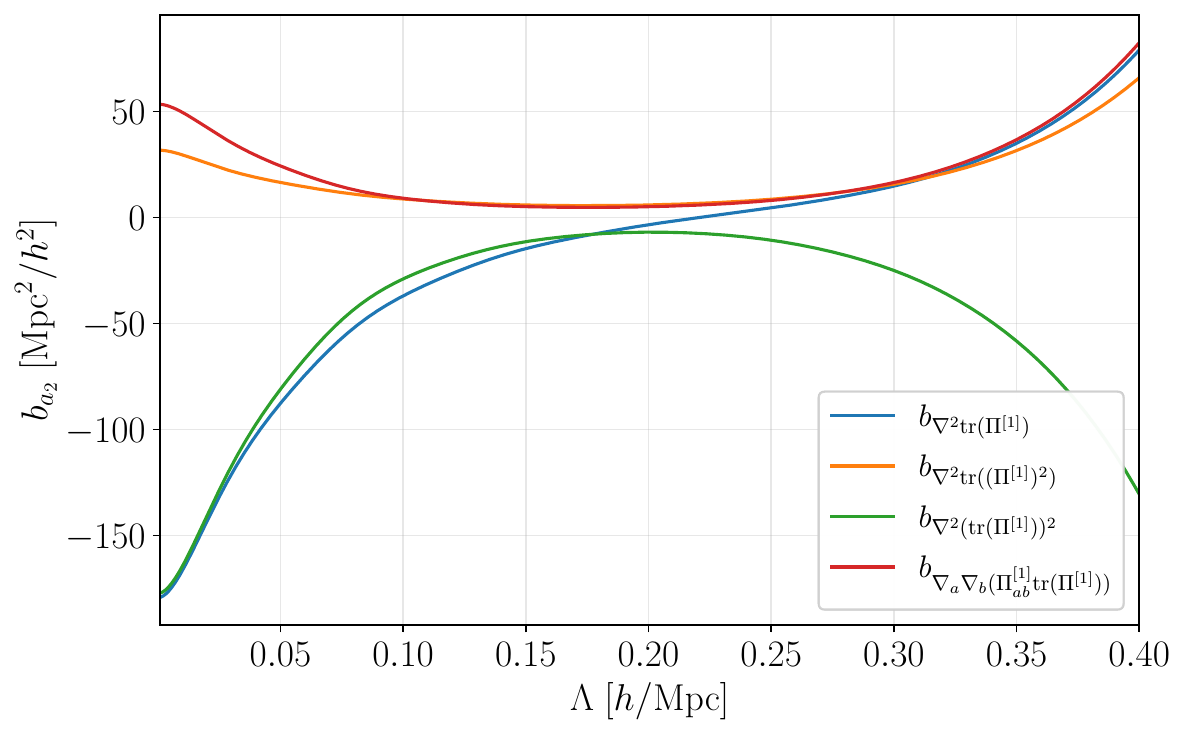}
    \caption{
    One-loop RG flow of LG bias coefficients (left) and NLG bias coefficients (right) as a function of the cutoff
  scale $\Lambda$.
  Solid lines correspond to the full evolution including $s_{a_2 b}$ mixing between the NLG and LG sectors,
  while dashed lines show the result with this mixing switched off.
  All curves are evolved from initial conditions at $\Lambda \to 0\,h/\mathrm{Mpc}$ with $b_{\delta}(\Lambda=0) = 1$ and $b_{\mathrm{tr}{\Pi^{[1]}}^2}(\Lambda=0) = 1$, with all remaining LG coefficients set to zero. The NLG initial values are chosen following Eq.~\eqref{eq:ba2IC2}.
    }
    \label{fig:plot_B_one_loop_with_without_snlplp}
\end{figure}

\reffig{plot_B_one_loop_with_without_snlplp} illustrates the qualitative impact of the LG--NLG mixing in the one-loop RG equations for a generic biased tracer. In principle, the initial conditions for all LG and NLG operators should be set by separate-universe measurements for the respective tracer. In the absence of such results,
for illustration, we set initial conditions at $\Lambda \to 0\,h/\mathrm{Mpc}$ with $b_{\delta}(\Lambda=0) = 1$ and $b_{\mathrm{tr}{\Pi^{[1]}}^2}(\Lambda=0) = 1$ for the LG operators, and all others set to zero. For the NLG operators, following the same idea of using the leading one-loop contribution to check whether the running parameters are of the order of their expected `natural' size on non-linear scales (see right panel), we fix similar to Eq.~\eqref{eq:ba2IC} 
\be \label{eq:ba2IC2}
b_{a_2}(\Lambda=0) = \left(b_{\delta}(\Lambda=0) s^{1{\rm L}}_{\delta a_2} + b_{\mathrm{tr}{\Pi^{[1]}}^2}(\Lambda=0) s^{1{\rm L}}_{{\mathrm{tr}{\Pi^{[1]}}^2} a_2}\right) \tilde{\sigma}^2_\infty +\Delta b_{a_2}\,,
\ee
with $s^{1{\rm L}}_{\delta a_2}$ and $s^{1{\rm L}}_{{\mathrm{tr}{\Pi^{[1]}}^2} a_2}$ given in \reftab{sab21L_A}.
For illustration, we set $\Delta b_{a_2}=0$ in \reffig{plot_B_one_loop_with_without_snlplp}.
In contrast to the matter case discussed above, the leading-gradient bias sector is no longer fixed by momentum conservation and once higher-derivative bias coefficients are present, the $s^{1{\rm L}}_{a_2 b}$ non-$\nabla^2$ block feeds back into the running of the leading-gradient coefficients. The solid curves show the resulting coupled evolution, while the dashed curves are obtained by artificially switching off this NLG-to-LG mixing. This comparison isolates the effect of the off-diagonal block in the anomalous-dimension matrix.
At small cutoffs the two sets of curves agree, as expected from the derivative expansion. The contribution of NLG operators to the running of LG coefficients is accompanied by the dimensionful factor $\Lambda^2 b_{a_2}$, and is therefore suppressed at sufficiently low $\Lambda$. Around $\Lambda\simeq 0.2\,h/{\rm Mpc}$, however, the solid and dashed curves begin to separate, showing that the feedback from higher-derivative bias operators can become numerically relevant before reaching deeply nonlinear scales. 

The NLG operators feature a strong running between the initial scale at $\Lambda=0$ and mildly non-linear scales. This is expected, since the `accidental' suppression of $\tilde{\sigma}_\infty^2$ from Eq. \eqref{eq:sigmatilde_limit} by $s^\text{1L}_{\delta \nabla^2 \delta} = -61/630$ in the matter case does not generalize to biased tracers, where the corresponding coefficient can be order unity, cf. \reftab{sab21L_A}. This indicates that those scales yield very sizeable non-linear corrections to the coefficients of the respective NLG operators, but is parametrically compatible with the one-loop expectation, see Eq.~\eqref{eq:ba2IC2}. At around the non-linear scale, the running is broadly compatible with `natural' values of order $1/k_\text{NL}^2$, while still featuring a sizeable scale-dependence. This indicates that priors on NLG parameters motivated by those arguments should be used with care. It also suggests a possible application of the RG technique in allowing to check whether bias parameters inferred from fits to data are compatible with the theoretically expected running when using several different values of the cutoff $\Lambda$.

\newpage

    \section{Conclusion}\label{sec:conclusion}

In this work we have developed a complete renormalization framework for biased tracers at next-to-leading gradient (NLG) order, extending the leading-gradient (LG) treatment of \hyperlink{cite.Bakx:2025cvu}{Paper~I} to include operators with two extra spatial derivatives. We constructed the complete non-redundant basis of scalar NLG bias operators at second gradient order, classifying them into operators with two overall derivatives, operators without overall derivatives (relevant for biased tracers only), and a set of `backreaction' operators. Together with the 29 LG operators up to fifth perturbative order, this provides the full operator content required for the renormalization of the two-loop galaxy power spectrum as well as the one-loop bispectrum and trispectrum.

We have generalized the renormalization condition of \cite{Assassi:2014fva} to simultaneously absorb both the leading ($k^0$) and the quadratic-in-wavenumber ($k^2$) UV-sensitive contributions of loop integrals. The generalized condition in Eq.~\eqref{eq:rencond2} fixes the full structure of the renormalization mixing matrix $Z_{AB}$, with blocks $Z_{ab}$, $Z_{ab_2}$, $Z_{a_2b}$, and $Z_{a_2b_2}$ coupling LG and NLG operators. We computed the complete one-loop anomalous-dimension matrices. Three distinct momentum integrals $\sigma_\Lambda^2$, $\tilde\sigma_\Lambda^2$, and $\hat\sigma_\Lambda^2$ enter the one-loop counterterms (cf. Eq.~\eqref{eq:Z1L}), induced by the mixing between operators of different derivative orders.  At two-loop order, we computed the renormalization matrices associated with the double-hard limits of the bias kernels. In addition to the function $g(r)$ from \hyperlink{cite.Bakx:2025cvu}{Paper~I}, three new weight functions $h_1(r)$, $h_2(r)$, and $h_3(r)$ appear in the NLG sector from Eq.~\eqref{eq:doublehard2}. All four functions are symmetric under $r\to 1/r$, vanish in the hierarchical limit $r\to 0$, and have most of their support at $r \sim 1$, encoding the selection of the intrinsic two-loop region $p\sim q$ via the method of regions. 

The operator product expansion framework for stochastic bias renormalization was extended from products of two to also products of three, and four LG operators. The corresponding contact-term renormalization conditions fix the renormalization constants $Z_{ab\cdots}$. We evaluate those that enter the power spectrum, bispectrum, and trispectrum at the relevant loop orders, respectively, by computing the corresponding hard limits of combinations of bias kernels. 

The RGEs governing the UV-cutoff dependence of the bare bias coefficients were derived for the full LG+NLG operator basis, including the coupled running between LG and NLG sectors. At one-loop order, the block form of the anomalous-dimension matrix in Eq.~\eqref{eq:gammaab1L} makes explicit how higher-derivative operators are generated from leading-gradient ones and vice versa. The two-loop anomalous dimensions separate intrinsic two-loop mode coupling from iterated one-loop contributions, with the commutation of limits providing the required cancellations.

The results of this paper provide all the ingredients needed to evaluate the renormalized two-loop galaxy power spectrum with second-gradient deterministic counterterms included. Moreover, we provide a concrete recipe for joint analyses that add the renormalized one-loop bispectrum and trispectrum fully consistently within the same framework. Together with \hyperlink{cite.Bakx:2025cvu}{Paper~I}, this establishes a complete and systematic renormalization procedure for tracers in the EFT up to fifth order in the galaxy bias expansion in real space. 

Let us reiterate this last point in more detail, following up on the discussion at the end of \refsec{counting}. To jointly model any set of $N$-point functions within the EFT at some given loop order 
using the galaxy bias renormalization approach,
one proceeds as follows: 
\begin{enumerate}
    \item List the deterministic LG and NLG bias operators that contribute to (any of the) $N$-point statistics at the relevant order in perturbation theory. We provide a complete list of all NLG operators required up to fifth order in Sec.~\ref{sec:basis} (when counting gradients equally as powers of $\delta_L$), while all LG operators are given in \hyperlink{cite.Bakx:2025cvu}{Paper~I}.
    \item Compute their renormalized kernels at the relevant loop order, using either Eq.~\eqref{eq:singlehardNLG} (one-loop) or Eq.~\eqref{eq:doublehard2} (two-loop). Renormalization of the deterministic part of loop corrections amounts to replacing the `bare' by those renormalized kernels inside the loop integrands.
    \item Remove any operators that are degenerate with other operators at  \textit{all} orders in perturbation theory at which they occur within the targeted set of $N$-point functions. This can be achieved by evaluating the corresponding renormalized kernels on the requisite wavenumber configurations; in practice, within the operator basis we use this is typically the case for operators for which all of the required renormalized kernels are identically zero (see the discussion at the end of \refsec{correlators}). We provide an overview of the required subsets of LG and NLG operators for various combinations of $N$-point functions in Tab.~\ref{tab:operator_counting}.
    \item Using this possibly reduced operator basis, compute the relevant `SPT' loop diagrams that contribute to the chosen $N$-point statistics and for all operator combinations, using any method of choice (e.g. \cite{Bakx:2024zgu,Anastasiou:2025jsy,Anastasiou:2022udy,Simonovic:2017mhp,Schmittfull:2016jsw}). 
    \item For each such loop diagram, perform renormalization of stochastic UV contributions by subtracting the hard limits of mode-coupling loops, as outlined in \refsec{stochastic}. 
    \item Using these fully renormalized operator loops, write the renormalized expressions for $N$-point functions with stochastic EFT terms as derived from the respective operator product expansion; for explicit forms for power spectrum, bispectrum and trispectrum, see Eqs.~\eqref{eq:pren}, \eqref{eq:bren} and \eqref{eq:tren}\footnote{We emphasize once more that these are valid at any loop order, and the renormalized expressions involve \textit{renormalized} lower-order operator products. Hence, step (5) must be performed before this step.}. 
\end{enumerate}
The above procedure yields expressions for $N$-point functions in terms of \textit{renormalized} loop integrals and \textit{renormalized} bias parameters, the latter of which are guaranteed to be identical across $N$-point functions. We stress that this 
would not be the case when
using bare bias parameters and bare loop integrals instead.

There are several immediate directions to pursue in future work. First, on the theoretical side, it would be interesting to identify a more principled `bottom-up' construction of all required EFT operators at NLG order from symmetry arguments alone. In particular, an understanding of the origin of the `backreaction' operators from Eqs. \eqref{eq:PiB12} and \eqref{eq:hatpin} which does not rely on the explicit use of the equations of motion for dark matter in the EFT would provide physical insight into the nature of the EFT approach.

Second, it is natural to extend the framework outlined here also to biased tracers in \textit{redshift space} (see e.g.~\cite{Kaiser:1987qv,Scoccimarro:1999ed,Scoccimarro:2004tg,Taruya:2010mx,Taruya:2013my,Vlah:2013lia,Lewandowski:2015ziq,Perko:2016puo,Chen:2020fxs,Taruya:2021ftd,DAmico:2022ukl,Taule:2023izt,Eggemeier:2025xwi}), which requires generalizing the operator product expansion also to products involving the galaxy density and galaxy velocity field. Moreover, in line with conventional power counting, NLG corrections also need to be included in the stochastic sector (see e.g.~Sec.~2.9 of \cite{Desjacques:2016bnm} and \cite{Eggemeier:2018qae}). This will allow the fifth-order galaxy bias expansion to be compared directly to observational data from currently ongoing and upcoming surveys.  

\section*{Acknowledgements}
The authors acknowledge Fabian Schmidt and Rom\'an Scoccimarro for useful conversations. This publication is part of the project `A rising tide: Galaxy intrinsic alignments as a new probe of cosmology and galaxy evolution' (with project number VI.Vidi.203.011) of the Talent programme Vidi which is (partly) financed by the Dutch Research Council (NWO). For the purpose of open access, a CC BY public copyright license is applied to any Author Accepted Manuscript version arising from this submission. We further acknowledge support by the Excellence Cluster ORIGINS, which is funded by the Deutsche Forschungsgemeinschaft (DFG, German Research
Foundation) under Germany's Excellence Strategy - EXC-2094 - 390783311. Z.V. acknowledges the support of the Croatian Science Foundation (grant number IP-2025-02-1338).

\begin{appendix}

\section{Additional details on the NLG bias basis}\label{app:nlg_basis}

In this appendix, we give more examples of redundancies between operators at the order in perturbation theory we work in (App. \ref{subsec:redundancy}) as well as the origin of the counterterms from Eqs. \eqref{eq:PiB12} and \eqref{eq:hatpin}, which can be seen to emerge when explicitly solving the equations of motion for dark matter in the EFT (App. \ref{subsec:extra_nlg_ops}). 

\subsection{Redundant operators}\label{subsec:redundancy}

Here we provide more details on redundancies for the NLG bias basis. In general, all additional second-gradient operators ${\cal O}'$ that can be expressed as linear combinations of the basis set of ${\cal O}_{a_2}$ up to third order, i.e. ${\cal O}'=\sum c_{a_2}{\cal O}_{a_2}+\,$4th order terms, can themselves effectively be viewed as (at least) fourth order operators. Indeed, one can define a linear combination ${\cal O}''\equiv {\cal O}'-\sum c_{a_2}{\cal O}_{a_2}$. Then ${\cal O}''$ starts only at 4th order in perturbations. We took this into account when constructing the minimal NLG basis.
Explicit examples are
\bea
\nabla_k\nabla_l\Pi^{[2]}_{kl} &=& \nabla_k\nabla_l[\Pi^{[1]}_{kl}\mbox{tr}(\Pi^{[1]})] -(3 \nabla^2\mbox{tr}[(\Pi^{[1]})^2])/14 +(3 \nabla^2[\mbox{tr}(\Pi^{[1]})]^2)/14 \nn\\
 && +\nabla_k\nabla_l[\Pi^{[1]}_{kl}\mbox{tr}[(\Pi^{[1]})^2]]/5-\nabla_k\nabla_l[\Pi^{[1]}_{kl}[\mbox{tr}(\Pi^{[1]})]^2]/5\nn\\
 && +(2\nabla_k\nabla_l[(\Pi^{[1]}\Pi^{[1]})_{kl}\mbox{tr}(\Pi^{[1]})])/5 -(2 \nabla_k\nabla_l[(\Pi^{[1]}\Pi^{[2]})_{kl}])/5 \nn \\
 &&-(4 \nabla^2\mbox{tr}[(\Pi^{[1]})^3])/45+(43 \nabla^2[\mbox{tr}[(\Pi^{[1]})^2]\mbox{tr}(\Pi^{[1]})])/735\nn\\
 && +(2 \nabla^2\mbox{tr}[\Pi^{[1]}\Pi^{[2]}])/21-(143 \nabla^2[\mbox{tr}(\Pi^{[1]})]^3)/2205 + \text{4th order terms}\,,\nn\\
 \nabla_k\nabla_l\Pi^{[3]}_{kl} &=& -((3 \nabla_k\nabla_l[\Pi^{[1]}_{kl}\mbox{tr}[(\Pi^{[1]})^2]])/20)+(3 \nabla_k\nabla_l[\Pi^{[1]}_{kl}[\mbox{tr}(\Pi^{[1]})]^2])/20\nn\\
 && -(3 \nabla_k\nabla_l[(\Pi^{[1]}\Pi^{[1]})_{kl}\mbox{tr}(\Pi^{[1]})])/10+(13  \nabla_k\nabla_l[(\Pi^{[1]}\Pi^{[2]})_{kl}])/10\nn\\
 && +(13 \nabla^2\mbox{tr}[(\Pi^{[1]})^3])/45+(29 \nabla^2[\mbox{tr}[(\Pi^{[1]})^2]\mbox{tr}(\Pi^{[1]})])/60\nn\\
 && -(7 \nabla^2\mbox{tr}[\Pi^{[1]}\Pi^{[2]}])/6+(71 \nabla^2[\mbox{tr}(\Pi^{[1]})]^3)/180 + \text{4th order terms}\,,\nn\\
 \nabla_k\nabla_l[\Pi^{[2]}_{kl}\mbox{tr}(\Pi^{[1]})] &=& -((2 \nabla_k\nabla_l[\Pi^{[1]}_{kl}\mbox{tr}[(\Pi^{[1]})^2]])/7)+(2 \nabla_k\nabla_l[\Pi^{[1]}_{kl}[\mbox{tr}(\Pi^{[1]})]^2])/7\nn\\
 && -\nabla_k\nabla_l[(\Pi^{[1]}\Pi^{[1]})_{kl}\mbox{tr}(\Pi^{[1]})]+2 \nabla_k\nabla_l[(\Pi^{[1]}\Pi^{[2]})_{kl}]/3\nn\\
 && +(2 \nabla^2[\mbox{tr}[(\Pi^{[1]})^2]\mbox{tr}(\Pi^{[1]})])/7-\nabla^2\mbox{tr}[\Pi^{[1]}\Pi^{[2]}]\nn\\
 && +(8 \nabla^2[\mbox{tr}(\Pi^{[1]})]^3)/21 + \text{4th order terms}\,.
\eea

\subsection{Extra operators}\label{subsec:extra_nlg_ops}

We observe in the main text that renormalization of the one-loop trispectrum and two-loop power spectrum requires considering `extra' operators at NLG order, see Eq.~\eqref{eq:extrabuildingblock} and the discussion below.
For the matter density field, these `extra' operators can be viewed as arising from backreaction of EFT corrections on the tidal tensor.
Here we provide some more details on this statement.

Consider the equation of motion for the density and momentum $P_i=(1+\delta)v_i$ derived from the underlying collisionless Boltzmann (i.e. Vlasov) equation for non-relativistic matter written in the form (see e.g. Eq.~70 in~\cite{Garny:2022kbk})
\be\label{eq:EoMmomentum}
\partial_\tau\delta+\nabla_iP_i=0,\quad \partial_\tau P_i+{\cal H}P_i+\nabla_jT_{ij}=0\,,
\ee
with\footnote{Note that $T_{ij}$ in Eq.~(70) in~\cite{Garny:2022kbk} is defined slightly differently,
\be
  T_{ij}\Big|_{\text{there}}= (1+\delta)(v_iv_j+\sigma_{ij}) + \Phi\delta^K_{ij} +\frac{1}{3{\cal H}^2\Omega_m}\left(\nabla_i\Phi\nabla_j\Phi-\Phi\nabla_i\nabla_j\Phi+\Phi\delta_{ij}^K\nabla^2\Phi\right)\,.
\ee
Both expressions give identical contributions in Eq. ~\eqref{eq:EoMmomentum}, i.e.~the $\Phi$-dependent part satisfies $\nabla_jT_{ij}^{\Phi}=(1+\delta)\nabla_i\Phi$.
The equivalence can be seen by using
\be
\nabla_j\Big[\nabla_i\Phi\nabla_j\Phi-\delta_{ij}^K(\nabla\Phi)^2+\Phi\nabla_i\nabla_j\Phi-\Phi\delta_{ij}^K\nabla^2\Phi\Big]=0
\ee
}
\be \label{eq:Tij}
  T_{ij}= (1+\delta)(v_iv_j+\sigma_{ij}) + \Phi\delta^K_{ij} +\frac{1}{3{\cal H}^2\Omega_m}\left(2\nabla_i\Phi\nabla_j\Phi-\delta_{ij}^K(\nabla\Phi)^2\right)\,,
\ee
where $\sigma_{ij}$ is the velocity dispersion tensor, and $\nabla^2\Phi=\frac32{\cal H}^2\Omega_m\delta$. This gives
\be\label{eq:EOMdelta}
  \partial_\tau^2\delta+{\cal H}\partial_\tau\delta+\nabla_i\nabla_jT_{ij}=0\,.
\ee
Corrections to $\delta$ beyond SPT enter via the term $S_\text{EFT} \equiv \nabla_i\nabla_j[(1+\delta)\sigma_{ij}]$. In the EFT approach, it can be expanded in terms of the second-gradient operators
with two overall gradients, 
\be\label{eq:SEFT}
S_\text{EFT} = \lambda^2\sum c_{a_2}{\cal O}_{a_2} +  {\cal O}(\lambda^4)\,,
\ee
with ${\cal O}_{a_2}$ from Eq.~\eqref{eq:NLGoverallgradients}, and some time-dependent EFT coefficients $c_{a_2}$ within the local-in-time, perturbative formulation.
We indicate the derivative expansion by a formal power counting parameter $\lambda$, that counts the number of gradients in the derivative expansion.\footnote{The actual small parameter in this expansion is of order $(kR_L)^2$ or $(k/k_{\sigma})^2$ for the higher-derivative bias for galaxies and the EFT corrections from
effective stress for matter, respectively, with dispersion scale $k_\sigma$ (see~\cite{Garny:2022kbk}) of order the non-linear scale $k_{\text{NL}}$, or Lagrangian halo radius $R_L$, respectively. The ``$k^2$'' arises from the
derivatives contained in ${\cal O}_{a_2}$, and the ``$1/k_\sigma^2$'' or ``$R_L^2$'' is encoded in the length-square-dimension of the coefficients $c_{a_2}$ in practice. Thus we use $\lambda$ as
a formal parameter to keep track of how many powers of $k/k_{\sigma}$ or $kR_L$ a given quantity contains, but set it to unity at the end.
} 
We are interested in corrections up to $\lambda^2$, i.e.~neglect even higher (fourth-gradient-order) terms in the derivative expansion. 
Note that  the fields $\delta$ and $v_i$ correspond to the ``coarse-grained'', long-wavelength fields, with short-wavelength UV degrees of freedom already integrated out.
The UV physics related to the latter is entirely encapsulated in the set of Wilson coefficients $c_{a_2}$ (as well as stochastic terms, which are not of interest for this discussion) within the EFT approach. 
In summary we write
\be\label{eq:S}
  S \equiv \nabla_i\nabla_jT_{ij} = \nabla_i\nabla_j[(1+\delta)v_iv_j + T_{ij}^{\Phi}] + S_\text{EFT}\,,
\ee
where $T_{ij}^{\Phi}$ is a shorthand notation for the $\Phi$-dependent terms in Eq.~\eqref{eq:Tij}, which satisfies $\nabla_jT_{ij}^{\Phi}=(1+\delta)\nabla_i\Phi$.

Using Eqs.~\eqref{eq:SEFT} and~\eqref{eq:S} implies an expansion of the resulting solution of the equation of motion Eq.~\eqref{eq:EOMdelta} in powers of $\lambda$.
We thus expand the density field as
\be  \delta=\delta_\text{SPT}+\lambda^2\delta_\text{EFT}+{\cal O}(\lambda^4)\,,
\ee
where the leading solution is the SPT prediction for the density
contrast (as described by the standard $F_n$ kernels order by order in perturbation theory), and $\delta_\text{EFT}$ denotes specifically the second-gradient EFT correction. 
A similar expansion holds for the velocity field $v_i$, its divergence $\theta$, the potential $\Phi$ and the momentum $P_i$. To obtain the $\lambda^2$ part it is sufficient to insert the $\lambda^0$ (i.e.~SPT) solution into the
operators ${\cal O}_{a_2}$, i.e. evaluate them using the standard SPT kernels $F_n$ and $G_n$, ${\cal O}_{a_2}={\cal O}_{a_2}|_{\delta_\text{SPT},\theta_\text{SPT}}$.
We further define $S_\text{SPT}=\nabla_i\nabla_j[(1+\delta)v_iv_j + T_{ij}^{\Phi}]_{\delta_\text{SPT},\theta_\text{SPT}}$ such that $\partial_\tau^2\delta_\text{SPT}+{\cal H}\partial_\tau\delta_\text{SPT}+S_\text{SPT}=0$. The second-gradient corrections are thus given by the solution of
\be
  \lambda^2\left(\partial_\tau^2\delta_\text{EFT}+{\cal H}\partial_\tau\delta_\text{EFT}\right)+\Delta S=-\lambda^2\sum c_{a_2}{\cal O}_{a_2}\,,
\ee
where 
\be
\Delta S \equiv \left(\nabla_i\nabla_j[(1+\delta)v_iv_j + T_{ij}^{\Phi}] \right) \big|_{{\cal O}(\lambda^2)}\,,
\ee
denotes the $\lambda^2$-part.
It is convenient to extract the only term that starts already at linear order in $T_{ij}^{\Phi}$, and write 
$\Delta S=\lambda^2(\nabla^2\Phi_\text{EFT}+(\Delta S)_\text{nonlin})$. Cancelling the common overall $\lambda^2$, we arrive at
\be
  \left(\partial_\tau^2+{\cal H}\partial_\tau+\frac32{\cal H}^2\Omega_m\right)\delta_\text{EFT} = -\sum c_{a_2}{\cal O}_{a_2} - (\Delta S)_\text{nonlin}\,,
\ee
where
\bea
  (\Delta S)_\text{nonlin} &=& \nabla_i\nabla_j\Big[2(1+\delta)v_i^\text{EFT}v_j + \delta_\text{EFT}v_iv_j  \nn\\
  && + \frac{2}{3{\cal H}^2\Omega_m}\left(2\nabla_i\Phi_\text{EFT}\nabla_j\Phi-\delta_{ij}^K\nabla_k\Phi_\text{EFT}\nabla_k\Phi\right)\Big]\,.
\eea
Here all fields without EFT index are evaluated within SPT.
We decompose the solution into  `direct' and `indirect' parts,
\be
  \delta_\text{EFT} = \delta_\text{EFT}^\text{direct} + \delta_\text{EFT}^\text{indirect} \,,
\ee
defined via
\be\label{eq:deltaEFTdirect}
  \left(\partial_\tau^2+{\cal H}\partial_\tau+\frac32{\cal H}^2\Omega_m\right)\delta_\text{EFT}^\text{direct}  = -\sum c_{a_2}{\cal O}_{a_2}\,,
\ee
\be\label{eq:deltaEFTindirect}
  \left(\partial_\tau^2+{\cal H}\partial_\tau+\frac32{\cal H}^2\Omega_m\right)\delta_\text{EFT}^\text{indirect}  = - (\Delta S)_\text{nonlin}\,.
\ee
The first contribution is directly sourced by the EFT corrections to the stress tensor, while the second one can be viewed as backreaction terms.

The direct part can be written equivalently as
\be
  \delta_\text{EFT}^\text{direct}  = \sum b_{a_2}{\cal O}_{a_2}\,,
\ee
with biases $b_{a_2}$. This can be shown order by order in perturbation theory, using factorization of time- and scale-dependence of the kernels describing ${\cal O}_{a_2}$. This means the direct contribution can be expanded in terms of `standard' NLG operators with two overall derivatives, as given in Eq.~\eqref{eq:NLGoverall}.

However, the `indirect' contribution generates terms beyond those captured by Eq.~\eqref{eq:NLGoverall}. This occurs starting at third order in perturbation theory. Inspecting Eq.~\eqref{eq:deltaEFTindirect}, the indirect contribution encompasses operators of the form 
\be\label{eq:extraO}
  \nabla_i\nabla_j\left[ \frac{\nabla_i\delta}{\nabla^2}\,\frac{\nabla_j}{\nabla^2}(\nabla_k\nabla_l{\cal O}_{kl})\right], \qquad
  \nabla^2\left[ \frac{\nabla_i\delta}{\nabla^2}\,\frac{\nabla_i}{\nabla^2}(\nabla_k\nabla_l{\cal O}_{kl})\right], \qquad
\ee
where $\nabla_k\nabla_l{\cal O}_{kl}$ stands for a second-overall-gradient operator included in the set of NLG operators ${\cal O}_{a_2}$.
There are further terms  in the `indirect' contribution involving the velocity field, but these turn out to be redundant with those from above at the order we are interested in (up to third order in perturbations).

We find that the EFT corrections of the form of Eq.~\eqref{eq:extraO}
can at third order in perturbation theory not be expressed as linear combinations of operators within the `standard' set Eq.~\eqref{eq:NLGoverallgradients}. This shows that the latter does not capture the most general modification of the density field on large scales due to corrections arising from UV physics. This means that the operator basis and the underlying set of building blocks $\Pi^{[n]}_{ij}$ are not sufficient  when considering NLG operators and needs to be extended. The most general allowed set of EFT operators should in principle be derivable from within the EFT framework, without having to consider particular UV completions, at any order in the gradient expansion. Deferring such a derivation to future work, we note that up to third order in perturbation theory all of the `missing' NLG EFT operators required to renormalize both the density field as any biased tracer are contained in the set of operators that results when
allowing the additional non-standard building block Eq.~\eqref{eq:extrabuildingblock} to contribute when constructing NLG operators.
At up to third order in perturbation theory this hypothesis yields the extra NLG operators given in Eq.~\eqref{eq:PiB12} and Eq.~\eqref{eq:hatpin} (with $n=3$) in the main text.
Ultimately, this hypothesis (or a suitable generalization) should be justified by a derivation of the most general structure of NLG operators within the EFT based on the set of assumed symmetries only. However, this goes beyond the scope of this work.

Note that allowing for the extra building blocks Eq.~\eqref{eq:extrabuildingblock} can equivalently be understood as the set of building blocks of the form $D_\eta^m\frac{\nabla_i\nabla_j}{\nabla^2}D_\eta^n\nabla_i\nabla_j\phi$ (Eq.~\eqref{eq:PiB12} contains operators of this form with $m=0$, and Eq.~\eqref{eq:hatpin} with $m=1$), while the set of `standard' building blocks $\Pi^{[n]}_{ij}$ can be understood as operators of the form $D_\eta^n\nabla_i\nabla_j\phi$. Thus, in both cases, `displacement' terms enter only via Lagrangian time derivatives, as expected. Note in particular that both operators from Eq.~\eqref{eq:extraO} can be rewritten as operators of this form (we checked this up to third order, as relevant in this work).

We stress that the non-standard building block Eq.~\eqref{eq:extrabuildingblock} is {\it not} required within any LG operator, but only contributes to the set of NLG operators. That is, the set of building blocks $\Pi^{[n]}_{ij}$ is sufficient to construct the most general operator basis at LG order. This applies, apart from the LG bias operator basis, also to the LG EFT expansion of the effective stress tensor, which is equivalent to a general LG tensor bias operator. Thus, for the matter density field, an alternative for deriving a set of `counterterms' is to solve Eqs.~\eqref{eq:deltaEFTdirect} and~\eqref{eq:deltaEFTindirect} perturbatively given a set of $c_{a_2}$ coefficients. This route is effectively taken in~\cite{Anastasiou:2025jsy} [see in particular Eq.~(2.61) in there].
The EFT expansion of the matter density is then inherited from the LG EFT expansion of the effective stress tensor, with the latter being expressible as a general linear combination of LG tensor bias operators (we checked that the operators included in the EFT expansion of the stress tensor up to third order in~\cite{Anastasiou:2025jsy} are equivalent to the tensor bias basis from~\cite{Vlah:2019byq}; see also footnote~\ref{eq:footnoteCompAnastasiou:2025jsy}). This construction relies on solving the equation of motion for the density field within the EFT (i.e.~the continuity equation together with the Euler equation containing the effective stress tensor), and requires assuming a specific time-dependence of the $c_{a_2}$ coefficients. A generalization of this approach to a general biased tracer is therefore not obvious. We therefore follow a different approach in this work, as discussed above and in Sec.~\ref{sec:basis}.

\section{Relation of commutation of limits to sequential loops} \label{app:consistency}

In \hyperlink{cite.Bakx:2025cvu}{Paper~I}, we showed that the leading-gradient double-hard functions $s_{ab}^\text{2L}(r)$ with $r=p/q$ can be expressed in terms of single-hard coefficients $s_{ab}^\text{1L}$ in the limit where $r \to 0$ (or equivalently, $r \to \infty$ by inversion symmetry). This is because the double-hard limit for $p \ll q$ coincides with taking two single-hard limits `sequentially'. This commutation-of-limits relation also generalizes to subleading gradient operators, as explained in Section \ref{subsec:comm_lim}.

More precisely, these relations come from the two ways of taking the limit $q\gg p\gg k_i$ via {\it (i)} first expanding in the double-hard limit $p,q\gg k_i$ with fixed ratio $r=p/q$, and then for $r\ll 1$, or {\it (ii)} taking sequential single-hard limits, i.e. first $q\gg p,k_i$ and then $p\gg k_i$. Given that we consider up to second-gradient operators, we expect that we can match those limits up to second order in the expansion parameters $r=p/q$, $k_i/p$ and $k_i/q$, respectively. Isotropy implies that no linear powers of these parameters may occur. Starting with case {\it (i)}, we have using Eq.~\eqref{eq:doublehard2}, Eq.~\eqref{eq:s2LNLG} and Eq.~\eqref{eq:sabTaylor}
\bea\label{eq:casei}
    \lefteqn{ K_a^{(n+4)}({\bm k}_1,\dots,{\bm k}_{n},{\bm p},-{\bm p},{\bm q},-{\bm q})_{\text{av}_{\hat p,\hat q}}^{p,q\gg k_i}\Big|_{r\ll 1}  }\nn\\
  &=& \frac{4n!}{(n+4)!}\Bigg[ \left( s_{ab}^{\text{2L}}(0)+\frac12 r^2\times (s_{ab}^\text{2L})''(0) +{\cal O}(r^4)\right) K_b^{(n)}({\bm k}_1,\dots,{\bm k}_{n})\nn\\
  && {}   + \left(\frac{1}{p^2}+\frac{1}{q^2}\right)\,  \left( s_{ab_2}^{\text{2L}}(0) +\frac12 r^2\times (s_{ab_2}^\text{2L})''(0) + {\cal O}(r^4)\right) K_{b_2}^{(n)}({\bm k}_1,\dots,{\bm k}_{n})
  +{\cal O}(k_i/p)^{4} \Bigg]\,.\nn\\
\eea
In our power counting $K_b^{(n)}({\bm k}_1,\dots,{\bm k}_{n})\sim{\cal O}(1)$ while $K_{b_2}^{(n)}({\bm k}_1,\dots,{\bm k}_{n})\sim{\cal O}(k_i^2)$, such that in the first line we have, apart from the leading term $s_{ab}^\text{2L}(0)$, an $r^2=p^2/q^2$ correction involving $(s_{ab}^\text{2L})''(0)$, while in the second line we have 
\be
  \frac{s_{ab_2}^{\text{2L}}(0)}{p^2}\times K_{b_2}^{(n)}\propto {\cal O}\left(\frac{k_i^2}{p^2}\right), \qquad
  \left(\frac{s_{ab_2}^{\text{2L}}(0)}{q^2} + \frac{\frac12 r^2\times (s_{ab_2}^\text{2L})''(0)}{p^2}\right)\times K_{b_2}^{(n)}\propto {\cal O}\left(\frac{k_i^2}{q^2}\right)\,.
\ee
This can be compared to case {\it (ii)}. Using Eq.~\eqref{eq:singlehardNLG}, Eq.~\eqref{eq:s1Lab2} and Eq.~\eqref{eq:s1La2b2} we have
\bea\label{eq:caseii}
    \lefteqn{ K_a^{(n+4)}({\bm k}_1,\dots,{\bm k}_{n},{\bm p},-{\bm p},{\bm q},-{\bm q})_{\text{av}_{\hat p,\hat q}}^{q\gg p,k_i}\Big|_{p\gg k_i}  }\nn\\
  &=& \frac{2(n+2)!}{(n+4)!}\frac{2n!}{(n+2)!}\Bigg[ s_{ac}^{\text{1L}} \left( s_{cb}^{\text{1L}}K_b^{(n)}({\bm k}_1,\dots,{\bm k}_{n}) + \frac{1}{p^2}s_{cb_2}^{\text{1L}}K_{b_2}^{(n)}({\bm k}_1,\dots,{\bm k}_{n}) +{\cal O}(k_i/p)^{4}\right) \nn\\
  && {}   +  \frac{1}{q^2}s_{ac_2}^{\text{1L}}\left( p^2 s_{c_2b}^{\text{1L}}K_b^{(n)}({\bm k}_1,\dots,{\bm k}_{n}) + s_{c_2b_2}^{\text{1L}}K_{b_2}^{(n)}({\bm k}_1,\dots,{\bm k}_{n}) +{\cal O}(k_i^4/p^{2}) \right) +{\cal O}(p/q)^4\Bigg]\,.\nn\\
\eea
Requiring that the leading terms match yields the first relation from Eq.~\eqref{eq:s2Ldoublesingle}, known already from~\cite{Bakx:2025cvu}.
The only contribution scaling as $k_i^2/p^2$ is the second term in the first line, and matching it with the corresponding $k_i^2/p^2$ contribution from Eq.~\eqref{eq:casei} yields
the second relation in Eq.~\eqref{eq:s2Ldoublesingle}. The relations from Eq.~\eqref{eq:s2Ldoublesingle2} ensure that the $p^2/q^2$ and $k_i^2/q^2$ contributions in Eq.~\eqref{eq:caseii}
match those from Eq.~\eqref{eq:casei}. Thus, as stated above, we find that the single/double-hard relations Eq.~\eqref{eq:s2Ldoublesingle} and Eq.~\eqref{eq:s2Ldoublesingle2}
follow from the interchangeability of limits $K_a^{(n+4)}|_{p,q\gg k_i}|_{p\ll q}$ versus $K_a^{(n+4)}|_{q\gg p,k_i}|_{p\gg k_i}$ at zeroth and second-gradient level, respectively.

\section{Tables}\label{app:tables}
To streamline notation for the purposes of implementation, we here introduce an operator ordering for $29$ LG and $20$ NLG operators that occur in the EFT up to the order we work here. Specifically, we use
\begin{equation}
\begin{aligned}
{\cal O}_{a}=\{&
1:{\rm tr}\big[\Pi^{[1]}\big],\quad
2:{\rm tr}\big[\big(\Pi^{[1]}\big)^2\big],\quad
3:\big({\rm tr}\big[\Pi^{[1]}\big]\big)^2,\\
&4:\big({\rm tr}\big[\Pi^{[1]}\big]\big)^3,\quad
5:{\rm tr}\big[\big(\Pi^{[1]}\big)^2\big]{\rm tr}\big[\Pi^{[1]}\big],\quad
6:{\rm tr}\big[\big(\Pi^{[1]}\big)^3\big],\\
&7:{\rm tr}\big[\Pi^{[1]}\Pi^{[2]}\big],\quad
8:\big({\rm tr}\big[\Pi^{[1]}\big]\big)^4,\quad
9:{\rm tr}\big[\big(\Pi^{[1]}\big)^3\big]{\rm tr}\big[\Pi^{[1]}\big],\\
&10:{\rm tr}\big[\big(\Pi^{[1]}\big)^2\big]\big({\rm tr}\big[\Pi^{[1]}\big]\big)^2,\quad
11:\big({\rm tr}\big[\big(\Pi^{[1]}\big)^2\big]\big)^2,\quad
12:{\rm tr}\big[\Pi^{[1]}\Pi^{[1]}\Pi^{[2]}\big],\\
&13:{\rm tr}\big[\Pi^{[1]}\big]{\rm tr}\big[\Pi^{[1]}\Pi^{[2]}\big],\quad
14:{\rm tr}\big[\Pi^{[1]}\Pi^{[3]}\big],\quad
15:{\rm tr}\big[\Pi^{[2]}\Pi^{[2]}\big],\\
&16:\big({\rm tr}\big[\Pi^{[1]}\big]\big)^5,\quad
17:{\rm tr}\big[\big(\Pi^{[1]}\big)^3\big]\big({\rm tr}\big[\Pi^{[1]}\big]\big)^2,\quad
18:{\rm tr}\big[\big(\Pi^{[1]}\big)^2\big]\big({\rm tr}\big[\Pi^{[1]}\big]\big)^3,\\
&19:{\rm tr}\big[\big(\Pi^{[1]}\big)^3\big]{\rm tr}\big[\big(\Pi^{[1]}\big)^2\big],\quad
20:{\rm tr}\big[\Pi^{[1]}\big]\big({\rm tr}\big[\big(\Pi^{[1]}\big)^2\big]\big)^2,\quad
21:\big({\rm tr}\big[\Pi^{[1]}\big]\big)^2{\rm tr}\big[\Pi^{[1]}\Pi^{[2]}\big],\\
&22:{\rm tr}\big[\Pi^{[1]}\Pi^{[1]}\big]{\rm tr}\big[\Pi^{[1]}\Pi^{[2]}\big],\quad
23:{\rm tr}\big[\Pi^{[1]}\big]{\rm tr}\big[\Pi^{[1]}\Pi^{[1]}\Pi^{[2]}\big],\quad
24:{\rm tr}\big[\Pi^{[1]}\Pi^{[2]}\Pi^{[2]}\big],\\
&25:{\rm tr}\big[\Pi^{[1]}\big]{\rm tr}\big[\Pi^{[2]}\Pi^{[2]}\big],\quad
26:{\rm tr}\big[\Pi^{[1]}\Pi^{[1]}\Pi^{[3]}\big],\quad
27:{\rm tr}\big[\Pi^{[1]}\big]{\rm tr}\big[\Pi^{[1]}\Pi^{[3]}\big],\\
&28:{\rm tr}\big[\Pi^{[2]}\Pi^{[3]}\big],\quad
29:{\rm tr}\big[\Pi^{[1]}\Pi^{[4]}\big]\}\, .
\end{aligned}
\end{equation}
for LG operators (where the ordering matches that of \hyperlink{cite.Bakx:2025cvu}{Paper~I}), and 
\begin{equation}
\begin{aligned}
{\cal O}_{a_2}=\{&
1:\nabla^2{\rm tr}(\Pi^{[1]}),\quad
2:\nabla^2{\rm tr}[(\Pi^{[1]})^2],\quad
3:\nabla^2[{\rm tr}(\Pi^{[1]})]^2,\\
&4:\nabla_k\nabla_l[\Pi^{[1]}_{kl}{\rm tr}(\Pi^{[1]})],\quad
5:\nabla^2{\rm tr}[\Pi^{[1]}\Pi^{[2]}],\quad
6:\nabla_k\nabla_l[(\Pi^{[1]}\Pi^{[2]})_{kl}],\\
&7:\nabla^2[{\rm tr}(\Pi^{[1]})]^3,\quad
8:\nabla^2[{\rm tr}[(\Pi^{[1]})^2]{\rm tr}(\Pi^{[1]})],\quad
9:\nabla^2{\rm tr}[(\Pi^{[1]})^3],\\
&10:\nabla_k\nabla_l[\Pi^{[1]}_{kl}[{\rm tr}(\Pi^{[1]})]^2],\quad
11:\nabla_k\nabla_l[\Pi^{[1]}_{kl}{\rm tr}[(\Pi^{[1]})^2]],\\
&12:\nabla_k\nabla_l[(\Pi^{[1]}\Pi^{[1]})_{kl}{\rm tr}(\Pi^{[1]})],\quad
13:{\rm tr}(\Pi^{[1]})\nabla^2{\rm tr}(\Pi^{[1]}),\\
&14:{\rm tr}[\Pi^{[2]}\nabla^2\Pi^{[1]}],\quad
15:[{\rm tr}(\Pi^{[1]})]^2\nabla^2{\rm tr}(\Pi^{[1]}),\quad
16:{\rm tr}(\Pi^{[1]}){\rm tr}[\Pi^{[1]}\nabla^2\Pi^{[1]}],\\
&17:[\nabla_k\Pi^{[1]}_{kl}]\times {\rm tr}[\Pi^{[1]}\nabla_l\Pi^{[1]}],\quad
18:\hat{\Pi}_3,\quad
19:\Pi_{B1},\quad
20:\Pi_{B2}\}\, .
\end{aligned}
\end{equation}
for NLG operators. Here, the first 12 are `standard' two-overall-derivative operators, 13 through 17 have no overall derivative, and 18 through 20 are the new operators discussed in \refsec{basis}. Note we use the same numbering in the rows and columns of the tables below, regardless of whether a number refers to the LG or NLG list. However, it is always clear from the indices on $s_{AB}$ which list is meant. The first index goes into the rows, while the second goes into the columns. 

\begin{table}[t]
\centering
\caption{Coefficients for $s_{ab}^{\text{1L}}$.}
\label{tab:sab1L}
\scriptsize
\setlength{\tabcolsep}{4pt}
\renewcommand{\arraystretch}{1.15}
\begin{tabular}{l|ccccccc}
\hline
$s_{ab}^{\text{1L}}$ 
& $1$ 
& $2$ 
& $3$ 
& $4$ 
& $5$ 
& $6$ 
& $7$ \\ 
\\ \hline \\

$1$ & 0 & 0 & 0 & 0 & 0 & 0 & 0 \\
\\ \hline \\
$2$ 
& $\frac{68}{21}$ & $\frac{254}{2205}$ & $\frac{2624}{735}$ 
& $\frac{141664}{101871}$ & $\frac{5856}{18865}$ 
& $\frac{457616}{2546775}$ & $-\frac{2644}{13475}$ \\

$3$ 
& $\frac{68}{21}$ & $\frac{254}{2205}$ & $\frac{2624}{735}$ 
& $\frac{141664}{101871}$ & $\frac{5856}{18865}$ 
& $\frac{457616}{2546775}$ & $-\frac{2644}{13475}$ \\
\\ \hline \\
$4$ 
& 3 & 0 & $\frac{68}{7}$ & $\frac{2624}{245}$ 
& $\frac{254}{735}$ & 0 & 0 \\

$5$ 
& $\frac{5}{3}$ & $\frac{116}{105}$ & $\frac{176}{35}$ 
& $\frac{1754}{343}$ & $\frac{46258}{15435}$ 
& $\frac{1784}{1715}$ & $-\frac{50}{63}$ \\

$6$ 
& 1 & $\frac{58}{35}$ & $\frac{94}{35}$ 
& $\frac{3971}{1715}$ & $\frac{4448}{1029}$ 
& $\frac{2676}{1715}$ & $-\frac{25}{21}$ \\

$7$ 
& $\frac{41}{21}$ & $\frac{76}{63}$ & $\frac{268}{49}$ 
& $\frac{2489609}{509355}$ & $\frac{61232}{18865}$ 
& $\frac{3262052}{2546775}$ & $-\frac{19231}{17325}$ \\
\\ \hline \\
$8$ 
& 0 & 0 & 6 & $\frac{136}{7}$ & 0 & 0 & 0 \\

$9$ 
& 0 & 1 & 1 & $\frac{94}{35}$ & $\frac{152}{35}$ & $\frac{58}{35}$ & 0 \\

$10$ 
& 0 & 1 & $\frac{7}{3}$ & $\frac{716}{105}$ & $\frac{572}{105}$ & 0 & 0 \\

$11$ 
& 0 & $\frac{38}{15}$ & $\frac{4}{15}$ & $\frac{32}{49}$ 
& $\frac{6184}{735}$ & $\frac{928}{735}$ & 0 \\

$12$ 
& 0 & $\frac{103}{105}$ & $\frac{17}{15}$ 
& $\frac{128668}{46305}$ & $\frac{21724}{5145}$ 
& $\frac{80018}{46305}$ & $\frac{22}{105}$ \\

$13$ 
& 0 & $\frac{4}{7}$ & $\frac{18}{7}$ 
& $\frac{324706}{46305}$ & $\frac{10832}{3087}$ 
& $\frac{207472}{231525}$ & $\frac{394}{1575}$ \\

$14$ 
& 0 & $\frac{268}{315}$ & $\frac{34}{15}$ 
& $\frac{940493}{169785}$ & $\frac{152275}{33957}$ 
& $\frac{1211306}{848925}$ & $-\frac{6614}{17325}$ \\

$15$ 
& 0 & $\frac{466}{735}$ & $\frac{2168}{735}$ 
& $\frac{348056}{46305}$ & $\frac{10120}{3087}$ 
& $\frac{9644}{9261}$ & $\frac{1516}{2205}$ \\
\\ \hline \\
$16$ 
& 0 & 0 & 0 & 10 & 0 & 0 & 0 \\

$17$ 
& 0 & 0 & 0 & 1 & 2 & 1 & 0 \\

$18$ 
& 0 & 0 & 0 & 3 & 3 & 0 & 0 \\

$19$ 
& 0 & 0 & 0 & 0 & $\frac{7}{5}$ & $\frac{9}{5}$ & 0 \\

$20$ 
& 0 & 0 & 0 & $\frac{4}{15}$ & $\frac{58}{15}$ & 0 & 0 \\

$21$ 
& 0 & 0 & 0 & $\frac{67}{21}$ & $\frac{8}{7}$ & 0 & 1 \\

$22$ 
& 0 & 0 & 0 & $\frac{38}{147}$ & $\frac{1819}{735}$ 
& $\frac{232}{735}$ & $\frac{19}{15}$ \\

$23$ 
& 0 & 0 & 0 & $\frac{17}{15}$ & $\frac{143}{105}$ 
& $\frac{10}{21}$ & $\frac{2}{3}$ \\

$24$ 
& 0 & 0 & 0 & $\frac{422}{343}$ & $\frac{1037}{1029}$ 
& $\frac{1493}{5145}$ & $\frac{128}{105}$ \\

$25$ 
& 0 & 0 & 0 & $\frac{856}{245}$ & $\frac{626}{735}$ 
& 0 & $\frac{20}{21}$ \\

$26$ 
& 0 & 0 & 0 & $\frac{6407}{6615}$ & $\frac{1189}{735}$ 
& $\frac{5371}{6615}$ & $\frac{1}{5}$ \\

$27$ 
& 0 & 0 & 0 & $\frac{2579}{945}$ & $\frac{100}{63}$ 
& $\frac{272}{675}$ & $\frac{41}{225}$ \\

$28$ 
& 0 & 0 & 0 & $\frac{142094}{46305}$ & $\frac{19994}{15435}$ 
& $\frac{2411}{9261}$ & $\frac{229}{315}$ \\

$29$ 
& 0 & 0 & 0 & $\frac{523888}{509355}$ & $\frac{98489}{113190}$ 
& $\frac{1332383}{5093550}$ & $\frac{746}{5775}$ \\

\hline
\end{tabular}%
\end{table}

\begin{table}[t]
\centering
\caption{Coefficients for $s_{a_2 b}^{\text{1L}}$.}
\label{tab:sa2b1L}
\scriptsize
\setlength{\tabcolsep}{4pt}
\renewcommand{\arraystretch}{1.15}
\begin{tabular}{l|ccccccc}
\hline
$s_{a_2 b}^{\text{1L}}$
& $1$
& $2$
& $3$
& $4$
& $5$
& $6$
& $7$ \\
\\ \hline \\

$1$ & 0 & 0 & 0 & 0 & 0 & 0 & 0 \\
\\ \hline \\
$2$ & 0 & 0 & 0 & 0 & 0 & 0 & 0 \\
$3$ & 0 & 0 & 0 & 0 & 0 & 0 & 0 \\
$4$ & 0 & 0 & 0 & 0 & 0 & 0 & 0 \\
\\ \hline \\
$5$ & 0 & 0 & 0 & 0 & 0 & 0 & 0 \\
$6$ & 0 & 0 & 0 & 0 & 0 & 0 & 0 \\
$7$ & 0 & 0 & 0 & 0 & 0 & 0 & 0 \\
$8$ & 0 & 0 & 0 & 0 & 0 & 0 & 0 \\
$9$ & 0 & 0 & 0 & 0 & 0 & 0 & 0 \\
$10$ & 0 & 0 & 0 & 0 & 0 & 0 & 0 \\
$11$ & 0 & 0 & 0 & 0 & 0 & 0 & 0 \\
$12$ & 0 & 0 & 0 & 0 & 0 & 0 & 0 \\
\\ \hline \\

$13$
& $\frac{82}{21}$
& $\frac{2396}{2205}$
& $\frac{1297}{245}$
& $\frac{1387394}{509355}$
& $\frac{553996}{169785}$
& $\frac{3903344}{2546775}$
& $-\frac{191194}{121275}$ \\

$14$
& $\frac{34}{21}$
& $\frac{46}{63}$
& $\frac{282}{49}$
& $\frac{3454958}{509355}$
& $\frac{233284}{56595}$
& $\frac{2588456}{2546775}$
& $-\frac{14908}{17325}$ \\

$15$
& 2
& 0
& $\frac{164}{21}$
& $\frac{2594}{245}$
& $\frac{4792}{2205}$
& 0
& 0 \\

$16$
& $\frac{4}{3}$
& $\frac{86}{105}$
& $\frac{74}{15}$
& $\frac{32741}{5145}$
& $\frac{18478}{5145}$
& $\frac{31972}{25725}$
& $-\frac{919}{1575}$ \\

$17$
& $-\frac{1}{3}$
& $-\frac{86}{105}$
& $-\frac{36}{35}$
& $-\frac{5504}{5145}$
& $-\frac{38662}{15435}$
& $-\frac{31972}{25725}$
& $\frac{919}{1575}$ \\
\\ \hline \\

$18$ & 0 & 0 & 0 & 0 & 0 & 0 & 0 \\
$19$ & 0 & 0 & 0 & 0 & 0 & 0 & 0 \\
$20$ & 0 & 0 & 0 & 0 & 0 & 0 & 0 \\

\hline
\end{tabular}%
\end{table}

\begin{table}[t]
\caption{Coefficients for $s_{a b_2}^{\text{1L}}$. Continues in \reftab{sab21L_B}.}
\label{tab:sab21L_A}
\centering
\scriptsize
\hspace*{-2cm}
\resizebox{1.3\textwidth}{!}{%
\begin{tabular}{l|cccccccccccc}
\hline
$s_{ab_2}^{\text{1L}}$
& $1$
& $2$
& $3$
& $4$
& $5$
& $6$
& $7$
& $8$
& $9$
& $10$
& $11$
& $12$ \\
\\ \hline \\
$1$ 
& $-\frac{61}{630}$ & $-\frac{67}{7546}$ & $\frac{6103}{339570}$ & $-\frac{1336}{18865}$ 
& $-\frac{55679}{926100}$ & $\frac{6585457}{44144100}$
& $\frac{173443573}{5562156600}$ & $\frac{1677259}{142619400}$ & $\frac{422783}{19864845}$ 
& $-\frac{930469}{15846600}$ & $\frac{2703971}{618017400}$ & $-\frac{412953}{4904900}$ \\
\\ \hline \\
$2$ 
& $-\frac{32}{21}$ & $\frac{9179}{30870}$ & $-\frac{25148}{15435}$ & $\frac{8584}{15435}$ 
& $\frac{127}{297}$ & $-\frac{854251}{509355}$ & $-\frac{360253}{611226}$
& $\frac{351019}{1018710}$ & $-\frac{27332}{218295}$ & $\frac{5071457}{7130970}$ 
& $\frac{293}{129654}$ & $\frac{480461}{509355}$ \\

$3$ 
& 0 & 0 & $-\frac{37}{294}$ & $\frac{32}{245}$ 
& $\frac{138538}{363825}$ & $-\frac{277076}{363825}$ & $\frac{328351}{7640325}$ & $-\frac{305156}{2546775}$
& $-\frac{961846}{7640325}$ & $\frac{411434}{2546775}$ & $\frac{184196}{2546775}$ & $\frac{576502}{2546775}$ \\
\\ \hline \\
$4$ 
& 0 & 0 & 0 & 0 & 0 & 0 & $\frac{1}{210}$ & 0 & 0 & 0 & 0 & 0 \\

$5$ 
& 0 & 0 & $-\frac{48}{49}$ & $\frac{48}{49}$ 
& $\frac{108772}{19845}$ & $-\frac{217544}{19845}$ & $-\frac{552764}{416745}$ & $-\frac{527677}{277830}$
& $-\frac{754048}{416745}$ & $\frac{610664}{138915}$ & $\frac{181892}{138915}$ & $\frac{629152}{138915}$ \\

$6$ 
& 0 & 0 & $-\frac{16}{49}$ & $\frac{16}{49}$ 
& $\frac{54386}{6615}$ & $-\frac{108772}{6615}$ & $-\frac{46189}{138915}$ & $-\frac{184153}{46305}$
& $-\frac{147809}{55566}$ & $\frac{189709}{46305}$ & $\frac{99397}{46305}$ & $\frac{78928}{9261}$ \\

$7$ 
& $-\frac{16}{21}$ & $\frac{2248}{15435}$ & $-\frac{30776}{15435}$ & $\frac{608}{3087}$ 
& $\frac{12348629}{2182950}$ & $-\frac{88526549}{7640325}$
& $-\frac{689673427}{320893650}$ & $-\frac{226096273}{106964550}$ & $-\frac{40970518}{22920975}$ 
& $\frac{408629587}{106964550}$ & $\frac{13931963}{9724050}$
& $\frac{45277789}{7640325}$ \\
\\ \hline \\
$8$ 
& 0 & 0 & 0 & 0 & 0 & 0 & 0 & 0 & 0 & 0 & 0 & 0 \\

$9$ 
& 0 & 0 & 0 & 0 & $-\frac{8}{7}$ & $\frac{16}{7}$ & $\frac{136}{147}$ & $-\frac{8}{49}$ & $\frac{8}{21}$ 
& $-\frac{16}{7}$ & $-\frac{16}{49}$ & $\frac{16}{49}$ \\

$10$ 
& 0 & 0 & 0 & 0 & 0 & 0 & 0 & 0 & 0 & 0 & 0 & 0 \\

$11$ 
& 0 & 0 & 0 & 0 & $\frac{3952}{189}$ & $-\frac{7904}{189}$ & $-\frac{16784}{3969}$ & $-\frac{12848}{1323}$ & $-\frac{27616}{3969}$
& $\frac{3104}{189}$ & $\frac{7856}{1323}$ & $\frac{25696}{1323}$ \\

$12$ 
& 0 & 0 & $-\frac{16}{147}$ & $\frac{16}{147}$ 
& $\frac{87964}{19845}$ & $-\frac{175928}{19845}$ & $\frac{61342}{416745}$ & $-\frac{82618}{27783}$
& $-\frac{575056}{416745}$ & $\frac{141026}{138915}$ & $\frac{161498}{138915}$ & $\frac{835648}{138915}$ \\

$13$ 
& 0 & 0 & $-\frac{24}{49}$ & $\frac{24}{49}$ 
& $\frac{5636}{2835}$ & $-\frac{11272}{2835}$ & $-\frac{411484}{416745}$ & $-\frac{127636}{138915}$
& $-\frac{269336}{416745}$ & $\frac{201808}{138915}$ & $\frac{12644}{27783}$ & $\frac{60328}{27783}$ \\

$14$ 
& 0 & $-\frac{4}{315}$ & $-\frac{1684}{2205}$ & $-\frac{44}{441}$ 
& $\frac{44967}{26950}$ & $-\frac{744703}{242550}$ & $-\frac{34652221}{23769900}$
& $-\frac{33675437}{23769900}$ & $-\frac{893021}{2546775}$ & $-\frac{5139797}{23769900}$ 
& $\frac{637087}{2160900}$ & $\frac{4759561}{1697850}$ \\

$15$ 
& 0 & $-\frac{160}{1029}$ & $-\frac{704}{1029}$ & $-\frac{32}{343}$ 
& $\frac{214384}{19845}$ & $-\frac{2995616}{138915}$ & $-\frac{9450548}{2917215}$
& $-\frac{4402724}{972405}$ & $-\frac{219952}{59535}$ & $\frac{193156}{27783}$ 
& $\frac{58484}{19845}$ & $\frac{268784}{27783}$ \\
\\ \hline \\
$16$ 
& 0 & 0 & 0 & 0 & 0 & 0 & 0 & 0 & 0 & 0 & 0 & 0 \\

$17$ 
& 0 & 0 & 0 & 0 & 0 & 0 & 0 & 0 & 0 & 0 & 0 & 0 \\

$18$ 
& 0 & 0 & 0 & 0 & 0 & 0 & 0 & 0 & 0 & 0 & 0 & 0 \\

$19$ 
& 0 & 0 & 0 & 0 & 0 & 0 & 0 & 0 & 0 & 0 & 0 & 0 \\

$20$ 
& 0 & 0 & 0 & 0 & 0 & 0 & 0 & 0 & 0 & 0 & 0 & 0 \\

$21$ 
& 0 & 0 & 0 & 0 & 0 & 0 & 0 & 0 & 0 & 0 & 0 & 0 \\

$22$ 
& 0 & 0 & 0 & 0 & $\frac{988}{189}$ & $-\frac{1976}{189}$ & $-\frac{4196}{3969}$ & $-\frac{3212}{1323}$ & $-\frac{6904}{3969}$ & $\frac{776}{189}$
& $\frac{1964}{1323}$ & $\frac{6424}{1323}$ \\

$23$ 
& 0 & 0 & 0 & 0 & $-\frac{8}{21}$ & $\frac{16}{21}$ & $\frac{136}{441}$ & $-\frac{8}{147}$ & $\frac{8}{63}$ 
& $-\frac{16}{21}$ & $-\frac{16}{147}$ & $\frac{16}{147}$ \\

$24$ 
& 0 & 0 & 0 & 0 & $\frac{1784}{1323}$ & $-\frac{3568}{1323}$ & $\frac{956}{27783}$ & $-\frac{8164}{9261}$ & $-\frac{13184}{27783}$ & $\frac{3460}{9261}$
& $\frac{3364}{9261}$ & $\frac{15968}{9261}$ \\

$25$ 
& 0 & 0 & 0 & 0 & $-\frac{128}{441}$ & $\frac{256}{441}$ & $-\frac{1952}{9261}$ & $-\frac{32}{3087}$ & $\frac{128}{1323}$ & $-\frac{640}{3087}$
& $-\frac{256}{3087}$ & $-\frac{128}{441}$ \\

$26$ 
& 0 & 0 & 0 & 0 & $\frac{326}{2835}$ & $-\frac{652}{2835}$ & $\frac{14933}{59535}$ & $-\frac{2459}{3969}$ & $-\frac{44}{59535}$ & $-\frac{2423}{2835}$
& $\frac{367}{19845}$ & $\frac{22952}{19845}$ \\

$27$ 
& 0 & 0 & 0 & 0 & $-\frac{1922}{2835}$ & $\frac{3844}{2835}$ & $-\frac{10796}{59535}$ & $-\frac{74}{19845}$ & $\frac{284}{1215}$ & $-\frac{1984}{2835}$
& $-\frac{710}{3969}$ & $-\frac{172}{3969}$ \\

$28$ 
& 0 & 0 & 0 & 0 & $\frac{42796}{19845}$ & $-\frac{90632}{19845}$ & $-\frac{397151}{416745}$ & $-\frac{152303}{138915}$ & $-\frac{320116}{416745}$
& $\frac{33493}{27783}$ & $\frac{68867}{138915}$ & $\frac{55964}{27783}$ \\

$29$ 
& 0 & 0 & 0 & 0 & $-\frac{58952}{218295}$ & $\frac{24785}{43659}$ & $-\frac{4453916}{32089365}$ & $-\frac{288283}{2139291}$ & $\frac{557306}{4584195}$
& $-\frac{1360424}{2139291}$ & $-\frac{80921}{972405}$ & $\frac{41999}{305613}$ \\
\hline
\end{tabular}%
}
\end{table}

\begin{table}[t]
\caption{Coefficients for $s_{a b_2}^{\text{1L}}$. See also \reftab{sab21L_A}.}
\label{tab:sab21L_B}
\centering
\scriptsize
\begin{tabular}{l|cccccccc}
\hline
$s_{ab_2}^{\text{1L}}$
& $13$
& $14$
& $15$
& $16$
& $17$
& $18$
& $19$
& $20$ \\
\\ \hline \\
$1$ 
& 0 & 0 & 0 & 0 & 0 & $\frac{668}{56595}$ & $-\frac{668}{66885}$ & $\frac{668}{66885}$ \\
\\ \hline \\
$2$ 
& $-\frac{256}{2205}$ & $\frac{21314}{72765}$ & $-\frac{23651}{56595}$ & $\frac{100354}{509355}$ & $-\frac{338}{3465}$ 
& 0 & $\frac{2672}{18865}$ & $-\frac{2672}{18865}$ \\

$3$ 
& $-\frac{256}{2205}$ & $\frac{21314}{72765}$ & $-\frac{23651}{56595}$ & $\frac{100354}{509355}$ & $-\frac{338}{3465}$ 
& 0 & 0 & 0 \\
\\ \hline \\
$4$ 
& 0 & 0 & $-\frac{64}{147}$ & $\frac{96}{245}$ & 0 & 0 & 0 & 0 \\

$5$ 
& $-\frac{128}{147}$ & $\frac{108772}{19845}$ & $-\frac{174332}{46305}$ & $-\frac{349792}{138915}$ & $-\frac{64592}{46305}$ 
& 0 & 0 & 0 \\

$6$ 
& $-\frac{64}{49}$ & $\frac{54386}{6615}$ & $-\frac{83806}{15435}$ & $-\frac{183968}{46305}$ & $-\frac{32296}{15435}$ 
& 0 & 0 & 0 \\

$7$ 
& $-\frac{1216}{2205}$ & $\frac{36874}{6237}$ & $-\frac{279098}{101871}$ & $-\frac{874112}{218295}$ & $-\frac{506488}{509355}$ 
& 0 & $\frac{4008}{18865}$ & $-\frac{4008}{18865}$ \\
\\ \hline \\
$8$ 
& 0 & 0 & 0 & 0 & 0 & 0 & 0 & 0 \\

$9$ 
& 0 & $-\frac{8}{7}$ & $-\frac{104}{49}$ & $\frac{144}{49}$ & 0 & 0 & 0 & 0 \\

$10$ 
& 0 & 0 & $-\frac{320}{147}$ & $\frac{96}{49}$ & 0 & 0 & 0 & 0 \\

$11$ 
& 0 & $\frac{3952}{189}$ & $-\frac{1808}{441}$ & $-\frac{26560}{1323}$ & $\frac{64}{441}$ 
& 0 & 0 & 0 \\

$12$ 
& $-\frac{64}{147}$ & $\frac{90124}{19845}$ & $-\frac{193388}{46305}$ & $-\frac{55040}{27783}$ & $-\frac{4352}{6615}$ 
& 0 & 0 & 0 \\

$13$ 
& $-\frac{64}{147}$ & $\frac{49172}{19845}$ & $-\frac{31868}{9261}$ & $-\frac{26816}{138915}$ & $-\frac{26128}{46305}$ 
& 0 & 0 & 0 \\

$14$ 
& $-\frac{128}{2205}$ & $\frac{37112}{24255}$ & $-\frac{292676}{169785}$ & $-\frac{65048}{169785}$ & $-\frac{22984}{169785}$ 
& 0 & $\frac{4008}{18865}$ & $-\frac{4008}{18865}$ \\

$15$ 
& $-\frac{128}{147}$ & $\frac{259888}{19845}$ & $-\frac{254336}{46305}$ & $-\frac{294848}{27783}$ & $-\frac{83168}{46305}$ 
& 0 & 0 & 0 \\
\\ \hline \\
$16$ 
& 0 & 0 & 0 & 0 & 0 & 0 & 0 & 0 \\

$17$ 
& 0 & 0 & 0 & 0 & 0 & 0 & 0 & 0 \\

$18$ 
& 0 & 0 & 0 & 0 & 0 & 0 & 0 & 0 \\

$19$ 
& 0 & 0 & 0 & 0 & 0 & 0 & 0 & 0 \\

$20$ 
& 0 & 0 & 0 & 0 & 0 & 0 & 0 & 0 \\

$21$ 
& 0 & 0 & $-\frac{160}{147}$ & $\frac{48}{49}$ & 0 & 0 & 0 & 0 \\

$22$ 
& 0 & $\frac{988}{189}$ & $-\frac{452}{441}$ & $-\frac{6640}{1323}$ & $\frac{16}{441}$ 
& 0 & 0 & 0 \\

$23$ 
& 0 & $-\frac{8}{21}$ & $-\frac{104}{147}$ & $\frac{48}{49}$ & 0 & 0 & 0 & 0 \\

$24$ 
& 0 & $\frac{1928}{1323}$ & $-\frac{4792}{3087}$ & $-\frac{5888}{9261}$ & $\frac{992}{3087}$ 
& 0 & 0 & 0 \\

$25$ 
& 0 & $\frac{304}{441}$ & $-\frac{592}{343}$ & $\frac{608}{3087}$ & $\frac{640}{1029}$ 
& 0 & 0 & 0 \\

$26$ 
& 0 & $\frac{326}{2835}$ & $-\frac{5062}{6615}$ & $\frac{824}{3969}$ & $-\frac{1216}{6615}$ 
& 0 & 0 & 0 \\

$27$ 
& 0 & $-\frac{1922}{2835}$ & $-\frac{892}{1323}$ & $\frac{20876}{19845}$ & $\frac{4}{945}$ 
& 0 & 0 & 0 \\

$28$ 
& 0 & $\frac{65512}{19845}$ & $-\frac{81404}{46305}$ & $-\frac{80408}{27783}$ & $\frac{808}{46305}$
& 0 & 0 & 0 \\

$29$ 
& 0 & $-\frac{397}{891}$ & $-\frac{49901}{509355}$ & $\frac{133568}{218295}$ & $\frac{236}{101871}$ 
& 0 & $\frac{668}{18865}$ & $-\frac{668}{18865}$ \\
\hline
\end{tabular}%
\end{table}

\begin{table}[t]
\caption{Coefficients for $s_{a_2 b_2}^{\text{1L}}$. Continues in \reftab{sa2b21L_B}.}
\label{tab:sa2b21L_A}
\centering
\scriptsize
\hspace*{-2cm}
\resizebox{1.3\textwidth}{!}{%
\begin{tabular}{l|cccccccccccc}
\hline
$s_{a_2 b_2}^{\text{1L}}$
& $1$
& $2$
& $3$
& $4$
& $5$
& $6$
& $7$
& $8$
& $9$
& $10$
& $11$
& $12$ \\
\\ \hline \\
$1$ & 0 & 0 & 0 & 0 & 0 & 0 & 0 & 0 & 0 & 0 & 0 & 0 \\
\\ \hline \\
$2$ & $\frac{68}{21}$ & $\frac{254}{2205}$ & $\frac{2624}{735}$ & 0 & $-\frac{2644}{13475}$ & 0 & $\frac{141664}{101871}$ & $\frac{5856}{18865}$ 
& $\frac{457616}{2546775}$ & 0 & 0 & 0 \\

$3$ & $\frac{68}{21}$ & $\frac{254}{2205}$ & $\frac{2624}{735}$ & 0 & $-\frac{2644}{13475}$ & 0 & $\frac{141664}{101871}$ & $\frac{5856}{18865}$ 
& $\frac{457616}{2546775}$ & 0 & 0 & 0 \\

$4$ & $\frac{152}{105}$ & $\frac{2551}{10290}$ & $\frac{4391}{10290}$ & $\frac{25643}{15435}$ & $\frac{9161}{11025}$ & $-\frac{2030647}{848925}$ 
& $-\frac{273131}{35654850}$ & $-\frac{672839}{1080450}$ & $-\frac{2256}{13475}$ & $\frac{1713113}{1697850}$ & $\frac{36137}{154350}$ & $\frac{257893}{94325}$ \\
\\ \hline \\
$5$ & $\frac{41}{21}$ & $\frac{76}{63}$ & $\frac{268}{49}$ & 0 & $-\frac{19231}{17325}$ & 0 & $\frac{2489609}{509355}$ & $\frac{61232}{18865}$ 
& $\frac{3262052}{2546775}$ & 0 & 0 & 0 \\

$6$ & $\frac{37}{35}$ & $\frac{1082}{1715}$ & $\frac{2528}{5145}$ & $\frac{51074}{15435}$ & $\frac{1658}{2205}$ & $-\frac{30256}{11319}$ & $\frac{2521438}{3565485}$ 
& $-\frac{97666}{108045}$ & $\frac{86147}{509355}$ & $\frac{18038}{8085}$ & $\frac{1772}{2205}$ & $\frac{979637}{169785}$ \\

$7$ & 3 & 0 & $\frac{68}{7}$ & 0 & 0 & 0 & $\frac{2624}{245}$ & $\frac{254}{735}$ & 0 & 0 & 0 & 0 \\

$8$ & $\frac{5}{3}$ & $\frac{116}{105}$ & $\frac{176}{35}$ & 0 & $-\frac{50}{63}$ & 0 & $\frac{1754}{343}$ & $\frac{46258}{15435}$ & $\frac{1784}{1715}$ 
& 0 & 0 & 0 \\

$9$ & 1 & $\frac{58}{35}$ & $\frac{94}{35}$ & 0 & $-\frac{25}{21}$ & 0 & $\frac{3971}{1715}$ & $\frac{4448}{1029}$ & $\frac{2676}{1715}$ 
& 0 & 0 & 0 \\

$10$ & $\frac{5}{3}$ & 0 & $\frac{188}{105}$ & $\frac{152}{35}$ & $\frac{50}{63}$ & $-\frac{100}{63}$ & $\frac{57478}{46305}$ & $-\frac{1300}{3087}$ 
& $-\frac{50}{189}$ & $\frac{33028}{5145}$ & $\frac{754}{2205}$ & $\frac{28306}{15435}$ \\

$11$ & $\frac{7}{5}$ & $\frac{116}{245}$ & $\frac{8}{735}$ & $\frac{688}{147}$ & 0 & $-\frac{100}{147}$ & $\frac{15332}{46305}$ & $-\frac{344}{5145}$ 
& $\frac{14306}{46305}$ & $\frac{62348}{15435}$ & $\frac{3418}{5145}$ & $\frac{2690}{1029}$ \\

$12$ & 1 & $\frac{58}{105}$ & $\frac{12}{35}$ & $\frac{362}{105}$ & $\frac{25}{63}$ & $-\frac{100}{63}$ & $\frac{36767}{46305}$ & $-\frac{11278}{15435}$ 
& $\frac{9931}{46305}$ & $\frac{12596}{5145}$ & $\frac{6778}{15435}$ & $\frac{77689}{15435}$ \\
\\ \hline \\
$13$ 
& $\frac{16}{7}$ & $\frac{1}{147}$ & $-\frac{1489}{1470}$ & $\frac{2258}{2205}$ 
& $-\frac{64942}{51975}$ & $\frac{885908}{363825}$ & $\frac{10142777}{7640325}$
& $\frac{964598}{2546775}$ & $\frac{499024}{1091475}$ & $-\frac{6332852}{2546775}$ 
& $-\frac{1061138}{2546775}$ & $-\frac{247468}{363825}$ \\

$14$ 
& $-\frac{3}{7}$ & $\frac{5608}{15435}$ & $-\frac{18136}{5145}$ & $\frac{19952}{15435}$ 
& $-\frac{1621283}{363825}$ & $\frac{21219028}{2546775}$
& $\frac{6590692}{53482275}$ & $\frac{1704482}{1980825}$ & $\frac{44486}{24255}$ 
& $-\frac{50192}{7425}$ & $-\frac{3047414}{2546775}$ & $\frac{402062}{2546775}$ \\

$15$ 
& 1 & 0 & 0 & 0 
& $-\frac{64}{135}$ & $\frac{128}{135}$ & $\frac{7}{81}$ & $\frac{1658}{6615}$ 
& $\frac{64}{405}$ & $-\frac{64}{189}$ & $-\frac{128}{945}$ & $-\frac{64}{135}$ \\

$16$ 
& $\frac{1}{3}$ & 0 & $-\frac{298}{105}$ & $\frac{298}{105}$ 
& $-\frac{1013}{567}$ & $\frac{2026}{567}$ & $\frac{32024}{59535}$ & $\frac{50249}{138915}$
& $\frac{35047}{59535}$ & $-\frac{372734}{138915}$ & $-\frac{9587}{19845}$ 
& $-\frac{12347}{27783}$ \\

$17$ 
& 0 & $-\frac{29}{1470}$ & $\frac{3223}{1470}$ & $-\frac{914}{735}$ 
& $-\frac{2453}{1890}$ & $\frac{34717}{13230}$ & $\frac{83693}{555660}$
& $\frac{122399}{185220}$ & $\frac{12716}{27783}$ & $\frac{44347}{185220}$ 
& $-\frac{33863}{185220}$ & $-\frac{29521}{18522}$ \\
\\ \hline \\
$18$ 
& $\frac{8}{21}$ & $\frac{7901}{30870}$ & $\frac{21299}{10290}$ & $-\frac{5039}{3087}$ 
& $-\frac{572107}{282975}$ & $\frac{2219981}{848925}$
& $\frac{724798}{540225}$ & $\frac{12936131}{5942475}$ & $\frac{2360872}{2546775}$ 
& $-\frac{250204}{282975}$ & $-\frac{10069}{11025}$ & $-\frac{2619466}{848925}$ \\

$19$ 
& $\frac{34}{21}$ & $\frac{26}{45}$ & $\frac{1494}{245}$ & 0 
& $-\frac{3884}{5775}$ & 0 & $\frac{14510921}{3565485}$ & $\frac{3159446}{1188495}$
& $\frac{1419464}{2546775}$ & 0 & 0 & 0 \\

$20$ 
& $\frac{62}{105}$ & $\frac{82}{343}$ & $\frac{386}{343}$ & $\frac{7972}{3087}$ 
& $\frac{5882}{11025}$ & $-\frac{153676}{94325}$ & $\frac{11928974}{17827425}$
& $-\frac{82624}{540225}$ & $-\frac{10982}{282975}$ & $\frac{9434011}{5942475}$ 
& $\frac{692414}{540225}$ & $\frac{1986694}{848925}$ \\

\hline
\end{tabular}%
}
\end{table}

\begin{table}[t]
\caption{Coefficients for $s_{a_2 b_2}^{\text{1L}}$. See also \reftab{sa2b21L_A}.}
\label{tab:sa2b21L_B}
\centering
\scriptsize
\begin{tabular}{l|cccccccc}
\hline
$s_{a_2 b_2}^{\text{1L}}$
& $13$
& $14$
& $15$
& $16$
& $17$
& $18$
& $19$
& $20$ \\
\\ \hline \\
$1$ & 0 & 0 & 0 & 0 & 0 & 0 & 0 & 0 \\
\\ \hline \\
$2$ & 0 & 0 & 0 & 0 & 0 & 0 & 0 & 0 \\

$3$ & 0 & 0 & 0 & 0 & 0 & 0 & 0 & 0 \\

$4$ & 0 & 0 & 0 & 0 & 0 & 0 & 0 & 0 \\
\\ \hline \\
$5$ & 0 & 0 & 0 & 0 & 0 & 0 & 0 & 0 \\

$6$ & 0 & 0 & 0 & 0 & 0 & 0 & 0 & 0 \\

$7$ & 0 & 0 & 0 & 0 & 0 & 0 & 0 & 0 \\

$8$ & 0 & 0 & 0 & 0 & 0 & 0 & 0 & 0 \\

$9$ & 0 & 0 & 0 & 0 & 0 & 0 & 0 & 0 \\

$10$ & 0 & 0 & 0 & 0 & 0 & 0 & 0 & 0 \\

$11$ & 0 & 0 & 0 & 0 & 0 & 0 & 0 & 0 \\

$12$ & 0 & 0 & 0 & 0 & 0 & 0 & 0 & 0 \\
\\ \hline \\
$13$ 
& $\frac{4231}{735}$ & $-\frac{144344}{72765}$ & $\frac{174887}{169785}$ & $\frac{3120176}{509355}$ 
& $-\frac{652}{3087}$ & 0 & 0 & 0 \\

$14$ 
& $\frac{164}{35}$ & $-\frac{377644}{72765}$ & $\frac{50723}{11319}$ 
& $\frac{1322936}{101871}$ & $-\frac{4052}{15435}$ & 0 & 0 & 0 \\

$15$ 
& $\frac{164}{21}$ & $-\frac{64}{135}$ & $\frac{829}{63}$ 
& $\frac{16684}{6615}$ & $-\frac{8}{147}$ & 0 & 0 & 0 \\

$16$ 
& $\frac{148}{35}$ & $-\frac{1238}{567}$ 
& $\frac{90689}{46305}$ & $\frac{1716646}{138915}$ & $-\frac{37294}{46305}$ 
& 0 & 0 & 0 \\

$17$ 
& $-\frac{2158}{735}$ 
& $-\frac{8773}{6615}$ & $-\frac{4223}{15435}$ & $-\frac{175391}{46305}$ 
& $\frac{11143}{5145}$ & 0 & 0 & 0 \\
\\ \hline \\
$18$ 
& 0 & 0 & 0 & 0 & 0 & $\frac{26107}{15435}$ & 0 & 0 \\

$19$ 
& 0 & 0 & 0 & 0 & 0 & 0 & $\frac{26107}{15435}$ & 0 \\

$20$ 
& 0 & 0 & 0 & 0 & 0 & 0 & 0 & $\frac{26107}{15435}$ \\
\hline
\end{tabular}%
\end{table}

\begin{table}[t]
\centering
\caption{Coefficients for $d_{a\delta}^{(5),0}$, $\tilde d_{a\delta}^{(5),1}$, $d_{a\nabla^2\delta}^{(5),0}$ and $\tilde d_{a\nabla^2\delta}^{(5),i}$.}
\label{tab:da5}
\scriptsize
\setlength{\tabcolsep}{4pt}
\renewcommand{\arraystretch}{0.8}
\begin{tabular}{l|cccccc}
\hline
& $d_{a\delta}^{(5),0}$ & $\tilde d_{a\delta}^{(5),1}$ & $d_{a\nabla^2\delta}^{(5),0}$ & $\tilde d_{a\nabla^2\delta}^{(5),1}$ & $\tilde d_{a\nabla^2\delta}^{(5),2}$ & $\tilde d_{a\nabla^2\delta}^{(5),3}$ \\
\\ \hline \\

$1$ 
& 0 & 0 & 0 & $-\frac{1931}{859950}$ & $\frac{79}{171990}$ & $\frac{256}{143325}$ \\
\\ \hline \\
$2$ 
& $\frac{862}{1575}$ & $\frac{376}{6615}$ & $-\frac{123542}{5457375}$ & $-\frac{31967}{242550}$ & $\frac{68479}{363825}$ & $-\frac{2656}{72765}$ \\

$3$ 
& $\frac{862}{1575}$ & $\frac{376}{6615}$ & $-\frac{123542}{5457375}$ & $-\frac{337}{22050}$ & $\frac{1499}{33075}$ & $-\frac{736}{72765}$ \\
\\ \hline \\
$4$ 
& $\frac{70739}{33075}$ & $\frac{4}{105}$ & $-\frac{61}{3150}$ & 0 & 0 & 0 \\

$5$ 
& $\frac{2917}{2205}$ & $\frac{716}{1323}$ & $-\frac{1823}{22050}$ & $-\frac{988}{3675}$ & $\frac{4904}{6615}$ & $-\frac{7592}{33075}$ \\

$6$ 
& $\frac{30263}{33075}$ & $\frac{1748}{2205}$ & $-\frac{2521}{22050}$ & $-\frac{227}{735}$ & $\frac{11219}{11025}$ & $-\frac{3796}{11025}$ \\

$7$ 
& $\frac{134957}{99225}$ & $\frac{148}{441}$ & $-\frac{859763}{10914750}$ & $-\frac{145667}{363825}$ & $\frac{264941}{363825}$ & $-\frac{68188}{363825}$ \\
\\ \hline \\
$8$ 
& $\frac{272}{105}$ & 0 & 0 & 0 & 0 & 0 \\

$9$ 
& $\frac{82}{105}$ & $\frac{5}{21}$ & $-\frac{32}{315}$ & 0 & 0 & 0 \\

$10$ 
& $\frac{6352}{4725}$ & $\frac{4}{21}$ & $-\frac{32}{315}$ & 0 & 0 & 0 \\

$11$ 
& $\frac{592}{675}$ & $\frac{8}{63}$ & $-\frac{1216}{4725}$ & 0 & 0 & 0 \\

$12$ 
& $\frac{16112}{19845}$ & $\frac{3706}{6615}$ & $-\frac{1216}{11025}$ & $-\frac{454}{2205}$ & $\frac{22438}{33075}$ & $-\frac{7592}{33075}$ \\

$13$ 
& $\frac{18166}{14175}$ & $\frac{373}{735}$ & $-\frac{35104}{496125}$ & $-\frac{988}{3675}$ & $\frac{4904}{6615}$ & $-\frac{7592}{33075}$ \\

$14$ 
& $\frac{12814}{11025}$ & $\frac{401}{2205}$ & $-\frac{121952}{1819125}$ & $-\frac{19351}{69300}$ & $\frac{118103}{291060}$ & $-\frac{39148}{363825}$ \\

$15$ 
& $\frac{27784}{19845}$ & $\frac{232}{315}$ & $-\frac{1408}{14175}$ & $-\frac{4388}{6615}$ & $\frac{5884}{3675}$ & $-\frac{14944}{33075}$ \\
\\ \hline \\
$16$ 
& 1 & 0 & 0 & 0 & 0 & 0 \\

$17$ 
& $\frac{11}{45}$ & 0 & 0 & 0 & 0 & 0 \\

$18$ 
& $\frac{7}{15}$ & 0 & 0 & 0 & 0 & 0 \\

$19$ 
& $\frac{31}{225}$ & 0 & 0 & 0 & 0 & 0 \\

$20$ 
& $\frac{163}{675}$ & 0 & 0 & 0 & 0 & 0 \\

$21$ 
& $\frac{47}{105}$ & $\frac{2}{21}$ & $-\frac{16}{315}$ & 0 & 0 & 0 \\

$22$ 
& $\frac{173}{675}$ & $\frac{2}{63}$ & $-\frac{304}{4725}$ & 0 & 0 & 0 \\

$23$ 
& $\frac{391}{1575}$ & $\frac{5}{63}$ & $-\frac{32}{945}$ & 0 & 0 & 0 \\

$24$ 
& $\frac{197}{735}$ & $\frac{82}{441}$ & $-\frac{2048}{33075}$ & $-\frac{16}{441}$ & $\frac{64}{441}$ & $-\frac{16}{147}$ \\

$25$ 
& $\frac{5057}{11025}$ & $\frac{110}{441}$ & $-\frac{64}{1323}$ & $-\frac{16}{147}$ & $\frac{16}{49}$ & $-\frac{32}{147}$ \\

$26$ 
& $\frac{643}{2835}$ & $\frac{104}{945}$ & $-\frac{16}{1575}$ & $-\frac{1}{15}$ & $\frac{131}{675}$ & $-\frac{4}{675}$ \\

$27$ 
& $\frac{5477}{14175}$ & $\frac{4}{35}$ & $-\frac{656}{70875}$ & $-\frac{2}{25}$ & $\frac{28}{135}$ & $-\frac{4}{675}$ \\

$28$ 
& $\frac{8629}{19845}$ & $\frac{491}{2205}$ & $-\frac{3664}{99225}$ & $-\frac{1397}{6615}$ & $\frac{5713}{11025}$ & $-\frac{6136}{33075}$ \\

$29$ 
& $\frac{33403}{198450}$ & $\frac{74}{6615}$ & $-\frac{11936}{1819125}$ & $-\frac{15593}{485100}$ & $\frac{35347}{1455300}$ & $\frac{4006}{363825}$ \\

\hline
\end{tabular}%
\end{table}

We present the coefficients that renormalize pairs of operators, leading to stochastic contributions, in \reftab{sabtrPi11L}, \reftab{sabtrPi1Pi11L} and \reftab{sabtrPi1trPi11L}. For triples of operators, see \reftab{sabtrPi1Pi1delta1L} and \reftab{sabtrPi1trPi1delta1L}. Note that part of these coefficients was already determined in \cite{Rubira:2024tea}, and our computations agree with those presented there up to $s^\text{1L}_{\delta^2\delta^2\delta}$ (the renormalization of the pair $\delta^2,\delta^2$ sourcing $\delta$) and $s^\text{1L}_{\delta^2\delta^2\delta^2\delta}$ (the renormalization of the triple $\delta^2,\delta^2,\delta^2$ sourcing $\delta$). For those, we checked that the difference originates from a missing term in \cite{Rubira:2024tea} from expanding the power spectra when changing variables.

\begin{table}[t]
\centering
\caption{Coefficients for $s_{ab\,\delta}^{\text{1L}}$.}
\label{tab:sabtrPi11L}
\scriptsize
\setlength{\tabcolsep}{4pt}
\renewcommand{\arraystretch}{1.15}
\begin{tabular}{l|ccccccc}
\hline
$s_{ab\,\delta}^{\text{1L}}$
& $1$ & $2$ & $3$
& $4$ & $5$
& $6$ & $7$ \\
\\ \hline \\
$1$ & 0 & 0 & 0 & 0 & 0 & 0 & 0 \\
\\ \hline \\
$2$ & 0 & $\frac{230}{21}$ & $\frac{230}{21}$ & 6 & $\frac{10}{3}$ & 2 & $\frac{82}{21}$ \\
$3$ & 0 & $\frac{230}{21}$ & $\frac{230}{21}$ & 6 & $\frac{10}{3}$ & 2 & $\frac{82}{21}$ \\
\\ \hline \\
$4$ & 0 & 6 & 6 & 0 & 0 & 0 & 0 \\
$5$ & 0 & $\frac{10}{3}$ & $\frac{10}{3}$ & 0 & 0 & 0 & 0 \\
$6$ & 0 & 2 & 2 & 0 & 0 & 0 & 0 \\
$7$ & 0 & $\frac{82}{21}$ & $\frac{82}{21}$ & 0 & 0 & 0 & 0 \\
\\ \hline \\
$8$ & 0 & 0 & 0 &  &  &  &  \\
$9$ & 0 & 0 & 0 &  &  &  &  \\
$10$ & 0 & 0 & 0 &  &  &  &  \\
$11$ & 0 & 0 & 0 &  &  &  &  \\
$12$ & 0 & 0 & 0 &  &  &  &  \\
$13$ & 0 & 0 & 0 &  &  &  &  \\
$14$ & 0 & 0 & 0 &  &  &  &  \\
$15$ & 0 & 0 & 0 &  &  &  &  \\
\hline
\end{tabular}%
\end{table}

\begin{table}[t]
\centering
\caption{Coefficients for $s_{ab\,{\rm tr}[ \Pi^{[1]}\Pi^{[1]}]}^{\text{1L}}$.}
\label{tab:sabtrPi1Pi11L}
\scriptsize
\setlength{\tabcolsep}{4pt}
\renewcommand{\arraystretch}{1.15}
\begin{tabular}{l|ccccccc}
\hline
$s_{ab\,{\rm tr}[ \Pi^{[1]}\Pi^{[1]}]}^{\text{1L}}$
& $1$ & $2$ & $3$
& $4$ & $5$
& $6$ & $7$ \\
\\ \hline \\
$1$ & 0 & 0 & 0 & 0 & 0 & 0 & 0 \\
\\ \hline \\
$2$ & 0 & $\frac{1784}{2205}$ & $\frac{1784}{2205}$ & 0 & $\frac{296}{105}$ & $\frac{148}{35}$ & $\frac{6376}{2205}$ \\
$3$ & 0 & $\frac{1784}{2205}$ & $\frac{1784}{2205}$ & 0 & $\frac{296}{105}$ & $\frac{148}{35}$ & $\frac{6376}{2205}$ \\
\\ \hline \\
$4$ & 0 & 0 & 0 & 0 & 0 & 0 & 0 \\
$5$ & 0 & $\frac{296}{105}$ & $\frac{296}{105}$ & 0 & $\frac{16}{15}$ & $\frac{8}{5}$ & $\frac{88}{105}$ \\
$6$ & 0 & $\frac{148}{35}$ & $\frac{148}{35}$ & 0 & $\frac{8}{5}$ & $\frac{12}{5}$ & $\frac{44}{35}$ \\
$7$ & 0 & $\frac{6376}{2205}$ & $\frac{6376}{2205}$ & 0 & $\frac{88}{105}$ & $\frac{44}{35}$ & $\frac{484}{735}$ \\
\\ \hline \\
$8$ & 0 & 0 & 0 &  &  &  &  \\
$9$ & 0 & 2 & 2 &  &  &  &  \\
$10$ & 0 & 2 & 2 &  &  &  &  \\
$11$ & 0 & $\frac{76}{15}$ & $\frac{76}{15}$ &  &  &  &  \\
$12$ & 0 & $\frac{206}{105}$ & $\frac{206}{105}$ &  &  &  &  \\
$13$ & 0 & $\frac{8}{7}$ & $\frac{8}{7}$ &  &  &  &  \\
$14$ & 0 & $\frac{536}{315}$ & $\frac{536}{315}$ &  &  &  &  \\
$15$ & 0 & $\frac{932}{735}$ & $\frac{932}{735}$ &  &  &  &  \\
\hline
\end{tabular}%
\end{table}

\begin{table}[t]
\centering
\caption{Coefficients for $s_{ab\,{\rm tr}[ \Pi^{[1]}]{\rm tr}[ \Pi^{[1]}]}^{\text{1L}}$.}
\label{tab:sabtrPi1trPi11L}
\scriptsize
\setlength{\tabcolsep}{4pt}
\renewcommand{\arraystretch}{1.15}
\begin{tabular}{l|ccccccc}
\hline
$s_{ab\,{\rm tr}[ \Pi^{[1]}]{\rm tr}[ \Pi^{[1]}]}^{\text{1L}}$
& $1$ & $2$ & $3$
& $4$ & $5$
& $6$ & $7$ \\
\\ \hline \\
$1$ & 0 & 0 & 0 & 0 & 0 & 0 & 0 \\
\\ \hline \\
$2$ & 0 & $\frac{17774}{735}$ & $\frac{17774}{735}$ & $\frac{230}{7}$ & $\frac{606}{35}$ & $\frac{334}{35}$ & $\frac{4782}{245}$ \\
$3$ & 0 & $\frac{17774}{735}$ & $\frac{17774}{735}$ & $\frac{230}{7}$ & $\frac{606}{35}$ & $\frac{334}{35}$ & $\frac{4782}{245}$ \\
\\ \hline \\
$4$ & 0 & $\frac{230}{7}$ & $\frac{230}{7}$ & 18 & 10 & 6 & $\frac{82}{7}$ \\
$5$ & 0 & $\frac{606}{35}$ & $\frac{606}{35}$ & 10 & $\frac{26}{5}$ & $\frac{14}{5}$ & $\frac{218}{35}$ \\
$6$ & 0 & $\frac{334}{35}$ & $\frac{334}{35}$ & 6 & $\frac{14}{5}$ & $\frac{6}{5}$ & $\frac{122}{35}$ \\
$7$ & 0 & $\frac{4782}{245}$ & $\frac{4782}{245}$ & $\frac{82}{7}$ & $\frac{218}{35}$ & $\frac{122}{35}$ & $\frac{1814}{245}$ \\
\\ \hline \\
$8$ & 0 & 12 & 12 &  &  &  &  \\
$9$ & 0 & 2 & 2 &  &  &  &  \\
$10$ & 0 & $\frac{14}{3}$ & $\frac{14}{3}$ &  &  &  &  \\
$11$ & 0 & $\frac{8}{15}$ & $\frac{8}{15}$ &  &  &  &  \\
$12$ & 0 & $\frac{34}{15}$ & $\frac{34}{15}$ &  &  &  &  \\
$13$ & 0 & $\frac{36}{7}$ & $\frac{36}{7}$ &  &  &  &  \\
$14$ & 0 & $\frac{68}{15}$ & $\frac{68}{15}$ &  &  &  &  \\
$15$ & 0 & $\frac{4336}{735}$ & $\frac{4336}{735}$ &  &  &  &  \\
\hline
\end{tabular}%
\end{table}

\begin{table}[t]
\centering
\caption{Coefficients for $s_{ab\,{\rm tr}[ \Pi^{[1]}\Pi^{[1]}]\,\delta}^{\text{1L}}$.}
\label{tab:sabtrPi1Pi1delta1L}
\scriptsize
\setlength{\tabcolsep}{4pt}
\renewcommand{\arraystretch}{1.15}
\begin{tabular}{l|ccccccc}
\hline
$s_{ab\,{\rm tr}[ \Pi^{[1]}\Pi^{[1]}]\,\delta}^{\text{1L}}$
& $1$ & $2$ & $3$
& $4$ & $5$
& $6$ & $7$ \\
\\ \hline \\
$1$ & 0 & 0 & 0 & 0 & 0 & 0 & 0 \\
\\ \hline \\
$2$ & 0 & $\frac{432}{7}$ & $\frac{432}{7}$ & 24 & $\frac{40}{3}$ & 8 & $\frac{328}{21}$ \\
$3$ & 0 & $\frac{432}{7}$ & $\frac{432}{7}$ & 24 & $\frac{40}{3}$ & 8 & $\frac{328}{21}$ \\
\hline
\end{tabular}%
\end{table}

\begin{table}[t]
\centering
\caption{Coefficients for $s_{ab\,{\rm tr}[ \Pi^{[1]}]{\rm tr}[ \Pi^{[1]}]\,\delta}^{\text{1L}}$.}
\label{tab:sabtrPi1trPi1delta1L}
\scriptsize
\setlength{\tabcolsep}{4pt}
\renewcommand{\arraystretch}{1.15}
\begin{tabular}{l|ccccccc}
\hline
$s_{ab\,{\rm tr}[ \Pi^{[1]}]{\rm tr}[ \Pi^{[1]}]\,\delta}^{\text{1L}}$
& $1$ & $2$ & $3$
& $4$ & $5$
& $6$ & $7$ \\
\\ \hline \\
$1$ & 0 & 0 & 0 & 0 & 0 & 0 & 0 \\
\\ \hline \\
$2$ & 0 & $\frac{432}{7}$ & $\frac{432}{7}$ & 24 & $\frac{40}{3}$ & 8 & $\frac{328}{21}$ \\
$3$ & 0 & $\frac{432}{7}$ & $\frac{432}{7}$ & 24 & $\frac{40}{3}$ & 8 & $\frac{328}{21}$ \\
\hline
\end{tabular}%
\end{table}

\end{appendix}

\bibliographystyle{JHEP}
\bibliography{ref}

\end{document}